\documentclass[epjST]{svjour}
\usepackage{graphics}
\usepackage{graphicx}
\usepackage{hyperref}
\usepackage{amsmath}
\begin{document}
\title{Approximate analytical description of the high latitude extinction}
\author{Alexei Nekrasov\inst{1, 2} \and Kirill Grishin\inst{1,2} \and Dana Kovaleva\inst{3}
 \and Oleg Malkov\inst{3}\fnmsep\thanks{\email{malkov@inasan.ru}}}
\institute{Physics Faculty, Moscow State University, Moscow 119992, Russia
 \and Sternberg Astronomical Institute, M.V.Lomonosov Moscow State University, Universitetsky prospect 13, Moscow, 119992, Russia
 \and Institute of Astronomy of the Russian Acad. Sci., Moscow 119017, Russia
 }
%
\abstract{
The distribution of visual interstellar extinction $A_V$ has been mapped in selected
areas over the Northern sky, using available LAMOST DR5 and Gaia DR2/EDR3 data.
$A_V$ was modelled as a barometric function of galactic
latitude and distance. The function parameters were then approximated
by spherical harmonics. The resulting analytical tridimensional model
of the interstellar extinction can be used to predict
$A_V$ values for stars with known parallaxes,
as well as the total Galactic extinction in a given location in the sky.
} 
\maketitle
\def\apj{Astrophys.~J. }
\def\aatr{Astron.~Astroph.~Trans. }
\def\aaps{Astron.~and Astrophys.~Suppl.~Ser. }
\def\pasp{Publ.~Astron.~Soc.~Pac. }
\def\gca{Geochim.~Cosmochim.~Acta. }
\def\aap{Astron.~Astrophys. }
\def\aspcs{ASP~Conf.~Ser. }
\def\asrep{Astron.~Rep. }
\def\azh{Astron.~Zh. }
\def\nat{Nature. }
\def\apjl{Astrophys.~J.~Lett. }
\def\apjs{Astrophys.~J.~Suppl.~Ser. }
\def\aj{Astron.~J. }
\def\mnras{Mon.~Not.~R.~Astron.~Soc. }
\def\araa{Ann.~Rev.~Astron.~Astrophys. }
\def\jcp{J.~Chem.~Phys. }
\def\apss{Astrophys.~Space.~Sci. }
\def\prl{Phys.~Rev.~Lett. }
\def\phrva{Phys.~Rev.~A. }
\def\phlb{Phys.~Let.~B. }
\def\pf{Phys.~Fluids. }
\def\jgr{J.~Geophys.~Res. }
\def\cemda{Celest.~Mech.~Dyn.~Astr. }
\def\jcoph{J.~Comp.~Phys. }
\def\cophc{Comput.~Phys.~Commun. }
\def\phpl{Physics~of~Plasmas. }
\def\pasj{Publ.~Astron.~Soc.~Jpn. }
\def\jrasc{J.~R.~Astron.~Soc.~Can. }
\def\cemec{Celest.~Mech. }
\def\pasau{Proc.~Astron.~Soc.~Aust. }
\def\puasau{Publ.~Astron.~Soc.~Aust. }
\def\jasa{J.~Acoust.~Soc.~Am. }
\def\jfm{J.~Fluid~Mech. }
\def\cajph{Can.~J.~Phys. }
\def\mitag{Mitt.~Astron.~Ges. }
\def\bain{Bull.~Astron.~Inst.~Neth. }
\def\epsl{Earth~Planet.~Sci.~Lett. }
\def\ibvs{Inf.~Bull.~Variable~Stars. }
\def\arep{Astr.~Rep. }
\def\phr{Phys.~Rep. }
\def\astl{Astron.~Letters. }
\def\sci{Science. }
\def\jqsrt{J.~Quant.~Spectrosc.~Radiat.~Transfer. }
\def\emp{Earth,~Moon~and~Planets. }
\def\icar{Icarus. }
\def\pss{Planet.~Space~Sci. }
\def\qjras{Q.~J.~R.~Astron.~Soc. }
\def\nimpa{Nucl.~Instrum.~Methods~Phys.~Res.,~Sect.~A. }
\def\soph{Sol.~Phys. }
\def\lnm{Lect.~Notes~in~Math. }
\def\an{Astron.~Nach. }
\def\aph{Astroparticle~Physics. }
\def\adspr{Adv.~Space~Res. }
\def\geoj{Geophys.~J. }
\def\bamass{Bull.~Am.~Astron.~Soc. }
\def\rmxaa{Rev.~Mex.~Astron.~Astrofis. }
\def\aapr{Astron.~Astrophys.~Rev. }
\def\acp{Atmosphere~Chem.~Phys. }
\def\ssrv{Space~Sci.~Rev. }
\def\jmph{J.~Math.~Phys. }
\def\rvmps{Rev.~Mod.~Phys.~Suppl. }
\def\rvmp{Rev.~Mod.~Phys. }
\def\prd{Phys.~Rev.~D. }
\def\nuphs{Nuc.~Phys.~B~Proc.~Suppl. }
\def\nuphb{Nuc.~Phys.~B. }
\def\skytel{Sky~Telesc. }

\section{Introduction}
\label{sec:intro}

The solution of many problems in astrophysics and stellar astronomy
is associated with the study of interstellar extinction.
In particular, the distribution of the interstellar matter in the solar vicinity
is of great interest since it may be related to phenomena on larger scales than
just a dimension of individual clouds. For extinction maps based on photometry and
spectroscopy of individual stars each sight line provides an upper distance limit for the dust.
Parenago~\cite{Parenago1940} suggested a formula relating visual extinction $A_V$ to distance $d$
and Galactic latitude $b$, based on a barometric (exponential) function. The classical model of a homogeneous,
semi-infinite absorbing layer with a density that is exponentially
distributed with height renders the so-called cosecant extinction law:
\begin{equation}
A_V (b, d) = \frac{a_0 \beta}{\sin \left|b\right|} \left( 1 - e^{-\frac{d \sin \left|b\right|}{\beta}} \right)\!.
\label{equ:parenago}
\end{equation}
The parameter $\beta$ is the
scale height, and $a_0$ is the extinction per unit length in the Galactic plane.

A direct method of determining interstellar extinction is the measurement
of color indices and the derivation of color excesses.
Color index is the difference in magnitudes found with two different filters.
Color excess is defined as the difference between the observed color index and the
intrinsic color index, the hypothetical true color index of the star, unaffected by extinction.
The distribution of colour excess and interstellar reddening material
in the solar neighbourhood has been studied by many authors.
According to some models, the parameters $a_0$ and $\beta$ have fixed values, 
while other authors propose more complex schemes where $a_0$ and $\beta$
values depend on the direction on the sky.
Also, there were a number of investigations, where 
the authors used for describing the distance dependence 
by piecewise linear or quadratic distance functions 
that are different for each area instead of the cosecant law.

Different values for the $a_0$ and $\beta$ parameters were proposed in different studies
(\cite{1976RC2...C......0D}, \cite{1972ApJ...178....1S}, \cite{1978ppim.book.....S},
\cite{1945PA.....53..441P}, \cite{1963AZh....40..900S}).
All of these extinction models were constructed for the entire sky
and used the cosecant law.
The value $0.^m15$ was proposed for $a_0$ in \cite{1976RC2...C......0D}
while \cite{1972ApJ...178....1S}
suggested
\mbox{$a_0=0.^m165(1.192-tg(b))$}
for $|b|<50^\circ$ and zero for higher latitudes.
For the $\beta$ scale, a value of 100 ps was proposed in \cite{1978ppim.book.....S}.
Pandey \& Mahra \cite{1987MNRAS.226..635P} in their comprehensive study of open clusters
obtained for V-band the average values
$a_0=1.35 \pm 0.12$ mag per kpc and $\beta=0.16 \pm 0.02$ kpc.
The latter value can be compared with the interstellar matter scale height value 125 (+17 -7) pc
found in \cite{2006A&A...453..635M}.
Attempts were made in \cite{1945PA.....53..441P} and, later, in \cite{1963AZh....40..900S}
to determine the parameters $a_0$ and $\beta$ for different regions, rather than for the entire sky as a whole.
Sharov \cite{1963AZh....40..900S} proposed a map of interstellar extinction, in which the values of both coefficients
were given for each of the 118 areas.
Similar studies were undertaken later
(\cite{1968AJ.....73..983F}, \cite{1980A&AS...42..251N}, \cite{1987MNRAS.226..635P}).
The difference of them from the previous ones was that here the authors
abandoned the cosecant law and proposed the dependence on distance
in the form of piecewise linear functions.
Also, these maps cover only low latitude areas.
Finally, the model constructed by Arenou et al.~\cite{1992A&A...258..104A} uses a quadratic distance function.
These studies were reviewed in \cite{2002Ap&SS.280..115M} and \cite{1997ARep...41...10K}.
Other published models, using spectral and photometric data,
were based on $10^4$-$10^5$ stars, or were constructed
for a very limited area in the sky (see, e.g., \cite{1978A&A....64..367L},
\cite{2003A&A...409..205D}, \cite{2014A&A...571A..11P}, \cite{2014MNRAS.443.2907S},
\cite{2015ApJ...810...25G}, \cite{2018A&A...616A.132L}).
The most recent studies are \cite{2019ApJ...887...93G}
who constructed a 3D dust map covering the sky north of a declination
of $\delta = -30^{\circ}$, out to a distance of a few kiloparsecs,
and based on Gaia DR2, Pan-STARRS~1, and 2MASS data; and
\cite{2019MNRAS.483.4277C} who presented a new three-dimensional
interstellar dust reddening map of the Galactic plane based on
Gaia DR2, 2MASS and WISE.

The important step in the construction of the extragalactic distance scale
is the estimation of the galactic extinction of extragalactic objects
in order to make proper allowance for the dimming of the primary distance
indicators in external galaxies by interstellar dust in our own Galaxy.
The total Galactic visual extinction (hereafter Galactic extinction, $A_{Gal}$)
is estimated from the Galactic dust reddening for a line of sight,
assuming a standard extinction law.
The reddening estimates were published by~\cite{1998ApJ...500..525S},
who combined results of IRAS and COBE/DIRBE, while
\cite{2011ApJ...737..103S} (hereafter SF2011)
provide new estimates of Galactic dust extinction from an analysis of the Sloan Digital Sky Survey.
Obviously, Galactic extinction can be derived from Parenago formula~\eqref{equ:parenago}
under the assumption that $d \to \infty$:
\begin{equation}
A_{Gal} (b) = \frac{a_0 \beta}{\sin \left|b\right|}.
\label{equ:agal}
\end{equation}

In the past few years, new data have become available so that the total
number of stars for which extinction and distance values can be derived
has increased to more than five million. This enables us to achieve
a considerable improvement in the extinction analysis and construct a simple approximation formula for quick estimation of interstellar extinction for both galactic and extragalactic studies.

The first main goal of the present study is the construction of a software
for the approximation of $A_V(d)$-relations for various areas in the sky,
and its application to a set of selected areas.
We assume that Parenago formula~\eqref{equ:parenago}
satisfactory reproduces the observed V-band interstellar extinction $A_V$
for high galactic latitudes. Once the $a_0, \beta$ parameters are determined
for each area, we construct analytical expressions for their
estimation over the entire sky.
Note that our results are applied to optical extinction. The extinction
is more uncertain at shorter wavelengths and can be described by the so-called
interstellar extinction law, IEL
(\cite{1989ApJ...345..245C},
\cite{1994ApJ...437..262O},
\cite{1994A&AS..105..311F},
\cite{2005ApJ...623..897L},
\cite{2007ApJ...663..320F},
\cite{2009ApJ...705.1320G}).

In this paper, we present results for high galactic latitudes, $|b|>20^{\circ}$.
Our model gives the general trend of the interstellar extinction
and does not take into account the local irregularities of the
absorbing material that concentrate toward the Galactic plane.
The data for the lower latitudes will be published in subsequent papers.

Section~\ref{sec:data} describes the selection of areas we use for the study
and the procedure of the determination of interstellar extinction we applied.
In Section~\ref{sec:1D} results of the construction of $A_V(d)$-relations
for the areas are presented.
Section~\ref{sec:2D} contains results of 2D-approximation of
extinction parameters
over the entire sky (except for the low-latitude regions)
and discussion of their possible physical implications.
In Section~\ref{sec:plans} our future plans are discussed, and
Section~\ref{sec:concl} gives the conclusion of the study.

\section{Observational data}
\label{sec:data}

The present study used the stars contained in
LAMOST DR5 \cite{2019yCat.5164....0L}
and
Gaia \cite{2016A&A...595A...1G}
DR2 \cite{2018A&A...616A...1G}
and EDR3 \cite{gaiacollaboration2020gaia}
surveys.
To determine the interstellar extinction for specific targets
we made a cross-identification of LAMOST and Gaia objects
in forty selected areas with $\left|b\right|>20^{\circ}$ evenly distributed over the sky.
We have used a radius matching of 1 arcsec (note that the identification peak is achieved at 0.1 arcsec).

The HEALPix (https://healpix.jpl.nasa.gov/) order 7 subdivision of the celestial sphere
was used, and, consequently, each resulting area is about 0.21 deg$^2$.
The resulting areas were indexed according to the Hierarchical Triangular mesh,
HTM (http://www.skyserver.org/htm/).
The scale of areas (respective to the order of subdivision) is selected so as to guarantee a sufficient number
of stars (dozens) in each area, but sufficiently small to minimize significant variations of extinction within 
an area.

\begin{figure}
\resizebox{0.75\columnwidth}{!}{%
  \includegraphics{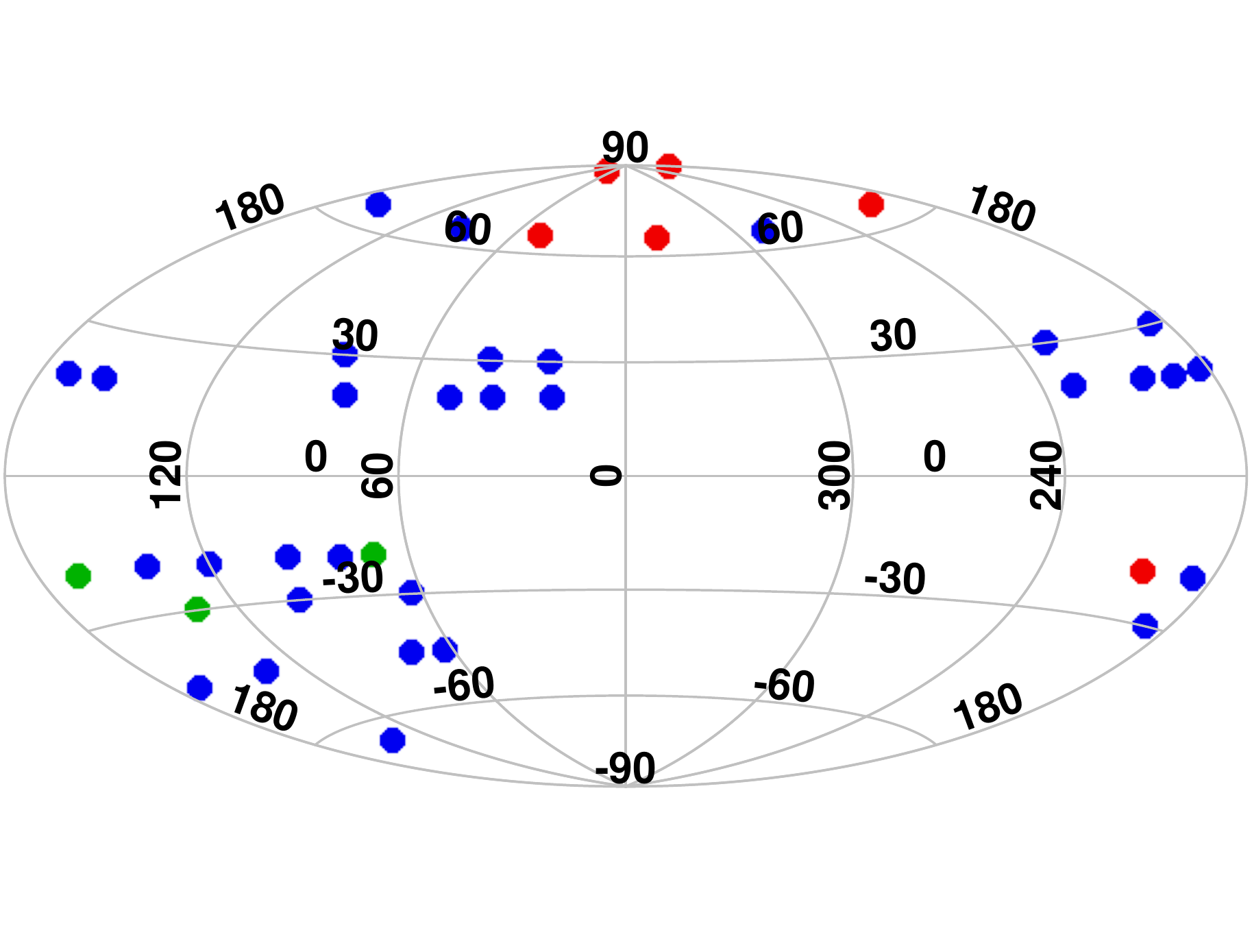} }
\caption{Selected areas. Galactic coordinates, Aitoff projection.
Our solutions for three ``green'' areas are shown in Figs.~\ref{fig:AN} and ~\ref{fig:KG}.
For six ``red'' areas no reasonable solution could be found, see discussion in
Section~\ref{sec:compare}.}
\label{fig:aitoff}       
\end{figure}

It should be mentioned that
Gaia is an all-sky survey, whereas the majority of the LAMOST survey data
are obtained in a declination range $-5^{\circ} < \delta < +60^{\circ}$.
The selected areas are
plotted in Fig.~\ref{fig:aitoff}.

The visual extinction $A_V$ for each star
have been computed using the following relation:
\begin{equation}
A_V = c_1/c_2 \times [(BP-RP)_0 - (BP-RP)].
\label{equ:Av}
\end{equation}
Here $c_2 \equiv A_G/A_V$, where $A_G$ is the interstellar extinction in the Gaia G-band.
Bono et al.~\cite{2019ApJ...870..115B} give 0.840 for $c_2$.
We use mean value of $c_1 \equiv A_G/E_{BP-RP}$ equal to 2.02.
This value is calculated for G2V star using extinction curve by~\cite{1989ApJ...345..245C}
with $R_V \equiv A_V/E_{B-V} =3.1$. Passbands characteristics were taken from~\cite{2010A&A...523A..48J}.
Distances to stars $d$ were taken from \cite{2018AJ....156...58B}.

 The broad Gaia G passband covers the range [330, 1050] nm, and its definition is optimized
 to collect the maximum of light for the astrometric measurements. The Gaia BP and RP photometry
 are derived from the integration of the blue and red photometer (BP and RP) low-resolution spectra
 covering the ranges [330, 680] nm and [630, 1050] nm \cite{2018A&A...616A...4E}.

In effect, the absorption coefficient $A_G/A_V$ is not a constant but rather a function of monochromatic 
absorption $A_0$ and temperature $T_{\rm eff}$, due to large width of Gaia G band (e.g.
\cite{2018A&A...614A..19D},
\cite{2018A&A...616A..10G},
\cite{2011MNRAS.411..435B},
\cite{2010A&A...523A..48J}). 
Here we, however, neglect this effect which may be estimated as follows. Using parameters from
\cite{2018A&A...616A..10G}, Table 1 to derive Gaia extinction coefficients as a function of colour 
and extinction, we obtain for the complete used data sample mean
$\langle k_G \rangle =A_G/A_0 \approx 0.82 \pm 0.04$.
In Eq.~\eqref{equ:Av}
$A_G$, however, collapses, leaving just $E(BP-RP)$ in the denominator. The colour excess
$E(BP-RP)$ demonstrates a certain relation with spectral class and absorption. The coefficient $c_1 \approx 2$ in 
Eq.~\eqref{equ:Av} becomes invalid for large absorption and for the stars later than K6 (see
Jordi et al. 2010 \cite{2010A&A...523A..48J}, Fig.~17 and calibration tables by Mamajek
\url{http://www.pas.rochester.edu/~emamajek/EEM_dwarf_UBVIJHK_colors_Teff.txt},
see also~\cite{2013ApJS..208....9P}), and actual values of $c_1$ are lower.
Not taking this effect into account may lead to overestimation of $A_V$.
It will be a subject of future investigation.

Only MS-stars with
Gaia (BP, RP) photometry
were used for the analysis.
Stars where DR2 and EDR3 parallaxes differ from each other by more than 0.2 mas, were dropped.
As a result, the areas each contain 19 to 228 stars with known $d$ and $A_V$ values,
altogether 3159 stars.
Luminosity class was estimated from LAMOST atmospheric parameters:
only stars with a gravity value of $\log g \ge 4$ were selected.
Intrinsic color index $(BP-RP)_0$ for MS-stars was estimated from
$T_{\rm eff}$(LAMOST) with Mamajek's relations
\url{http://www.pas.rochester.edu/~emamajek/EEM_dwarf_UBVIJHK_colors_Teff.txt},
see also~\cite{2013ApJS..208....9P}.

\section{Approximation for interstellar extinction in the selected areas}
\label{sec:1D}

We assume that
Parenago formula~\eqref{equ:parenago} satisfactorily
reproduces the observed interstellar extinction for relatively
high galactic latitudes.
To estimate $a_0, \beta$ parameters in Eq.~\eqref{equ:parenago} for each area,
we have developed two independent procedures, described below.

For several objects, interstellar extinction values
calculated according to Eq.~\eqref{equ:Av} are negative.
Nevertheless, they have been taken into account in the calculations.

\begin{figure}[h]
\resizebox{1.0\columnwidth}{!}{%
  \includegraphics{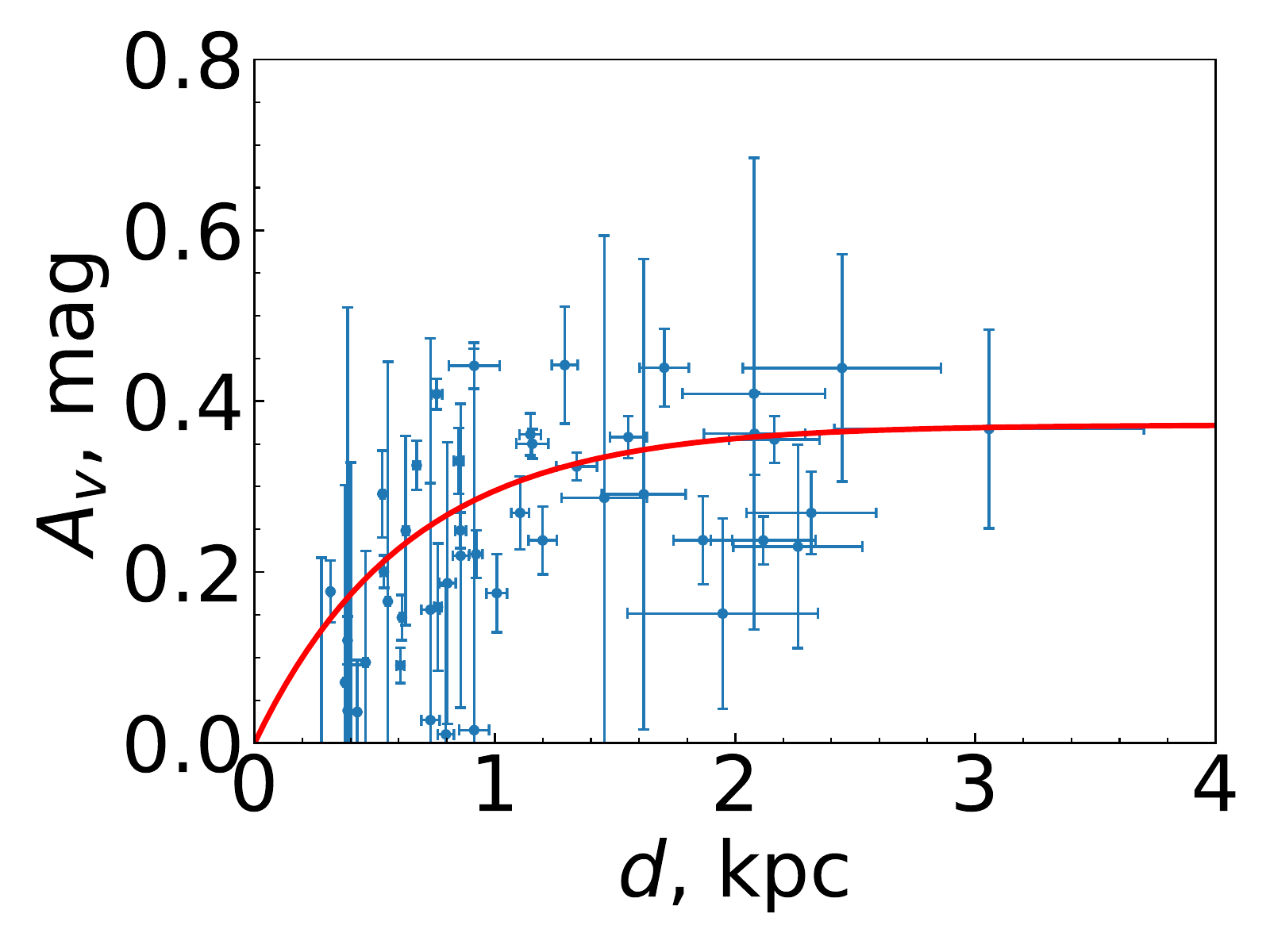}
  \includegraphics{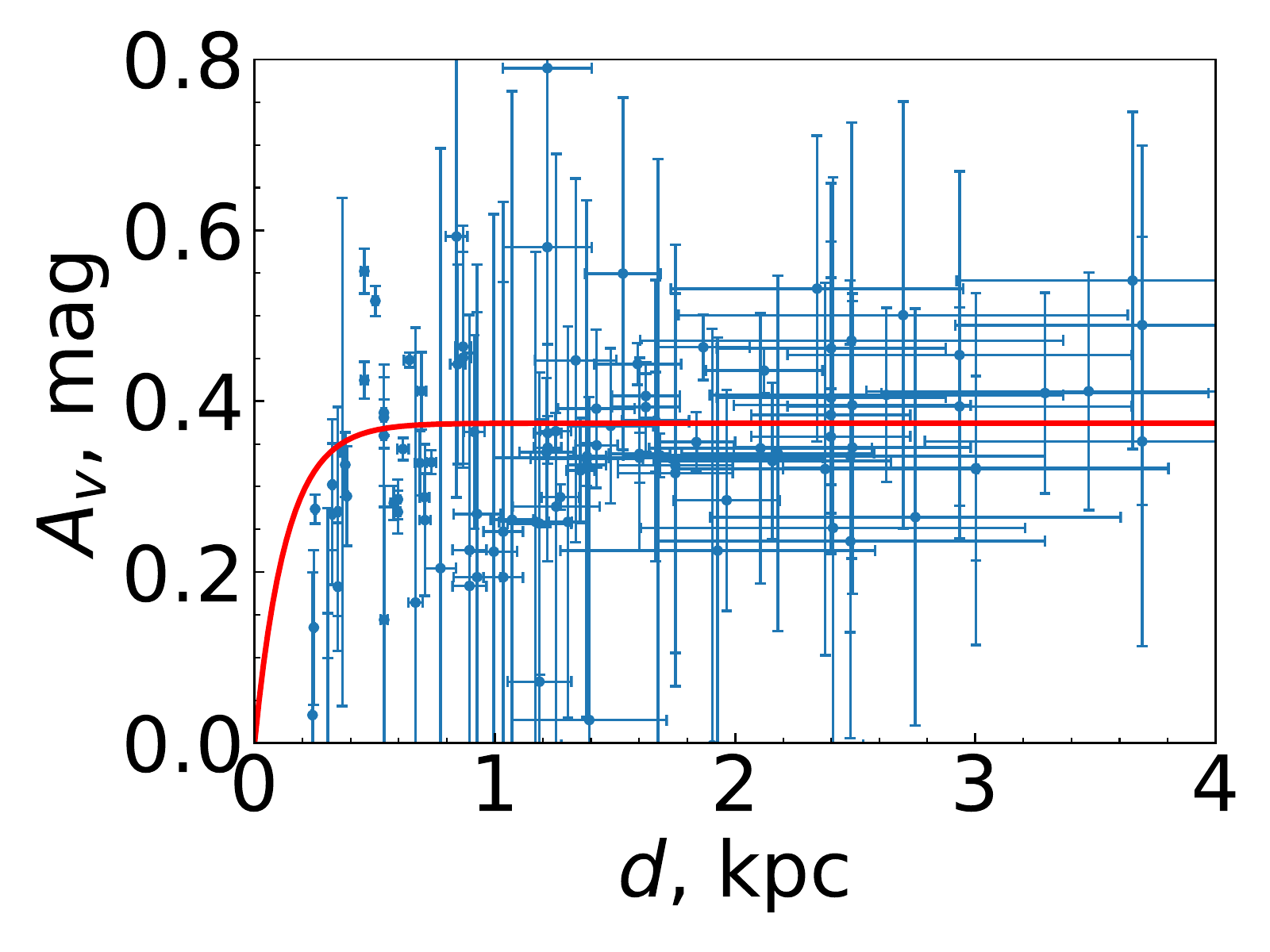}
  \includegraphics{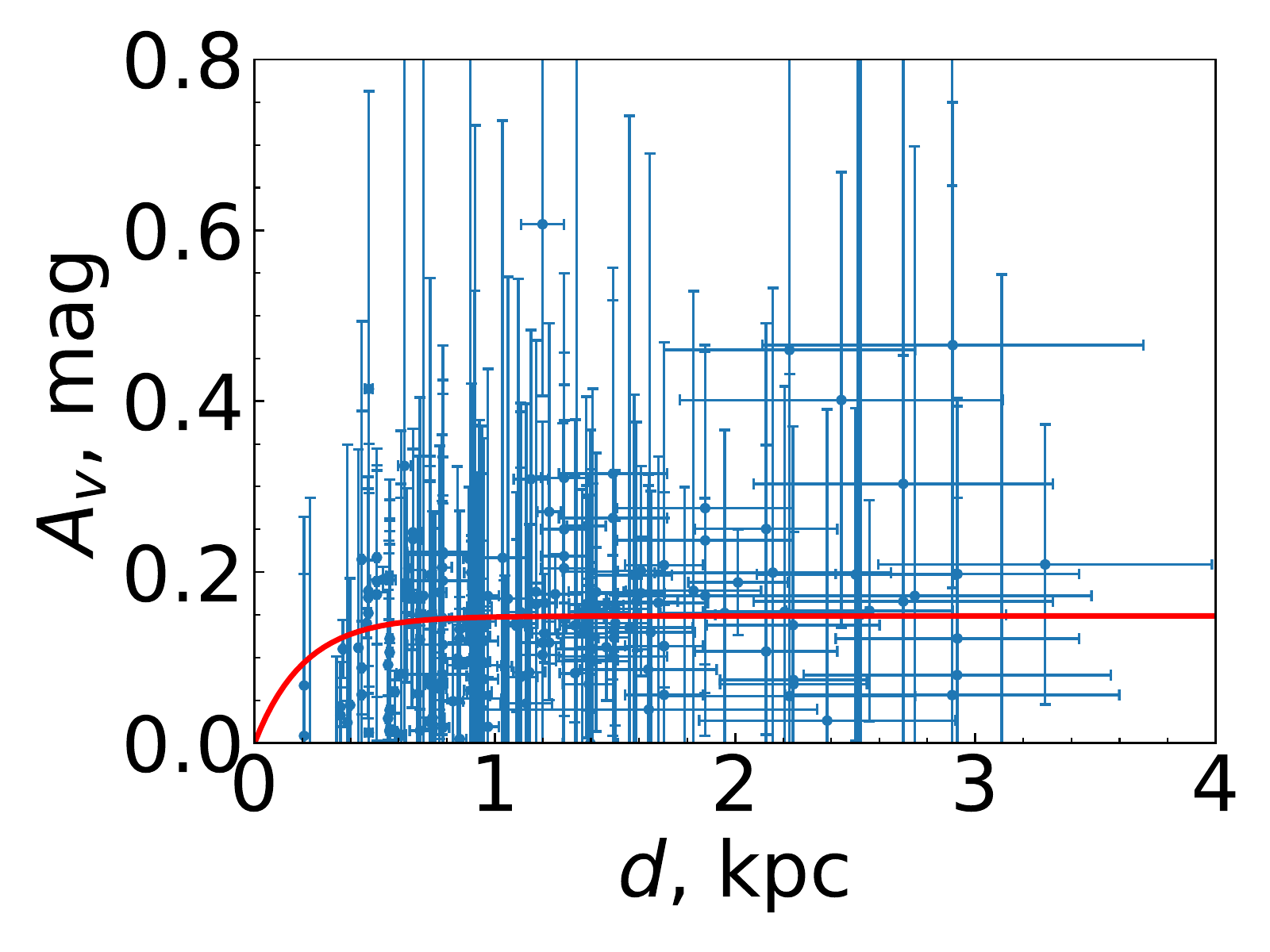}
 }
\caption{Examples of best-fit $\chi^2$ minimization solutions for three areas (from left to right: 84622, 100961, 160342). Blue dots with error bars are values for stars from data, red line is best-fit curve.
See discussion in the text.}
\label{fig:AN}
\end{figure}

\begin{figure}[h]
\resizebox{1.0\columnwidth}{!}{%
  \includegraphics{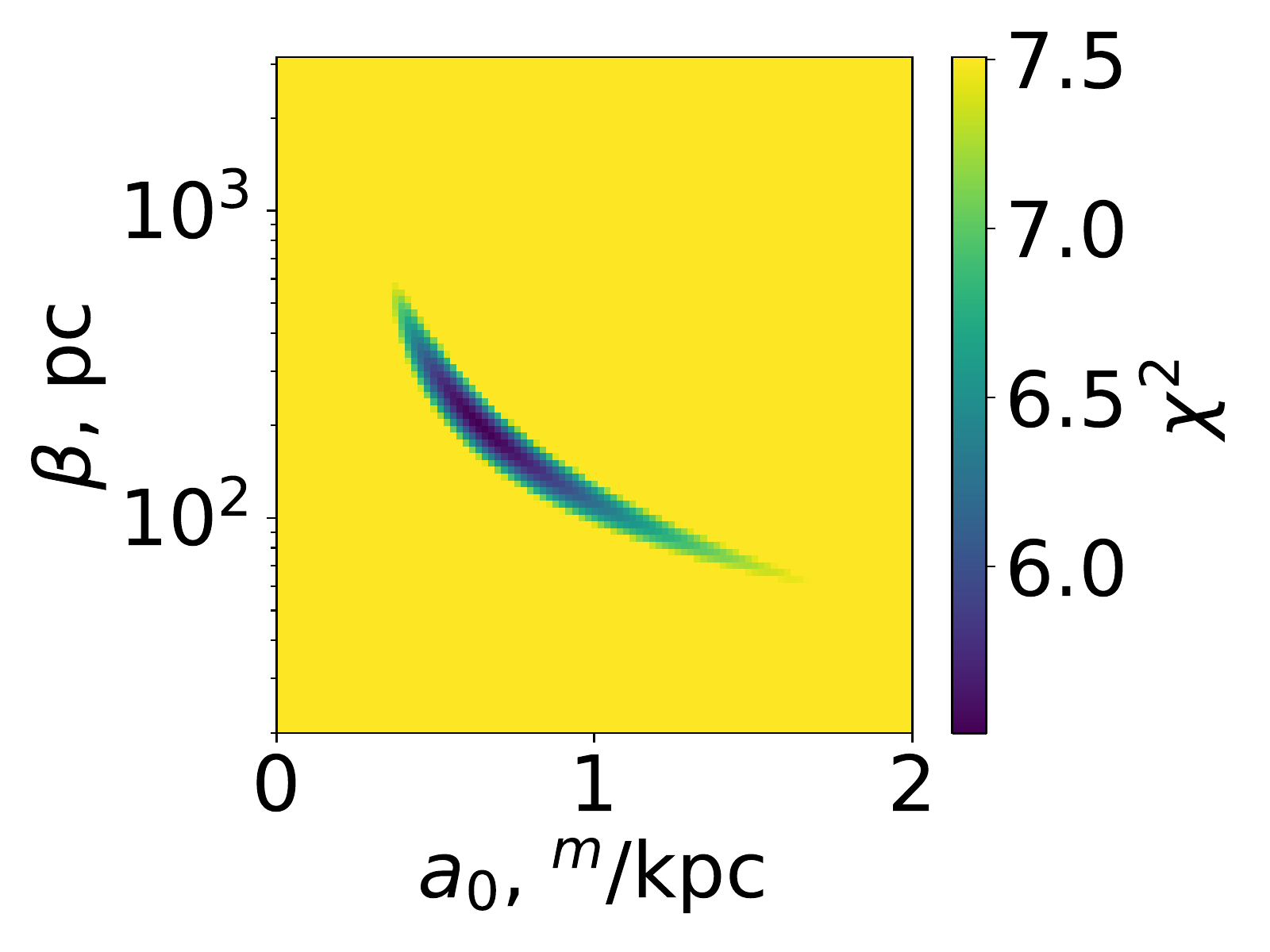}
  \includegraphics{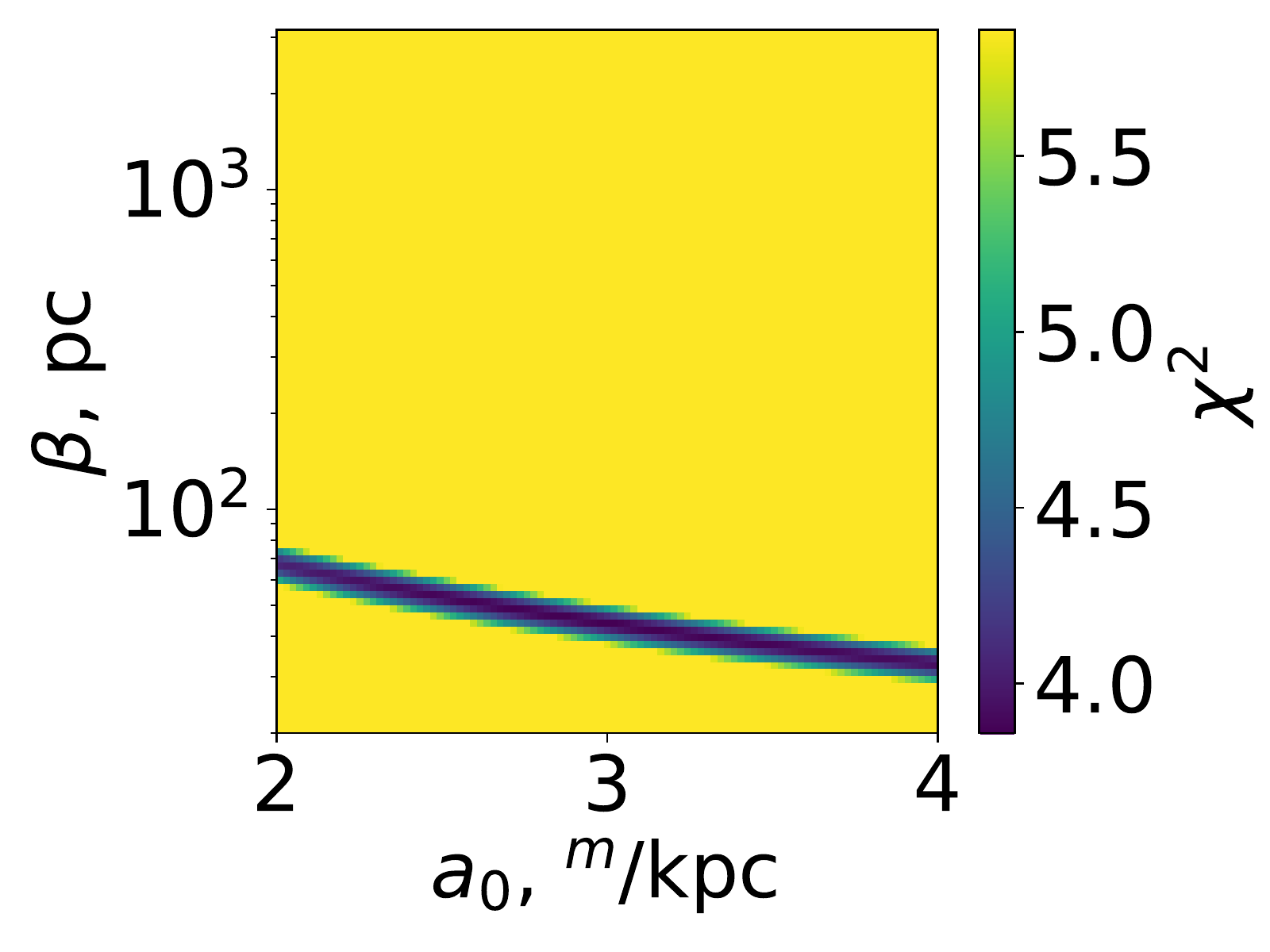}
  \includegraphics{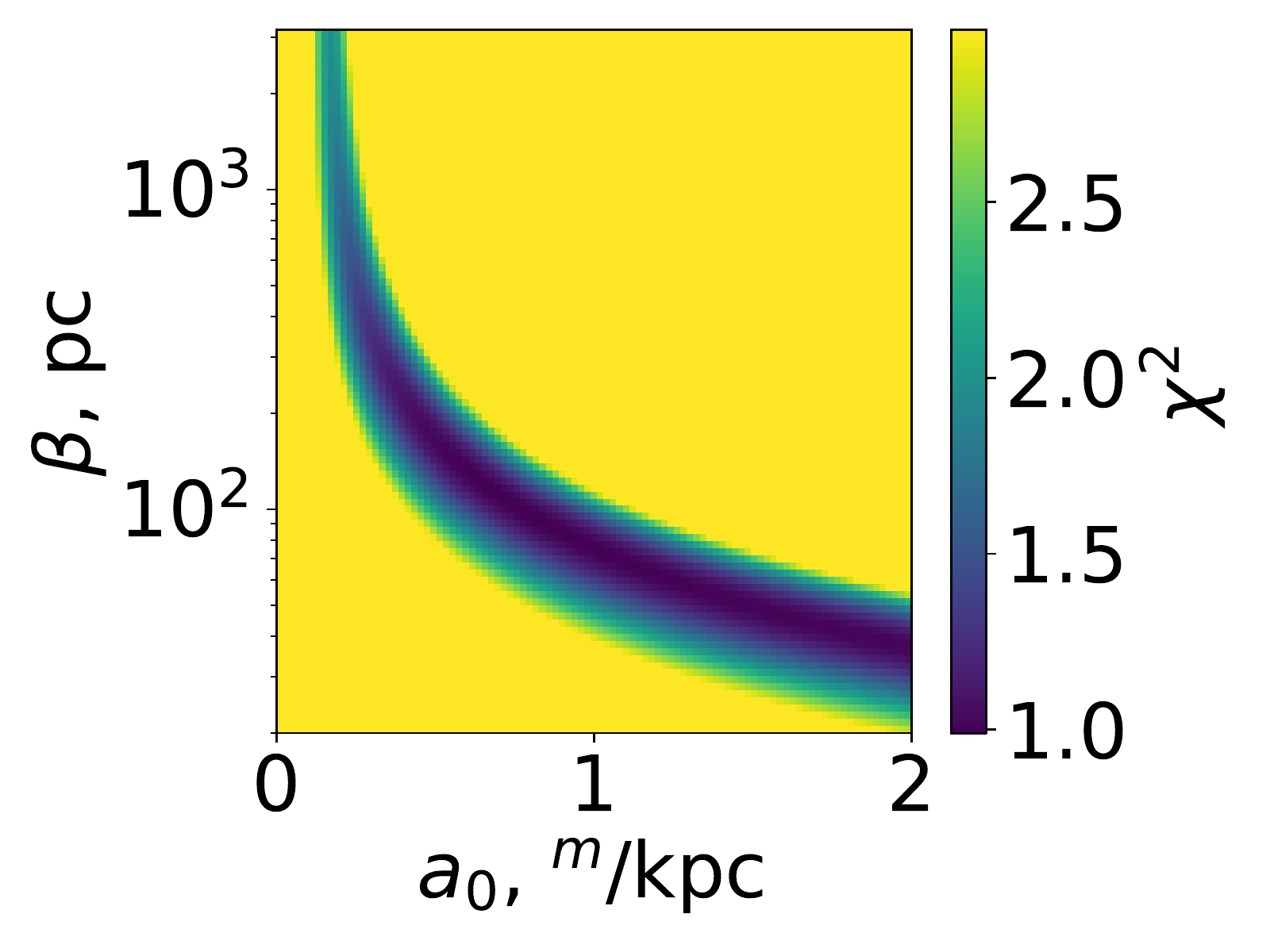}
 }
\caption{Examples of $\chi^2$ scan solutions for the same areas as for Fig.~\ref{fig:AN}.}
\label{fig:KG}
\end{figure}

\subsection{Best-fit $\chi^2$ minimization}
\label{sec:AN}

We use LMFIT~\cite{newville_matthew_2014_11813} package for Python programming language to calculate best-fit minimization parameters $a_0, \beta$ for our datasets $A_V(d)$ for selected areas $( l, b )$ in the sky. We minimize $\chi^2$ functional, which is defined as follows:
\begin{equation}
\chi^2 ( a_0, \beta) = \sum\limits_{n=1}^{N} \left( \frac{A_V (d_n) - A_{V, n}}{\varepsilon_n \left(A_V, d\right)}\right)^2\!,
\label{equ:chi2eq}
\end{equation}
where $N$ is number of points at the selected area, $d_n, A_{V, n}$ are values of distance and extinction for point number $n$ in the area, $A_V (d)$ is value of extinction from Parenago formula~\eqref{equ:parenago}. Values $\varepsilon_n (A_V, d)$ are the standard error values of extinction and distance for input data. Examples of solutions $A_V (d)$ with best-fit parameters $a_0, \beta$ for some areas are illustrated in Fig.~\ref{fig:AN}.

The group of stars where the reddening falls below the fit
(Fig.~\ref{fig:AN}, left panel) is located in the same part of an area in the sky and
at a similar distance from the Sun, which may indicate a fluctuation of extinction due to the non-uniformity of 
distribution of interstellar matter. On the other hand, this same part of the area contains other stars at 
a similar distance which are on or above the fit. Due to the low number of stars in both groups, and the uncertainty
of distances, this group may also be formed by a coincidence.
However, it should be mentioned that Lallement et al. 2019~\cite{2019A&A...625A.135L} found
many irregularities in the distribution of interstellar dust
in this direction, at distances greater than 500-1000 pc.

The uncertainties in extinction are defined by the procedure of its estimation, involving an interplay between
the observational error of effective temperature and slope of calibration $T_{\rm eff}$ - intrinsic colour $(BP-RP)_0$.
Observational errors of effective temperature in use are defined by LAMOST procedures and are mainly related
to the quality and characteristics of LAMOST spectra for a given source. The slope of
$T_{\rm eff}$ -- $(BP-RP)_0$ relation 
flattens toward higher temperatures, and within considered dataset increase of $T_{\rm eff}$ for 1000 K leads to 
decrease of error in $A_V$ for approximately 30\% at fixed error in $T_{\rm eff}$.
The influence of errors in $T_{\rm eff}$ is 
much more significant: at fixed $T_{\rm eff}$, error in $A_V$ increases in direct proportion to the increase
of error in $T_{\rm eff}$.

In the considered datasets, effective temperatures of stars measured by LAMOST mainly lay between approximately 4000 
and 8000 K, while errors of effective temperature change from about 10 to 1000 K (overall mean 120 K, median 
100 K). This leads to an overall mean error in $A_V$=0.15 mag (median 0.10 mag). In various areas typical errors in 
$A_V$ may differ due to different typical errors in $T_{\rm eff}$, obviously related to LAMOST processing.
In particular, for the three areas in Fig.~\ref{fig:AN}:
area  84622: mean error in $T_{\rm eff}$ is  99 K, mean error in $A_V$ is 0.12 mag;
area 100961: mean error in $T_{\rm eff}$ is 160 K, mean error in $A_V$ is 0.16 mag;
area 160342: mean error in $T_{\rm eff}$ is 198 K, mean error in $A_V$ is 0.22 mag. 
Such uncertainties in mass estimates of visual extinction for individual stars are typical (see, e.g., 
\cite{2019A&A...628A..94A}
resulting in median precision 0.20 mag in V-band extinction). 

Figures for all areas are placed in Appendix B.

\subsection{$\chi^2$ scan}
\label{sec:KG}

Independently from best-fit minimization we calculate $\chi^2$ values obtained from the formula~\eqref{equ:chi2eq} on $(a_0, \beta)$ grid and find the minimum value of $\chi^2$ on the grid. Values $a_0, \beta$ of the grid corresponding to the minimum $\chi^2$ value are considered the best approximation values. We also estimate the standard error values of the parameters $a_0, \beta$ determination by $1\sigma$-contour on $\chi^2$ map. Examples of $\chi^2 (a_0, \beta)$ solutions for some areas are illustrated in Fig.~\ref{fig:KG}. Note that the shape of the blue area around the minimum value gives a representation about degeneration of the approximated parameters for a given area.
Figures for all areas are placed in the Appendix B.

\begin{figure}[h]
\resizebox{1.0\columnwidth}{!}{%
  \includegraphics{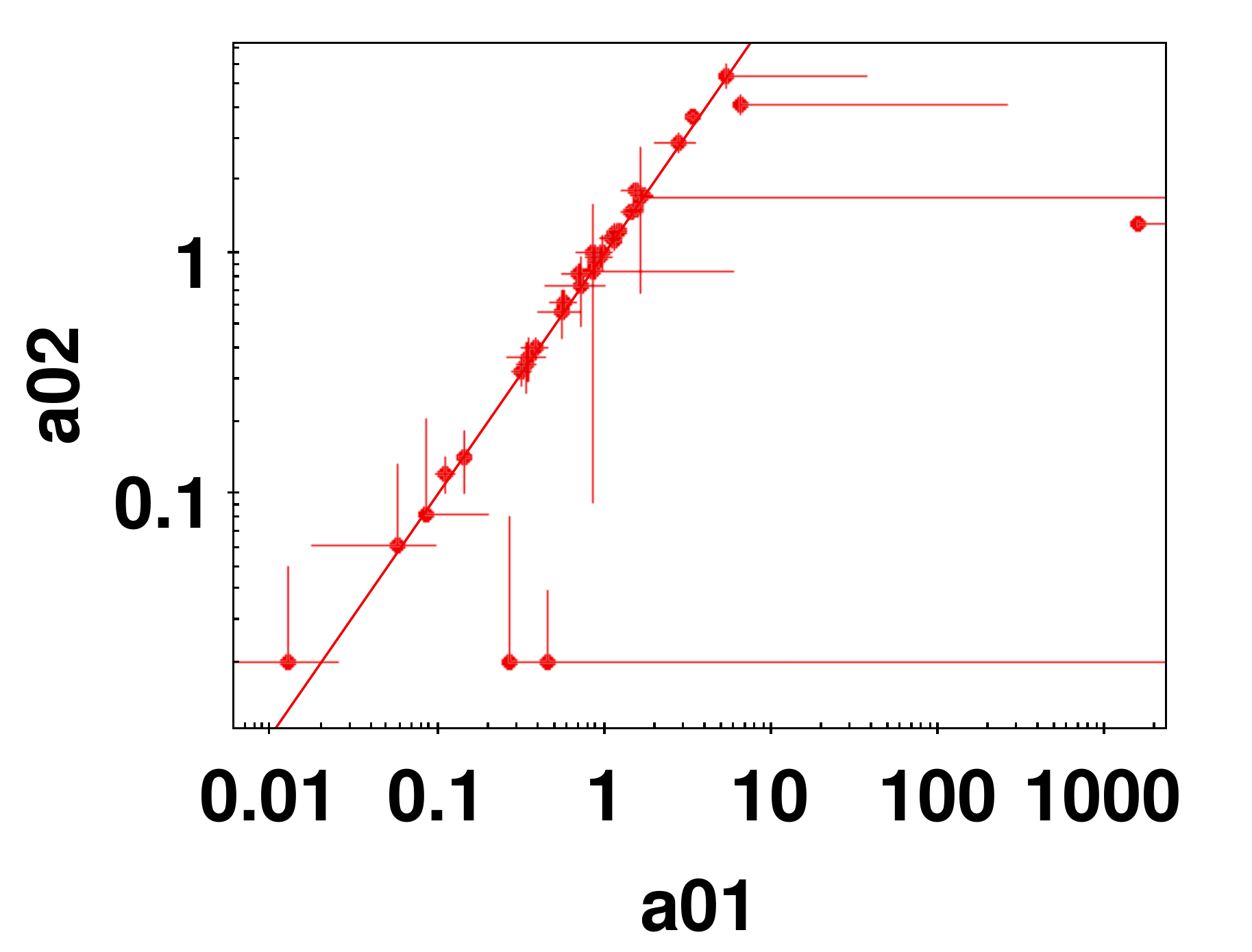}
  \includegraphics{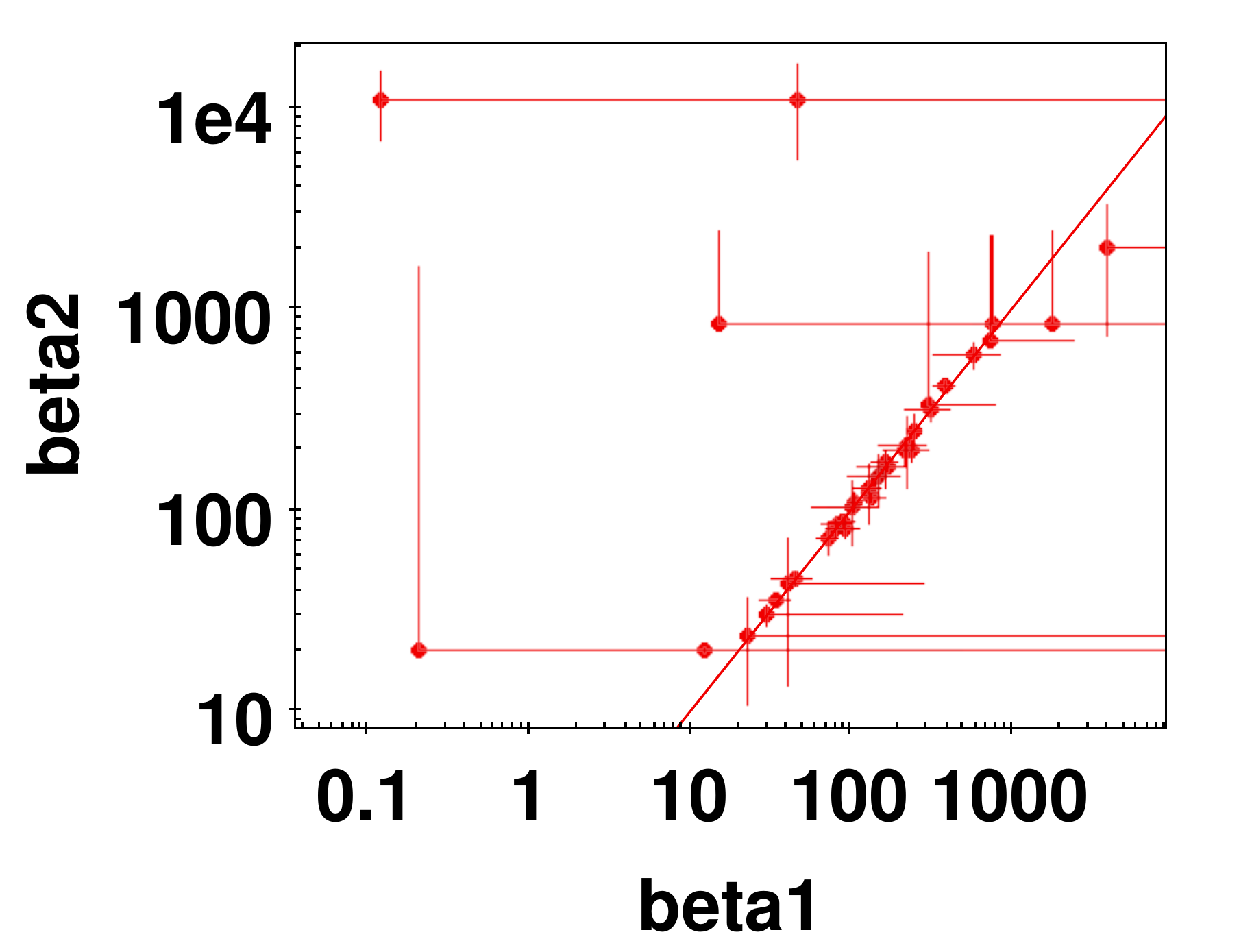}
  \includegraphics{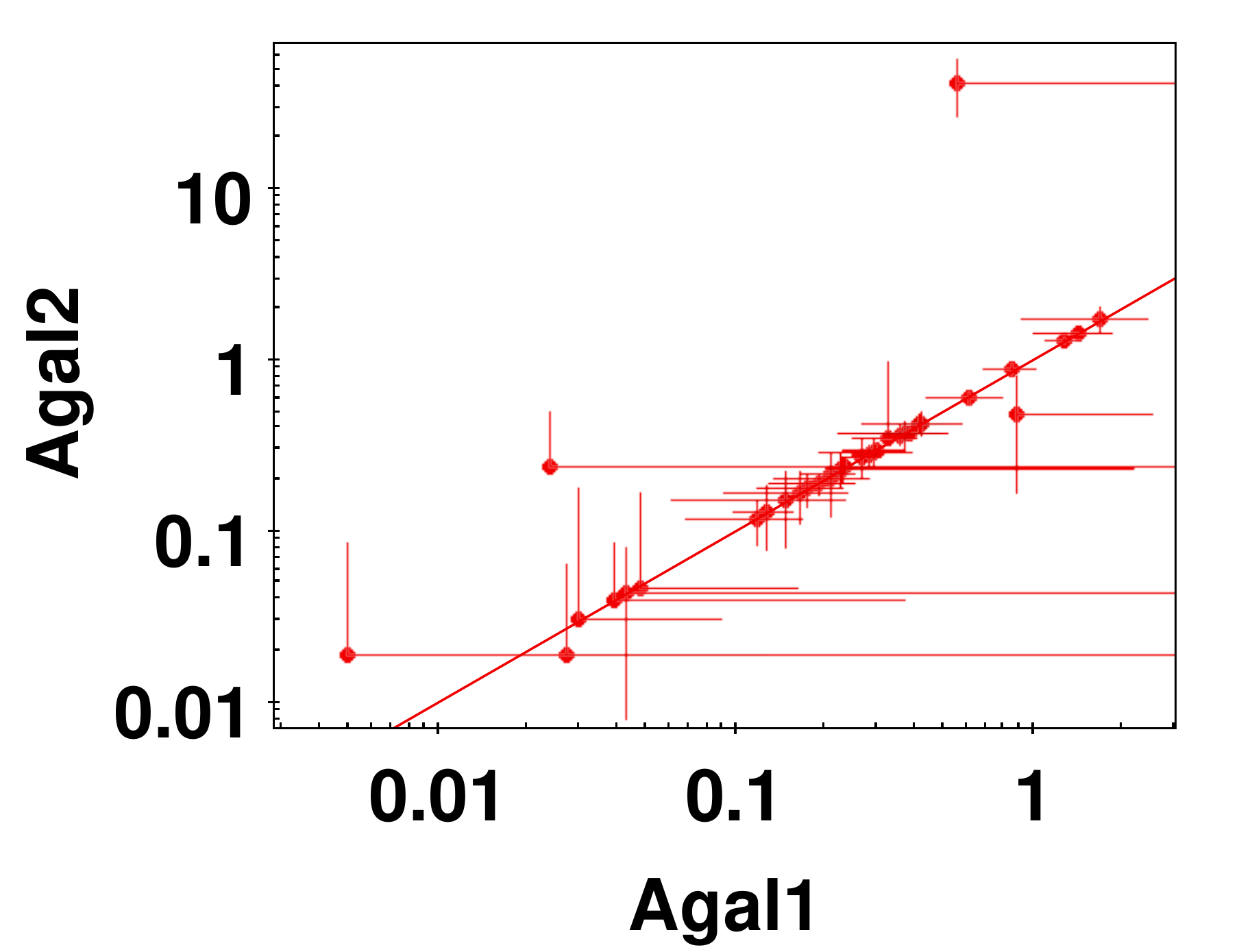}
 }
\caption{Comparison of the results obtained with best-fit $\chi^2$ minimization (index 1)
and $\chi^2$ scan (index 2) approaches, from left to right,
for $a_0$ (mag/kpc), $\beta$ (pc), $A_{Gal}$ (mag), respectively.
The one-to-one relation is shown as a solid line, for reference.}
\label{fig:compare}
\end{figure}

\subsection{Comparison of two approaches and final values}
\label{sec:compare}

Both approaches, described in Sections~\ref{sec:AN} and~\ref{sec:KG}
were applied to all 40 areas. Values of the $a_0$ and $\beta$ parameters, obtained
with both methods, are plotted in Fig.~\ref{fig:compare}.
Total Galactic extinction values ($A_{Gal}$), calculated from $a_0$ and $\beta$
with Eq.~\eqref{equ:agal} are plotted as well.
For six areas no reasonable solution could be found with the described methods
(these areas can be seen in Fig.~\ref{fig:compare} as outliers).

The values of $\chi^2$ were estimated for both methods. To compile the final list
of the $a_0$, $\beta$ parameters for our areas, we have taken
values obtained with one of the two approaches
(see Sections~\ref{sec:AN} and~\ref{sec:KG}), which demonstrates the minimum $\chi^2$.
These data, including the Galactic extinction $A_{Gal}$, calculated with Eq.~\eqref{equ:agal},
are presented in Table~\ref{tab:areas}.

\begin{figure}
\resizebox{0.9\columnwidth}{!}{%
  \includegraphics{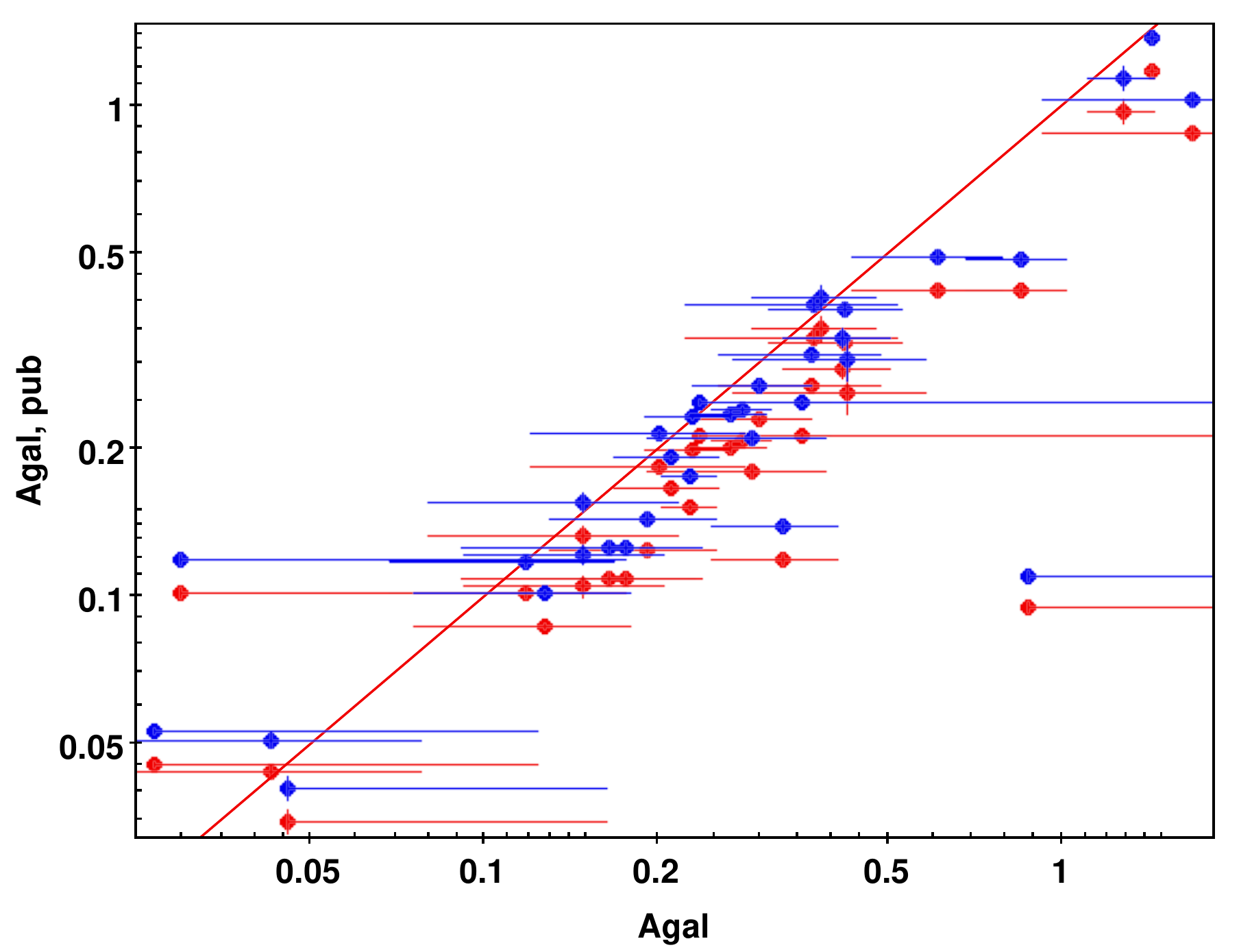}
 }
\caption{Galactic extinction $A_{Gal}$, comparison with \cite{2011ApJ...737..103S}
(red dots) and \cite{1998ApJ...500..525S} (blue dots) models.
The one-to-one relation is shown as a solid line, for reference.}
\label{fig:Agal}
\end{figure}

We have compared the calculated Galactic extinction with values, predicted
by \cite{2011ApJ...737..103S} and \cite{1998ApJ...500..525S} (Fig.~\ref{fig:Agal}, values $A_{Gal}$ from \cite{2011ApJ...737..103S} and \cite{1998ApJ...500..525S}
were obtained with Galactic extinction calculator
https://irsa.ipac.caltech.edu/applications/DUST/).
One can see that \cite{1998ApJ...500..525S}'s data better approximate
our results than \cite{2011ApJ...737..103S}'s data.
Our estimates of $A_{Gal}$ are on average 15\% higher than those predicted in  \cite{1998ApJ...500..525S}.
The reason for the differences seems to be not high enough accuracy of the relations used in Eq.(\ref{equ:Av}).
In particular, this may be due to the fact that the coefficients $c_1$ and $c_2$
in Eq.~\eqref{equ:Av} are not constants (see discussion in Section~\ref{sec:data}).
This problem will be investigated by us later.

\begin{table}
\caption{Interstellar extinction in the selected areas}
\begin{tabular}{r|rr|r|lll|l}
\hline\noalign{\smallskip}
     &   &   &  & \multicolumn{3}{|c|}{This paper} & \multicolumn{1}{c}{SF2011~\cite{2011ApJ...737..103S}} \\
Area     & l      & b         & N  & $a_0$               & $\beta$                  & $A_{Gal}$           & $A_{Gal}$ \\
\noalign{\smallskip}\hline\noalign{\smallskip}
  1713   & 48.16  & 20.11     & 20 & 1.72  $\pm$ 0.273   & 84.365    $\pm$ 17.347   & 0.422  $\pm$ 0.11   & 0.3282 $\pm$ 0.0118   \\
  2482   & 36.21  & 20.11     & 43 & 1.554 $\pm$ 0.268   & 135.75    $\pm$ 31.224   & 0.613  $\pm$ 0.176  & 0.4183 $\pm$ 0.0124   \\
  3733   & 39.02  & 30.0      & 46 & 0.896 $\pm$ 0.081   & 150.371   $\pm$ 18.659   & 0.269  $\pm$ 0.041  & 0.2007 $\pm$ 0.0037   \\
  8789   & 22.15  & 30.0      & 33 & 1.108 $\pm$ 0.151   & 172.873   $\pm$ 34.206   & 0.383  $\pm$ 0.092  & 0.3483 $\pm$ 0.0189   \\
  17777  & 155.04 & 20.11     & 44 & 1.142 $\pm$ 0.087   & 107.93    $\pm$ 9.877    & 0.358  $\pm$ 0.043  & 0.212  $\pm$ 0.0028   \\
  28666  & 90.66  & 64.95     & 72 & 0.061 $\pm$ 0.071   & 681.292   $\pm$ 1571.163 & 0.046  $\pm$ 0.118  & 0.0346 $\pm$ 0.0019   \\
  29945  & 155.51 & 64.95     & 76 & 0.013 $\pm$ 0.013   & 1798.0    $\pm$ 6346.7   & 0.027  $\pm$ 0.098  & 0.0453 $\pm$ 0.0006   \\
  34455  & 229.92 & 19.79     & 43 & 0.392 $\pm$ 0.066   & 152.425   $\pm$ 41.533   & 0.176  $\pm$ 0.056  & 0.1075 $\pm$ 0.0022   \\
  35483  & 206.02 & 19.79     &116 & 0.343 $\pm$ 0.081   & 125.893   $\pm$ 41.57    & 0.128  $\pm$ 0.052  & 0.0866 $\pm$ 0.0012   \\
  36249  & 228.87 & 30.0      &112 & 0.113 $\pm$ 0.015   & 3904.2    $\pm$ 7294.1   & 0.883  $\pm$ 1.654  & 0.0945 $\pm$ 0.0022   \\
  64053  & 285.22 & 64.95     & 41 & 1.667 $\pm$ 1.0     & 23.263    $\pm$ 12.645   & 0.043  $\pm$ 0.035  & 0.0435 $\pm$ 0.0012   \\
  79217  & 20.04  & 20.11     & 21 & 14.081$\pm$ 0.202   & 35.03     $\pm$ 0.0      & 1.435  $\pm$ 0.021  & 1.1739 $\pm$ 0.0229   \\
  82277  & 98.09  & $-$30.0   & 73 & 1.242 $\pm$ 0.143   & 168.756   $\pm$ 29.785   & 0.419  $\pm$ 0.088  & 0.2886 $\pm$ 0.014    \\
  83604  & 94.92  & $-$19.79  & 57 & 1.118 $\pm$ 0.167   & 90.967    $\pm$ 16.123   & 0.3    $\pm$ 0.07   & 0.2292 $\pm$ 0.005    \\
  84386  & 79.8   & $-$20.11  & 56 & 6.589 $\pm$ 256.178 & 12.294    $\pm$ 480.196  & 0.236  $\pm$ 12.987 & 0.2119 $\pm$ 0.0022   \\
  84622  & 69.96  & $-$20.11  & 49 & 0.586 $\pm$ 0.1     & 218.42    $\pm$ 56.704   & 0.372  $\pm$ 0.116  & 0.2664 $\pm$ 0.0034   \\
  96855  & 78.05  & 19.79     & 99 & 0.323 $\pm$ 0.04    & 210.001   $\pm$ 80.871   & 0.201  $\pm$ 0.081  & 0.1835 $\pm$ 0.0028   \\
  97942  & 83.32  & 30.0      & 45 & 0.727 $\pm$ 0.222   & 102.591   $\pm$ 35.665   & 0.149  $\pm$ 0.069  & 0.1325 $\pm$ 0.0059   \\
  98650  & 188.79 & $-$30.0   & 36 & 1.463 $\pm$ 0.182   & 577.226   $\pm$ 250.491  & 1.689  $\pm$ 0.762  & 0.8784 $\pm$ 0.026    \\
  99901  & 185.98 & $-$20.11  & 69 & 3.364 $\pm$ 0.235   & 130.833   $\pm$ 14.606   & 1.28   $\pm$ 0.169  & 0.9715 $\pm$ 0.0505   \\
  100961 & 164.88 & $-$20.11  &121 & 2.847 $\pm$ 0.787   & 45.192    $\pm$ 12.992   & 0.374  $\pm$ 0.149  & 0.3349 $\pm$ 0.0105   \\
  112048 & 194.06 & 19.79     &224 & 0.144 $\pm$ 0.01    & 771.115   $\pm$ 183.167  & 0.329  $\pm$ 0.081  & 0.1187 $\pm$ 0.0028   \\
  112305 & 183.16 & 20.11     &143 & 0.316 $\pm$ 0.02    & 248.257   $\pm$ 22.276   & 0.228  $\pm$ 0.025  & 0.1503 $\pm$ 0.0012   \\
  113239 & 168.05 & 19.79     & 93 & 1.16  $\pm$ 0.082   & 81.896    $\pm$ 7.634    & 0.281  $\pm$ 0.033  & 0.2066 $\pm$ 0.0031   \\
  114025 & 186.68 & 30.0      & 39 & 0.555 $\pm$ 0.152   & 149.965   $\pm$ 52.302   & 0.166  $\pm$ 0.074  & 0.1078 $\pm$ 0.0034   \\
  136719 & 75.25  & $-$45.39  & 71 & 0.852 $\pm$ 0.178   & 244.611   $\pm$ 66.037   & 0.293  $\pm$ 0.1    & 0.1791 $\pm$ 0.0062   \\
  137487 & 63.15  & $-$45.39  & 73 & 5.444 $\pm$ 0.636   & 30.045    $\pm$ 3.757    & 0.23   $\pm$ 0.039  & 0.1978 $\pm$ 0.0025   \\
  138857 & 62.93  & $-$30.0   & 67 & 0.353 $\pm$ 0.094   & 168.121   $\pm$ 55.478   & 0.119  $\pm$ 0.05   & 0.1008 $\pm$ 0.0022   \\
  149151 & 143.06 & $-$65.32  &119 & 0.081 $\pm$ 0.121   & 332.83    $\pm$ 1571.163 & 0.03   $\pm$ 0.147  & 0.1015 $\pm$ 0.0019   \\
  151503 & 135.0  & $-$45.39  & 55 & 0.955 $\pm$ 0.176   & 317.56    $\pm$ 100.983  & 0.426  $\pm$ 0.157  & 0.2597 $\pm$ 0.0245   \\
  152806 & 169.88 & $-$44.99  & 47 & 1.526 $\pm$ 0.177   & 395.27    $\pm$ 62.142   & 0.853  $\pm$ 0.167  & 0.4179 $\pm$ 0.0105   \\
  160342 & 133.95 & $-$30.0   &228 & 0.722 $\pm$ 0.182   & 102.905   $\pm$ 28.656   & 0.149  $\pm$ 0.056  & 0.1045 $\pm$ 0.0053   \\
  161428 & 139.92 & $-$19.79  &125 & 0.711 $\pm$ 0.15    & 92.061    $\pm$ 22.274   & 0.193  $\pm$ 0.062  & 0.1235 $\pm$ 0.0047   \\
  162401 & 119.88 & $-$20.11  &104 & 0.983 $\pm$ 0.13    & 74.29     $\pm$ 12.014   & 0.212  $\pm$ 0.044  & 0.1647 $\pm$ 0.0019   \\
\noalign{\smallskip}\hline
\end{tabular}
\label{tab:areas}
\end{table}

\section{Approximation for $a_0, \beta$ parameters over the entire sky}
\label{sec:2D}

\begin{figure}
\resizebox{1.0\columnwidth}{!}{%
  \includegraphics{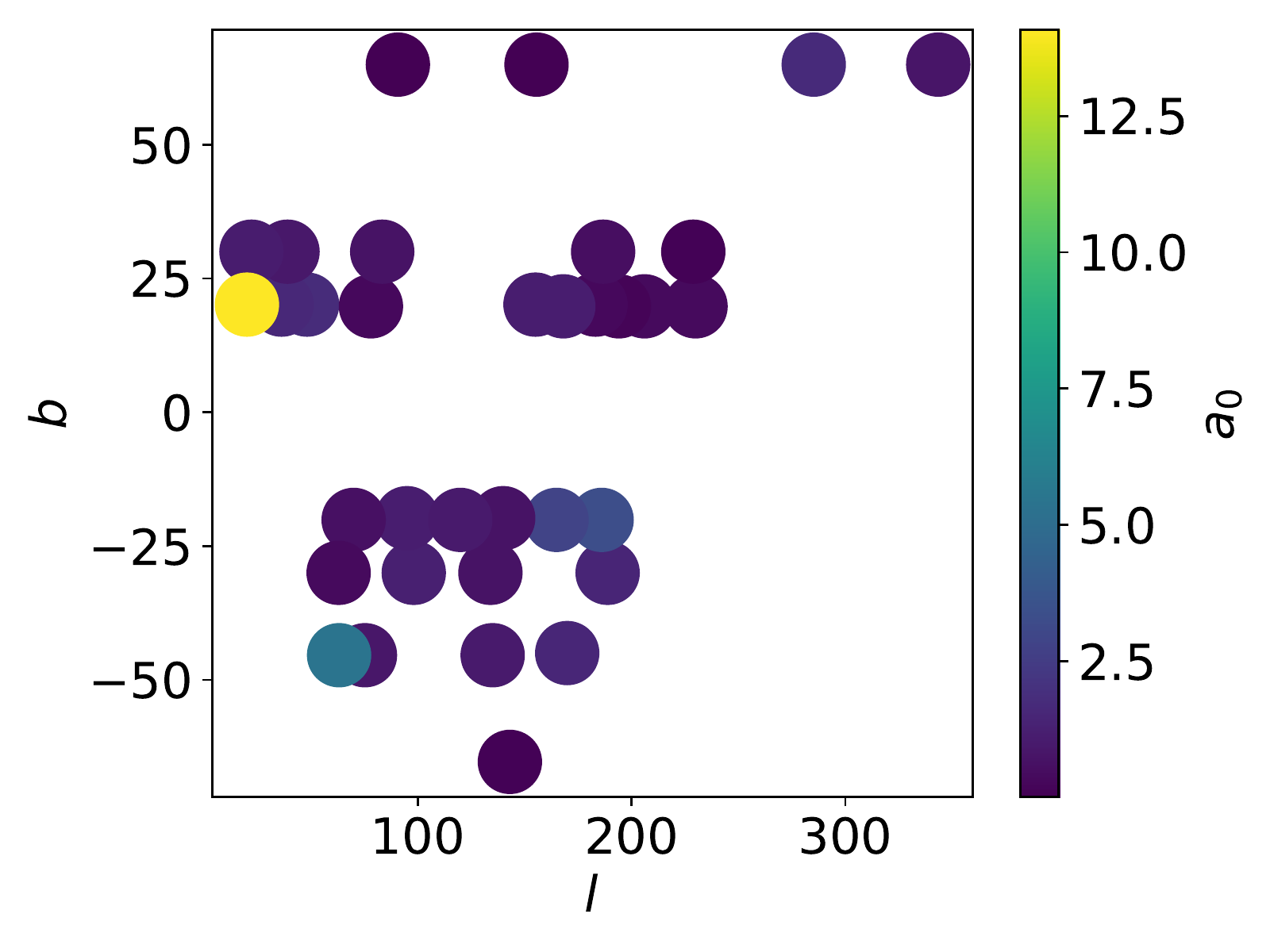}
  \includegraphics{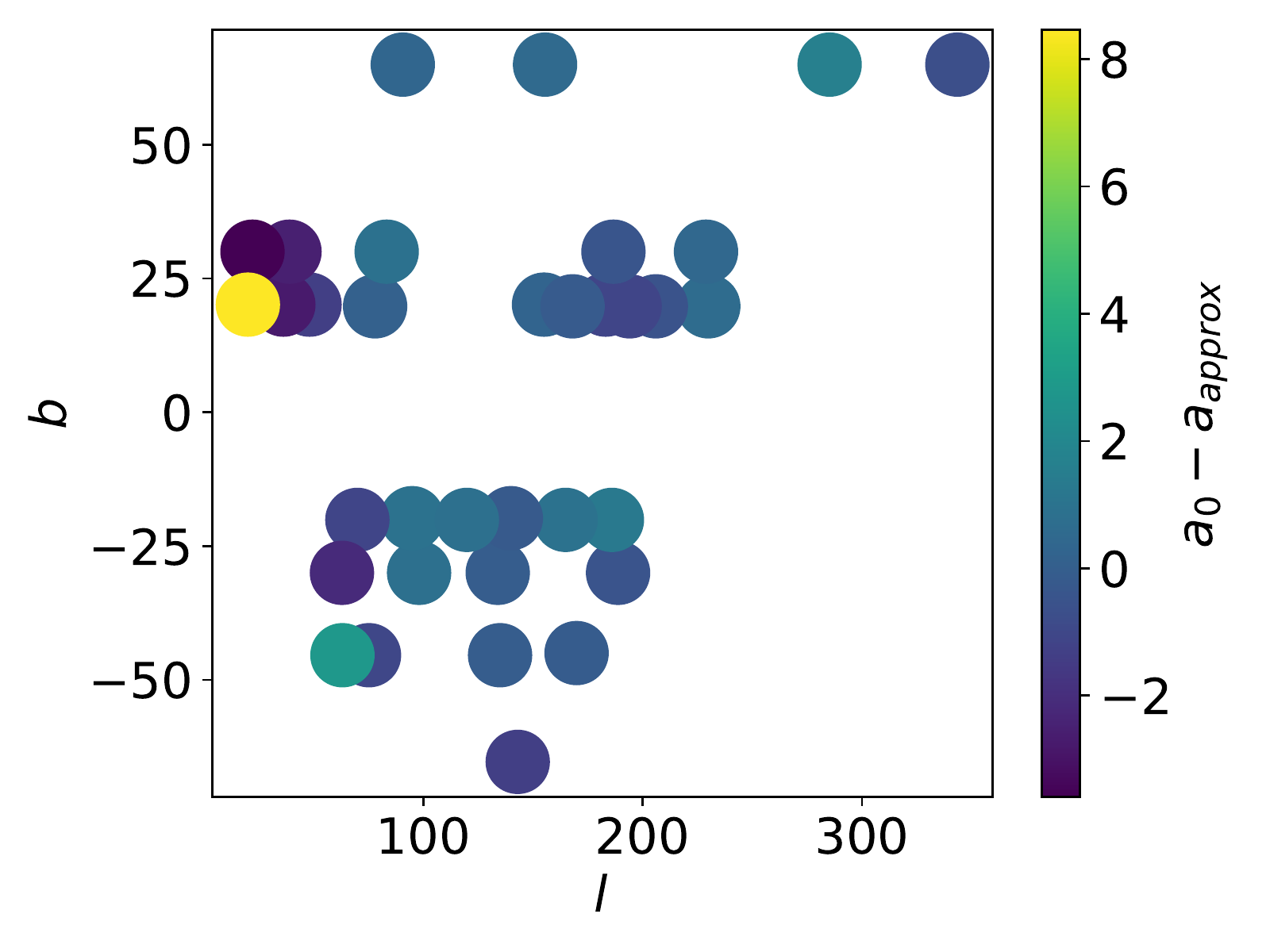}
 }
\resizebox{1.0\columnwidth}{!}{%
  \includegraphics{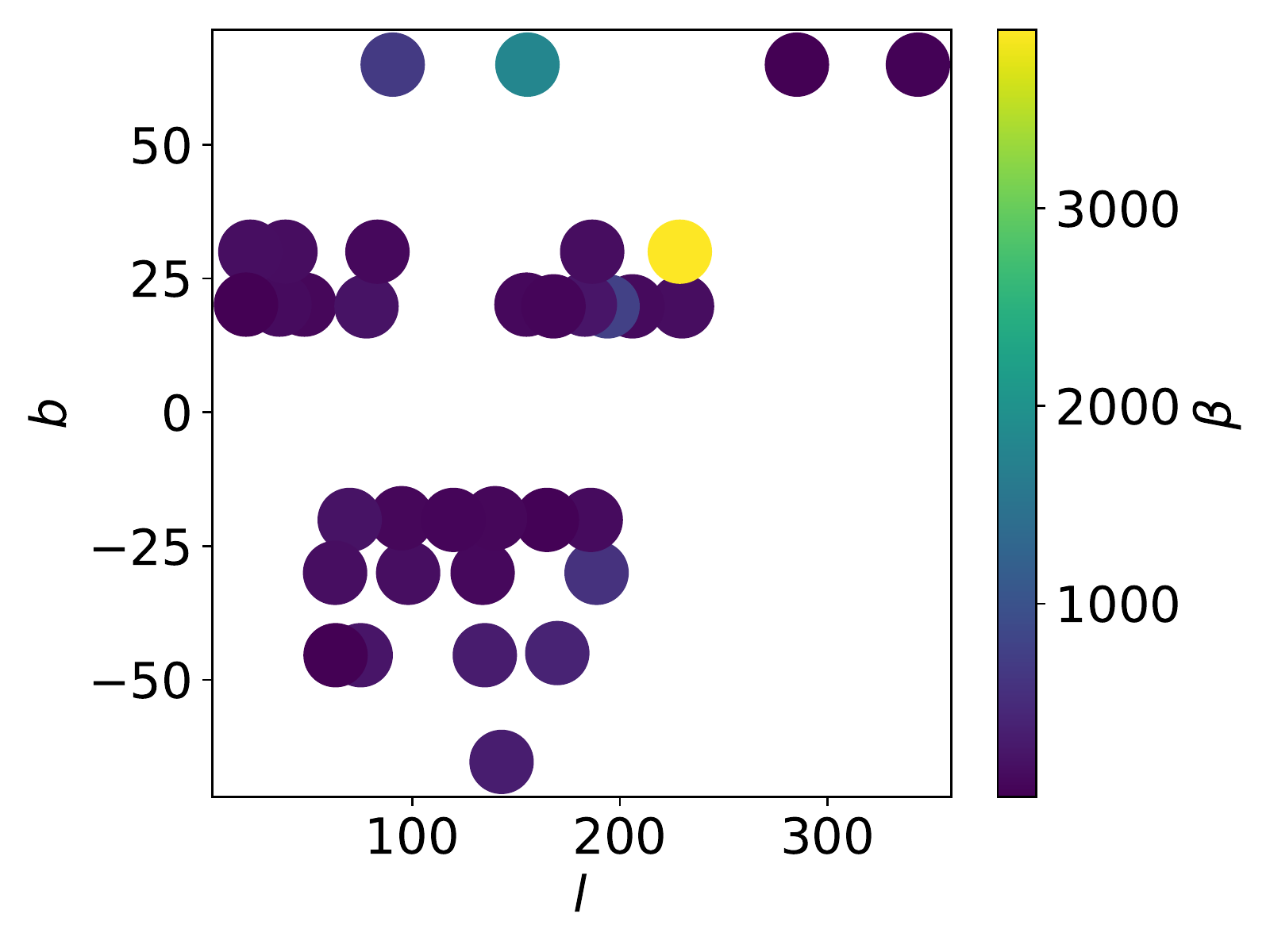}
  \includegraphics{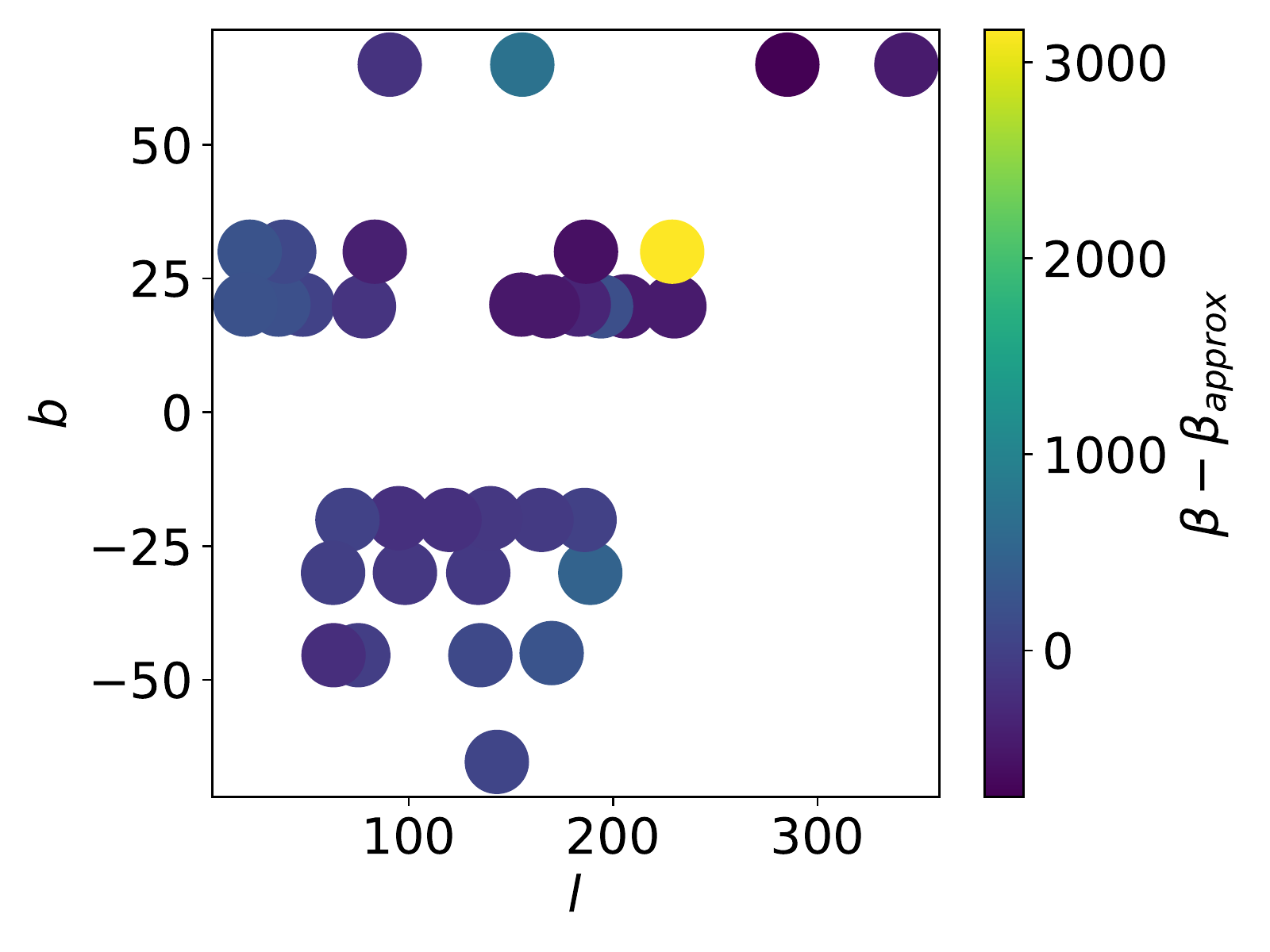}
 }
\resizebox{1.0\columnwidth}{!}{%
  \includegraphics{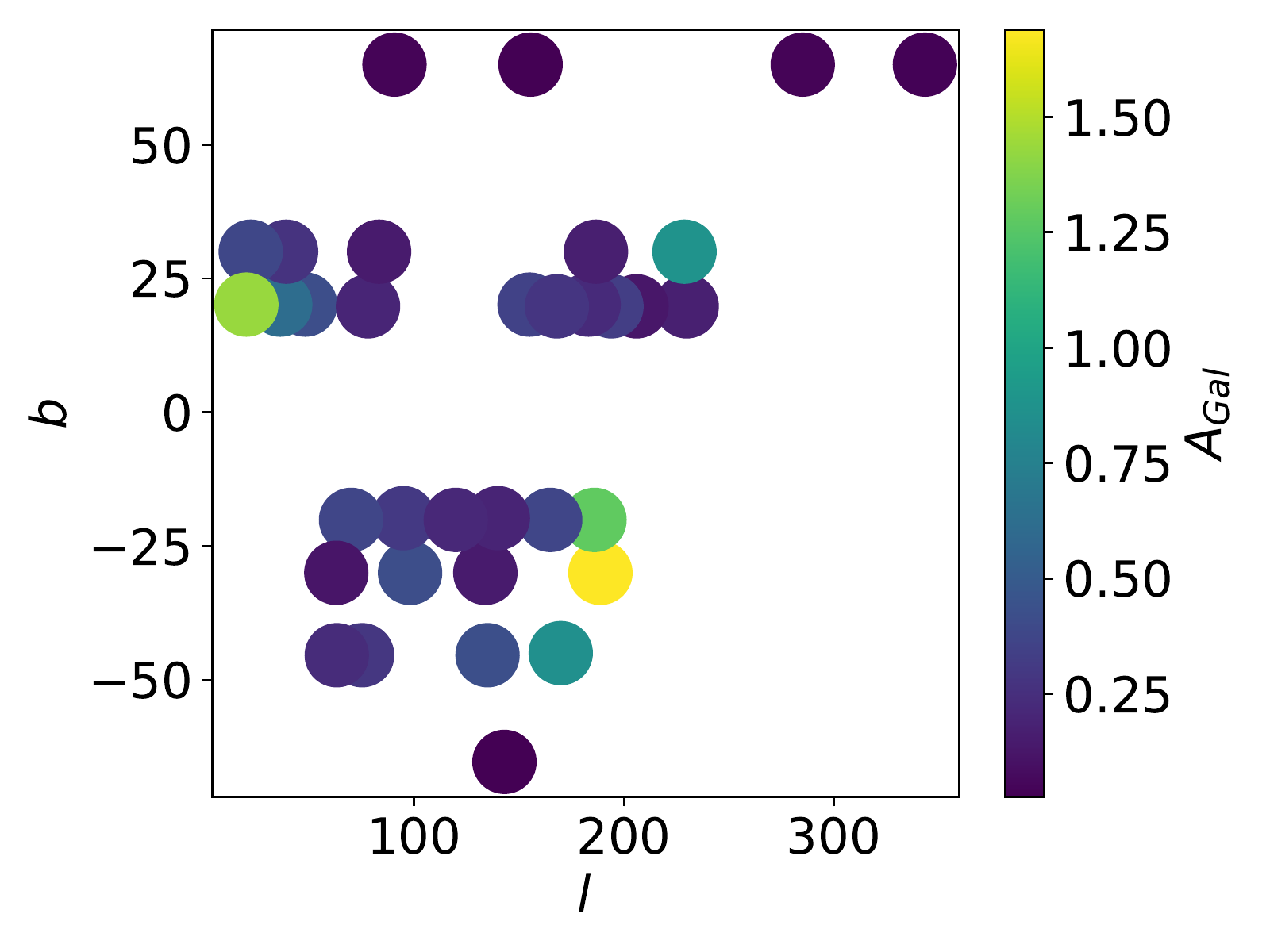}
  \includegraphics{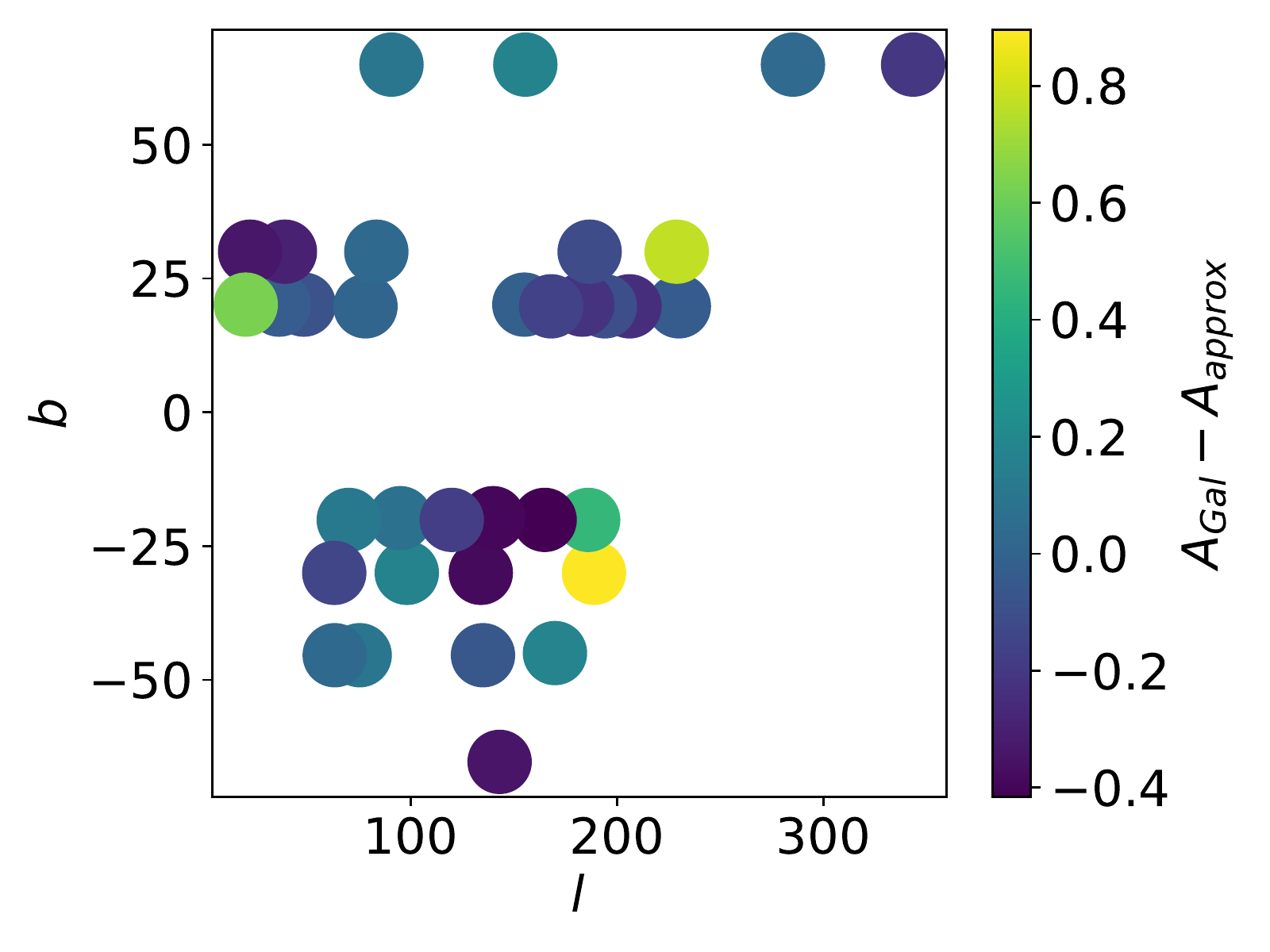}
 }
\caption{Results of approximation. Values (left column) and residuals (right column)
for $a_0$, $\beta$, $A_{Gal}$ (from top to bottom), respectively.
Residuals are computed as the values from Table \ref{tab:areas} minus
the values calculated with Eqs. \ref{equ:spher-a0}-\ref{equ:spher-Agal}.
}
\label{fig:spher}
\end{figure}

Our final step is the approximation for $a_0, \beta$ parameters over the entire sky
by spherical harmonics (see Appendix A for details).
Thirty four of the selected areas
were used for the approximation, they are
listed in Table~\ref{tab:areas} (N is number of stars in the area).
The results are presented in Table~\ref{tab:spher},
Eqs.~\eqref{equ:spher-a0}-\eqref{equ:spher-Agal}
and Fig.~\ref{fig:spher}.

These relationships will apparently apply better at higher extinction
where the dust is averaged over a long line, and it will work less well at high latitudes.
This should be checked when we extend our consideration to low latitudes and
to (on average more distant) non-MS stars (see Section~\ref{sec:plans}).

\begin{table}
\caption{The coefficients in the approximation of spherical harmonics}
\begin{tabular}{l|lll}
\hline\noalign{\smallskip}
       & $a_0$             & $\beta$ & $A_{Gal}$  \\
\noalign{\smallskip}\hline\noalign{\smallskip}
$A_{00}$ & 6.484 $\pm$ 1.532  &  1151.305 $\pm$ 495.928 &  1.225 $\pm$ 0.239      \\
$A_{10}$ & $-$2.388 $\pm$ 1.677 &  578.567 $\pm$ 542.883  &  $-$0.279 $\pm$ 0.262 \\
$A_{11}$ & $-$7.645 $\pm$ 2.412 &  874.441 $\pm$ 780.663  &  $-$0.221 $\pm$ 0.377 \\
$A_{20}$ & $-$1.345 $\pm$ 1.640 &  516.379 $\pm$ 530.792  &  $-$0.470 $\pm$ 0.256 \\
$A_{21}$ & 0.274 $\pm$ 2.099  &  554.834 $\pm$ 679.292  &  $-$0.570 $\pm$ 0.328   \\
$A_{22}$ & 6.331 $\pm$ 2.271  &  $-$455.485 $\pm$ 735.097 &  0.793 $\pm$ 0.355    \\
\noalign{\smallskip}\hline
\end{tabular}
\label{tab:spher}
\end{table}

\begin{multline}
a_0 = +6.484 Y_0^0 \left(l, \frac{\pi}{2} - b\right) - 2.388 Y_1^0 \left(l, \frac{\pi}{2} - b\right) - 7.645 Y_1^1 \left(l, \frac{\pi}{2} - b\right) - \\
- 1.345 Y_2^0 \left(l, \frac{\pi}{2} - b\right) + 0.274 Y_2^1 \left(l, \frac{\pi}{2} - b\right) + 6.331 Y_2^2 \left(l, \frac{\pi}{2} - b\right)
\label{equ:spher-a0}
\end{multline}

\begin{multline}
\beta = +1151.305 Y_0^0 \left(l, \frac{\pi}{2} - b\right) + 578.567 Y_1^0 \left(l, \frac{\pi}{2} - b\right) + 874.441 Y_1^1 \left(l, \frac{\pi}{2} - b\right) + \\
+ 516.379 Y_2^0 \left(l, \frac{\pi}{2} - b\right) + 554.834 Y_2^1 \left(l, \frac{\pi}{2} - b\right) - 455.485 Y_2^2 \left(l, \frac{\pi}{2} - b\right)
\label{equ:spher-beta}
\end{multline}

\begin{multline}
A_{Gal} = +1.225 Y_0^0 \left(l, \frac{\pi}{2} - b\right) - 0.279 Y_1^0 \left(l, \frac{\pi}{2} - b\right) - 0.221 Y_1^1 \left(l, \frac{\pi}{2} - b\right) - \\
- 0.470 Y_2^0 \left(l, \frac{\pi}{2} - b\right) - 0.570 Y_2^1 \left(l, \frac{\pi}{2} - b\right) + 0.793 Y_2^2 \left(l, \frac{\pi}{2} - b\right)
\label{equ:spher-Agal}
\end{multline}

\section{Future plans}
\label{sec:plans}

This pilot study explores the possibility to construct an equation for quick estimation
of interstellar extinction values. In our further work, we plan
to include non-MS stars from LAMOST/Gaia and
to extend our results to lower galactic latitudes.

Beside that, we plan to extend our procedure to the Southern sky,
using RAVE survey ($\sim$460,000 objects) \cite{2017AJ....153...75K}
and upcoming 4MOST and MOONS surveys.
4MOST (2022+) \cite{2019Msngr.175....3D}, 4-metre Multi-Object Spectrograph Telescope
is a second-generation instrument built for ESO's 4.1-metre
Visible and Infrared Survey Telescope for Astronomy (VISTA)
 at the Paranal Observatory (Chile).
MOONS (2020+) \cite{2016ASPC..507..109C}, is the Multi-Object Optical
and Near-infrared Spectrograph for the ESO VLT (8.2 m).

For the Northern sky, besides LAMOST, other
spectroscopic surveys can be used. Among them are
APOGEE ($\sim$450,000 objects) \cite{2018MNRAS.476.2117R},
SEGUE ($\sim$350,000 objects) \cite{2009AJ....137.4377Y},
and upcoming WEAVE (2021+) \cite{2016sf2a.conf..271S},
WHT Enhanced Area Velocity Explorer, a multi-object survey spectrograph
for the 4.2-m William Herschel Telescope (WHT) at the
Observatorio del Roque de los Muchachos (La Palma, Canary Islands).

\section{Conclusions}
\label{sec:concl}

Spectroscopic surveys can serve as exceptional sources
not only of stellar parameter values but also of nature
of interstellar dust and its distribution in the Milky Way.
Stellar atmospheric parameters, when used with trigonometric
parallaxes, provide us with an exceptional opportunity to
estimate independently both the distance $d$ and interstellar extinction $A_V$.

We have cross-identified objects in LAMOST DR5 spectroscopic survey
and Gaia DR2/EDR3 surveys. For 40 test areas located
at high galactic latitudes ($\left|b\right| > 20^{\circ}$) we have constructed
$A_V(d)$ relations and approximated them by the cosecant law~\eqref{equ:parenago}.
We have determined then $a_0$ and $\beta$ values for each area
and made a 2D approximation for the entire sky (except low galactic latitudes
and the southern polar cap). We have also estimated
the Galactic extinction for the test areas by spherical
harmonics (Eqs. \ref{equ:spher-a0}-\ref{equ:spher-Agal}), and the comparison
of our results with data from \cite{2011ApJ...737..103S} demonstrates a good agreement.

The Eqs.~\eqref{equ:spher-a0}, \eqref{equ:spher-beta}
can be used for calculation of parameters $a_0$ and $\beta$
and, consequently, for estimation of interstellar extinction from~(\ref{equ:parenago}).
These results are valid for high Galactic latitudes ($\left|b\right| > 20^{\circ}$)
and distances within about 6 to 8 kpc from the Sun.
The Eq.~\eqref{equ:spher-Agal} allows one to estimate the Galactic extinction
for high Galactic latitudes.

\begin{acknowledgement}
We are grateful to our reviewers whose constructive comments greatly helped us to improve the paper.
We thank Alexei Sytov for valuable remarks and suggestions.
The work was partly supported by NSFC/RFBR grant 20-52-53009. KG research was supported by the Russian Science Foundation (RScF) grant No. 17-72-20119. AN research has been supported by the Interdisciplinary Scientific
and Educational School of Moscow University ``Fundamental and Applied
Space Research''.
This research has made use of NASA's Astrophysics Data System,
and use of TOPCAT, an interactive graphical viewer and editor
for tabular data \cite{2005ASPC..347...29T}.
\end{acknowledgement}

{\bf Author Contributions:} The authors made equal contribution to this work; O.M. wrote the paper.


%

\bibliographystyle{inasan}
\bibliography{iem}

\begin{appendix}
\large{Appendix A. Spherical harmonics}

We use orthonormalized spherical harmonics $Y_n^m$ to calculate 2D approximation of extinction parameters. Only real parts of harmonics are used, hence we use harmonics of non-negative order $0 \le m \le n$. Therefore spherical harmonics are calculated by the following formulas:
\begin{equation*}
Y_0^0 (l, b) = \frac{1}{2} \sqrt{\frac{1}{\pi}}, \quad Y_1^0 (l, b) = \frac{1}{2} \sqrt{\frac{3}{\pi}} \cos b,
\end{equation*}

\begin{equation*}
Y_1^1 (l, b) = \frac{1}{2} \sqrt{\frac{3}{\pi}} \cos l \sin b, \quad Y_2^0 (l, b) = \frac{1}{4} \sqrt{\frac{5}{ \pi}} \left( 3 \cos^2 b - 1\right),
\end{equation*}

\begin{equation*}
Y_2^1 (l, b) = \frac{1}{2} \sqrt{\frac{15}{\pi}} \cos l \sin b \cos b, \quad Y_2^2 (l, b) = \frac{1}{4} \sqrt{\frac{15}{\pi}} \cos 2 l \sin^2 b.
\end{equation*}

\end{appendix}

\bigskip

\begin{appendix}
\large{Appendix B. Figures. Best-fit $\chi^2$ minimization and $\chi^2$ scan solutions for the areas
used for determination of the interstellar extinction.}

\begin{figure}[h]
\resizebox{1.0\columnwidth}{!}{%
  \includegraphics{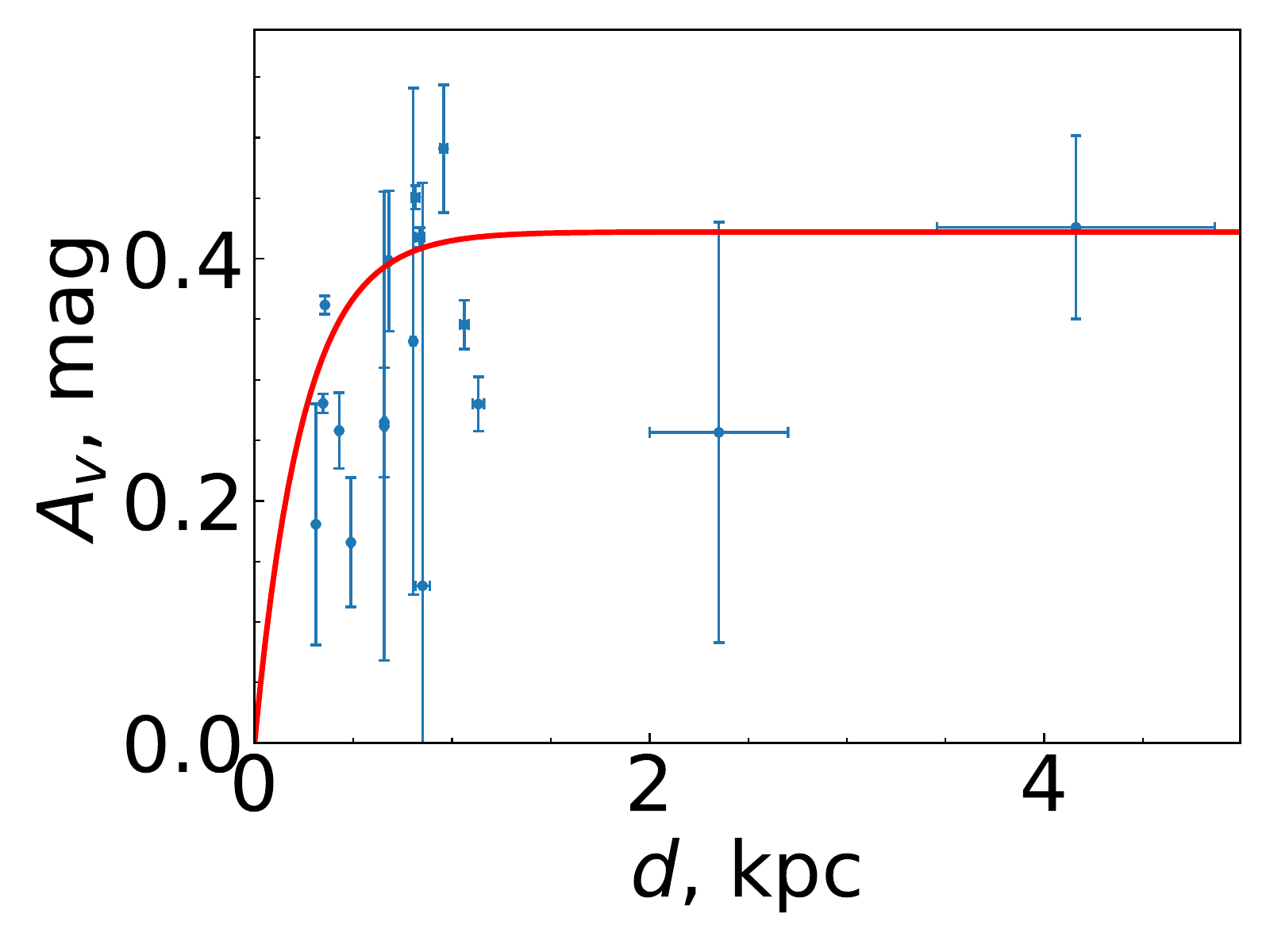}
  \includegraphics{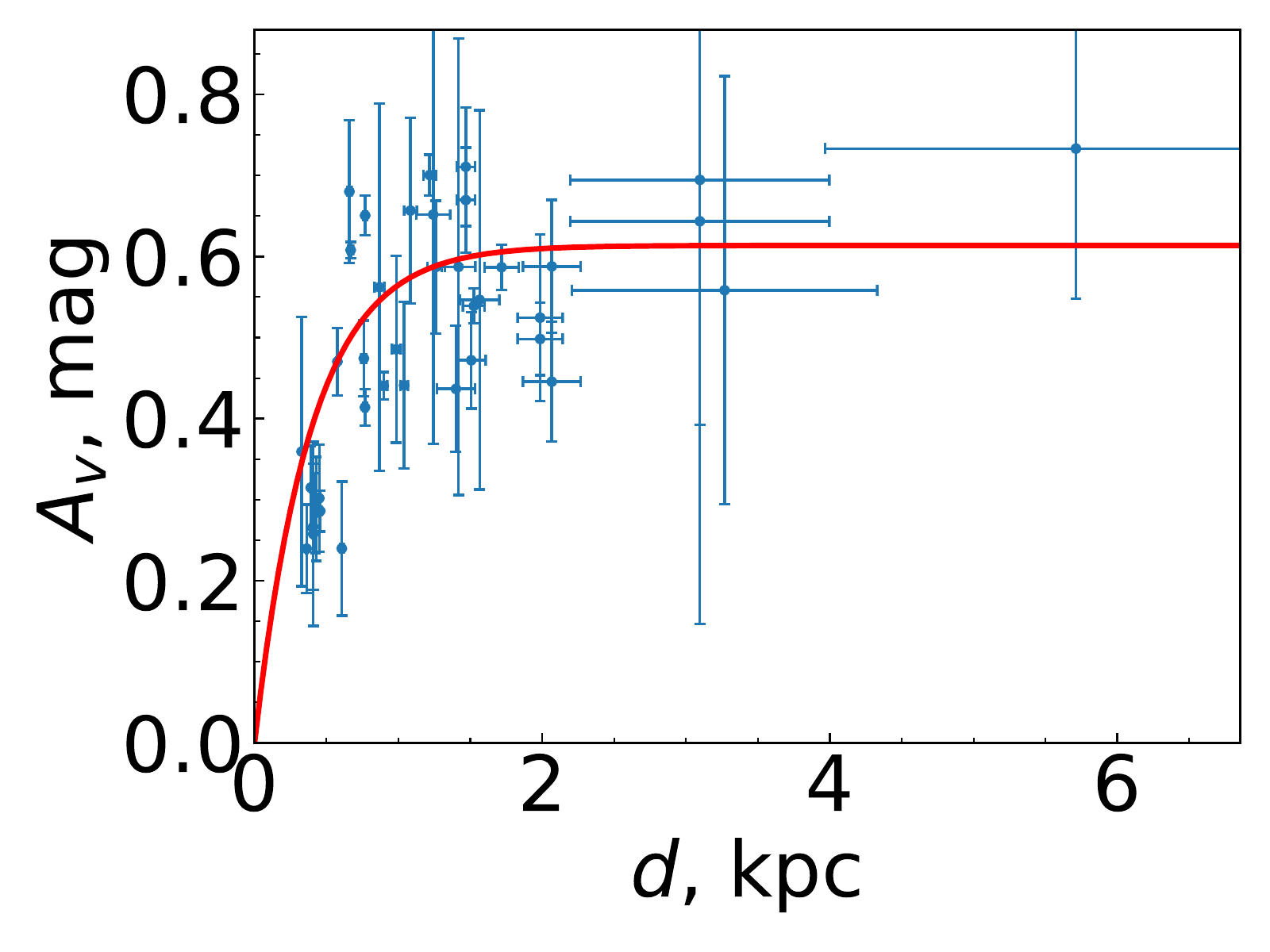}
  \includegraphics{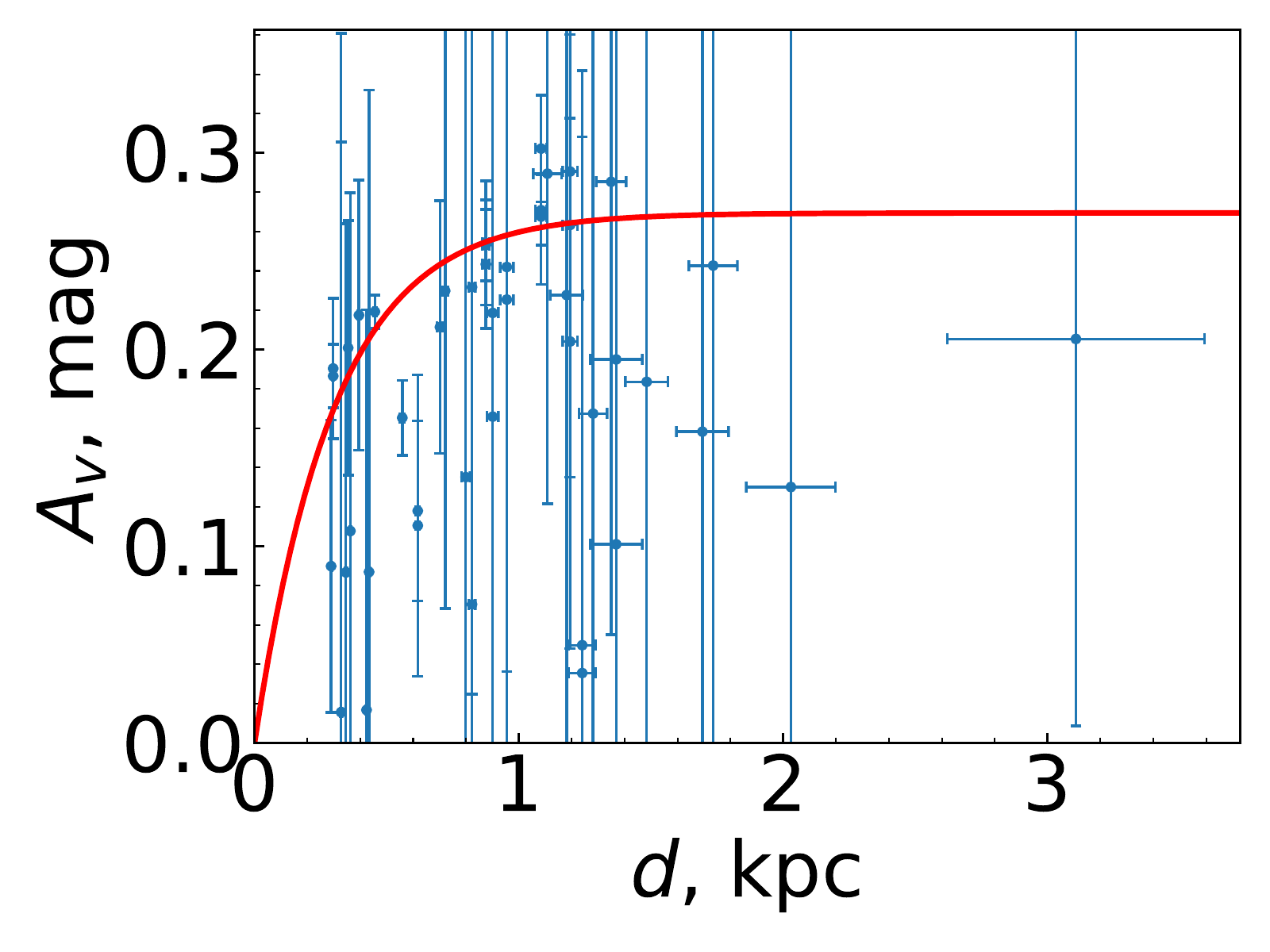}
 }
\caption{Best-fit for 1713, 2482, 3733.}
\end{figure}

\begin{figure}[h]
\resizebox{1.0\columnwidth}{!}{%
  \includegraphics{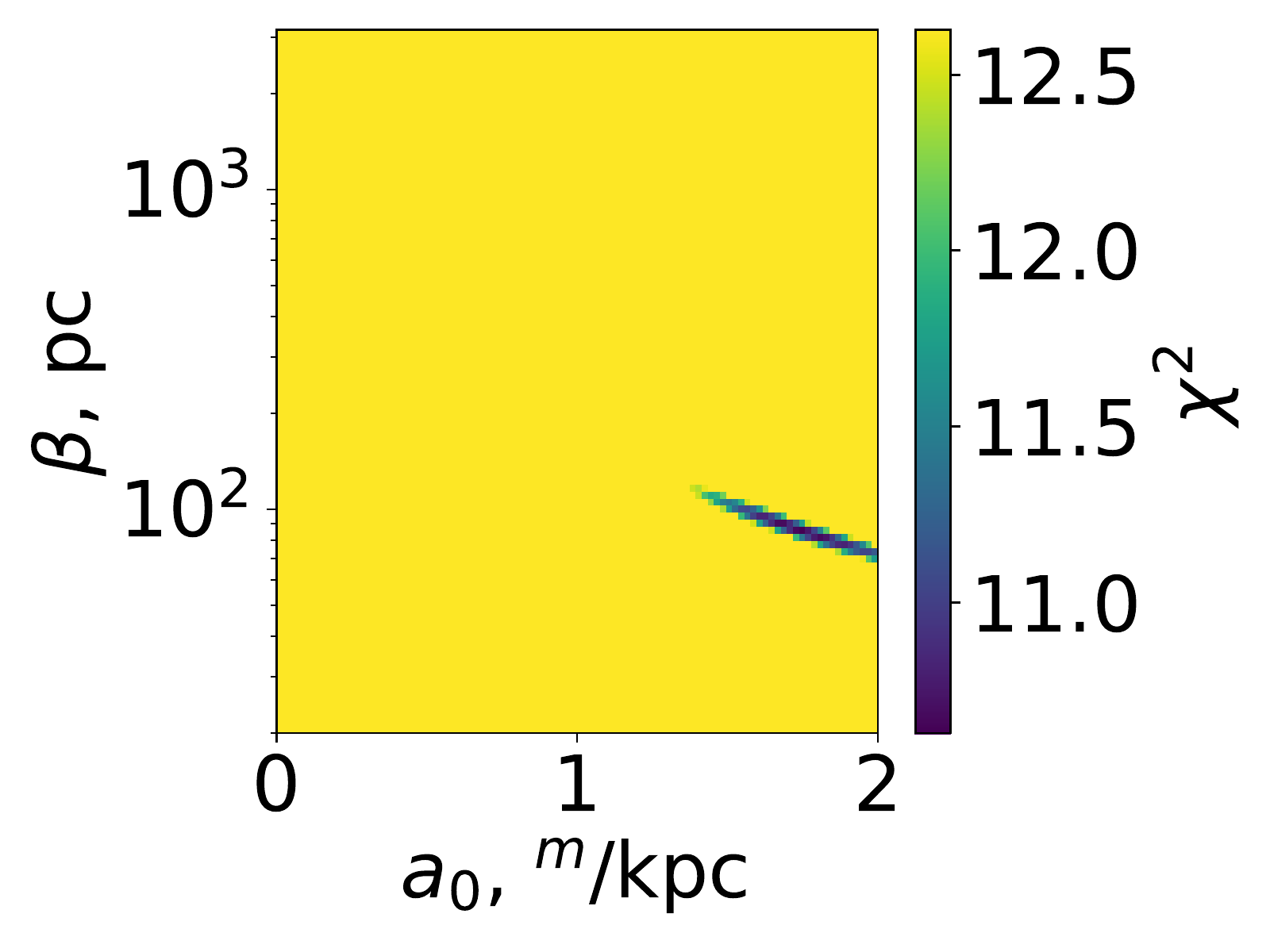}
  \includegraphics{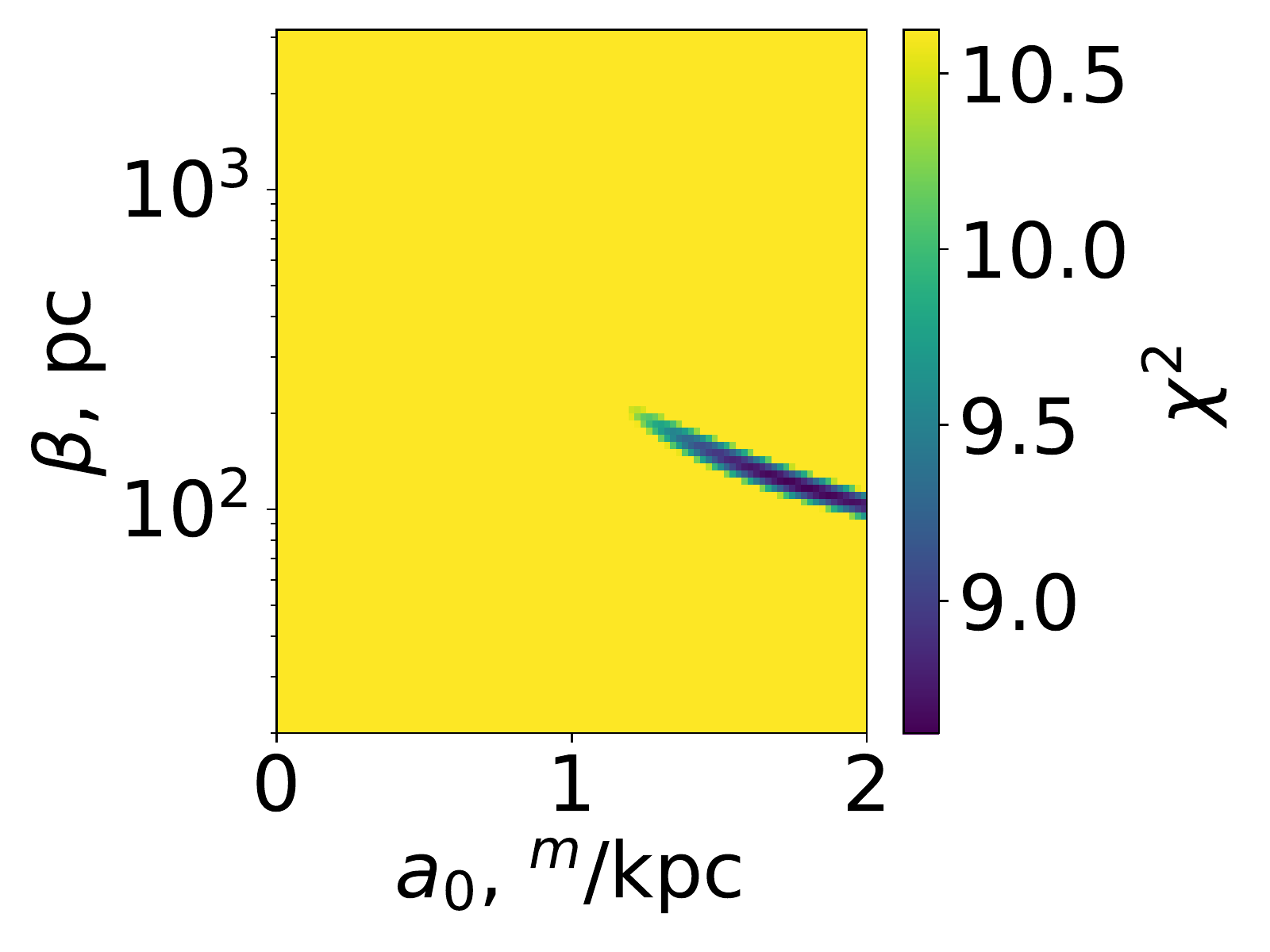}
  \includegraphics{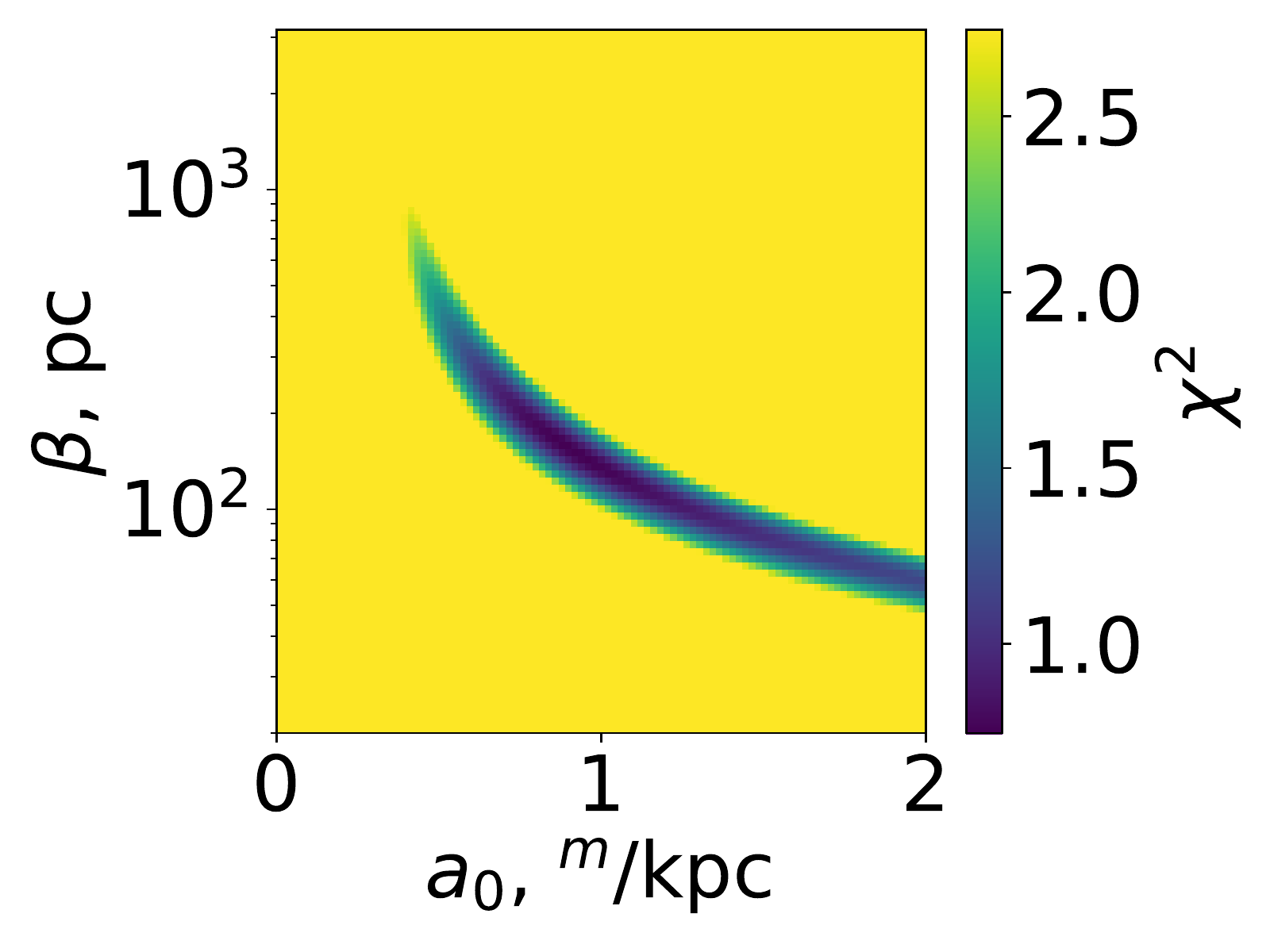}
 }
\caption{$\chi^2$ scan for 1713, 2482, 3733.}
\end{figure}

\begin{figure}[h]
\resizebox{1.0\columnwidth}{!}{%
  \includegraphics{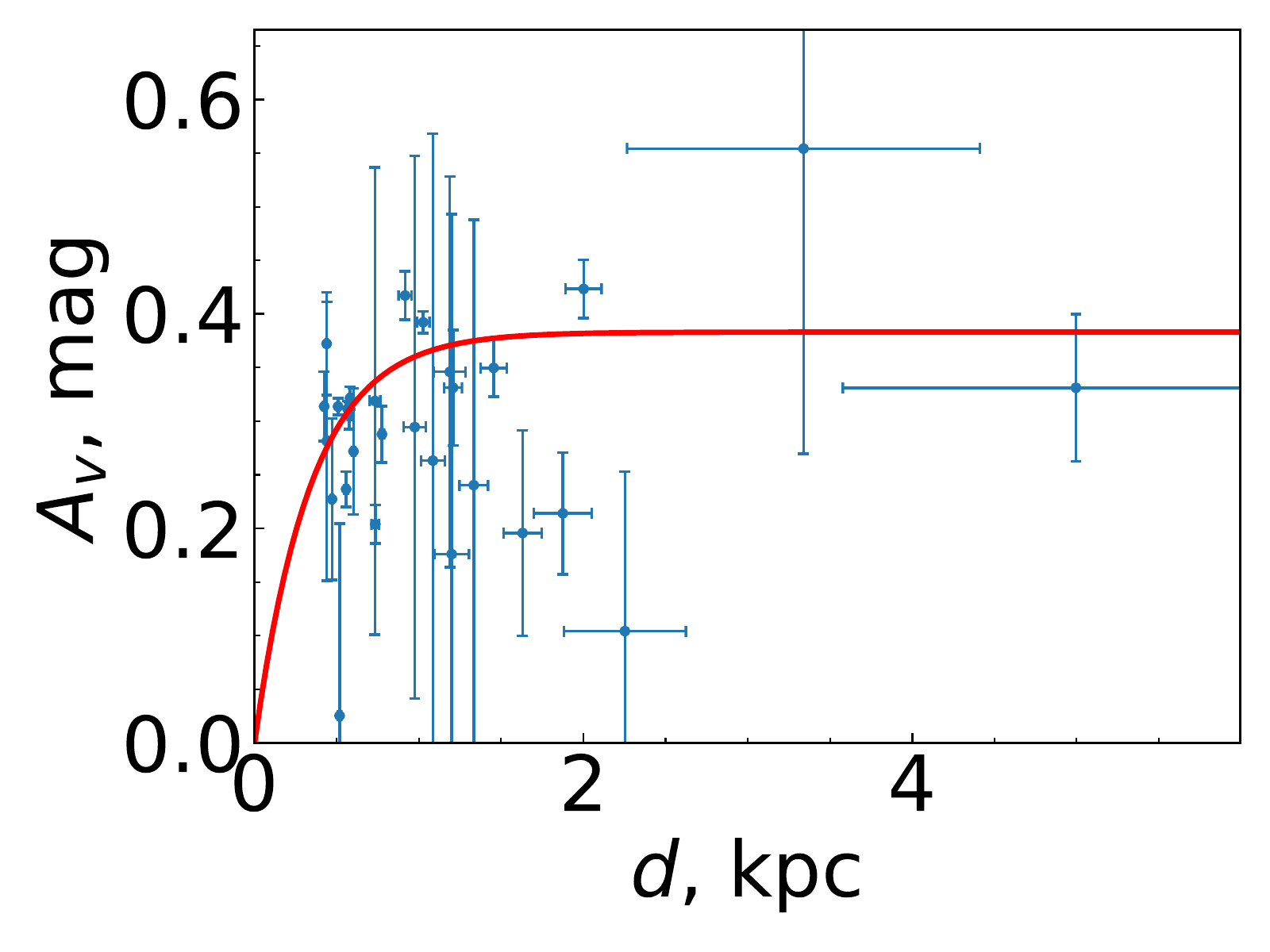}
  \includegraphics{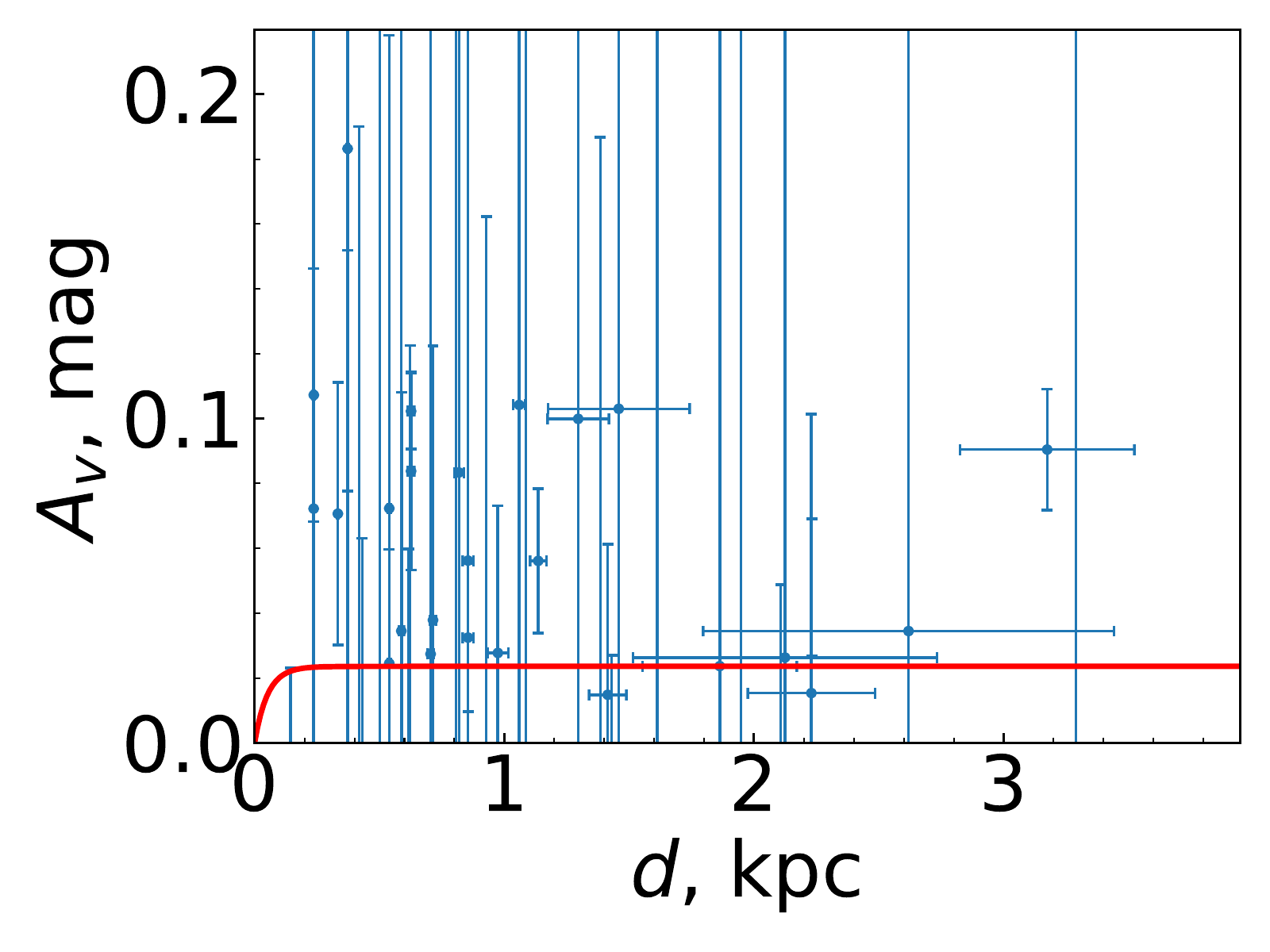}
  \includegraphics{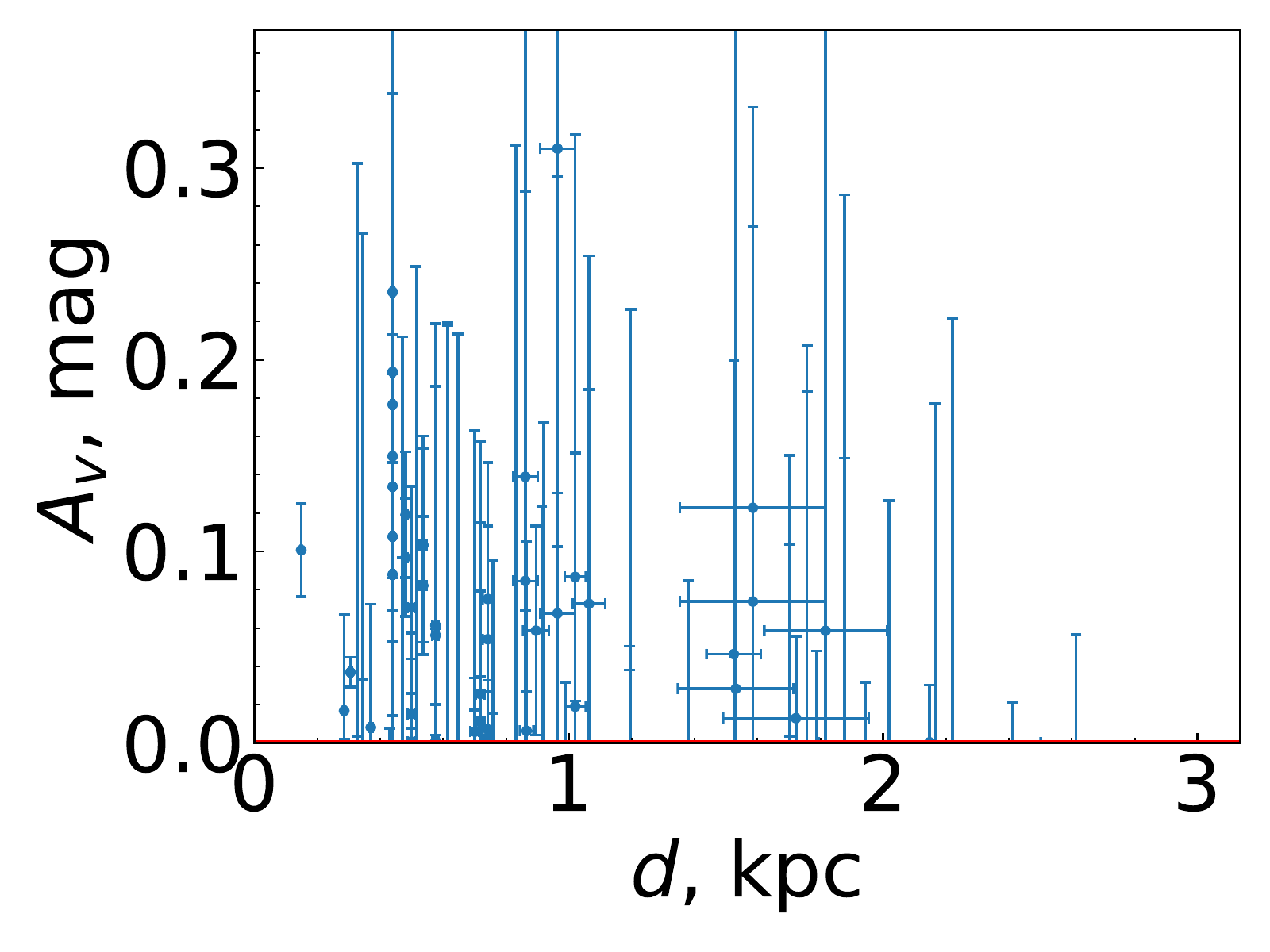}
 }
\caption{Best-fit for 8789, 13305, 16342.}
\end{figure}

\begin{figure}[h]
\resizebox{1.0\columnwidth}{!}{%
  \includegraphics{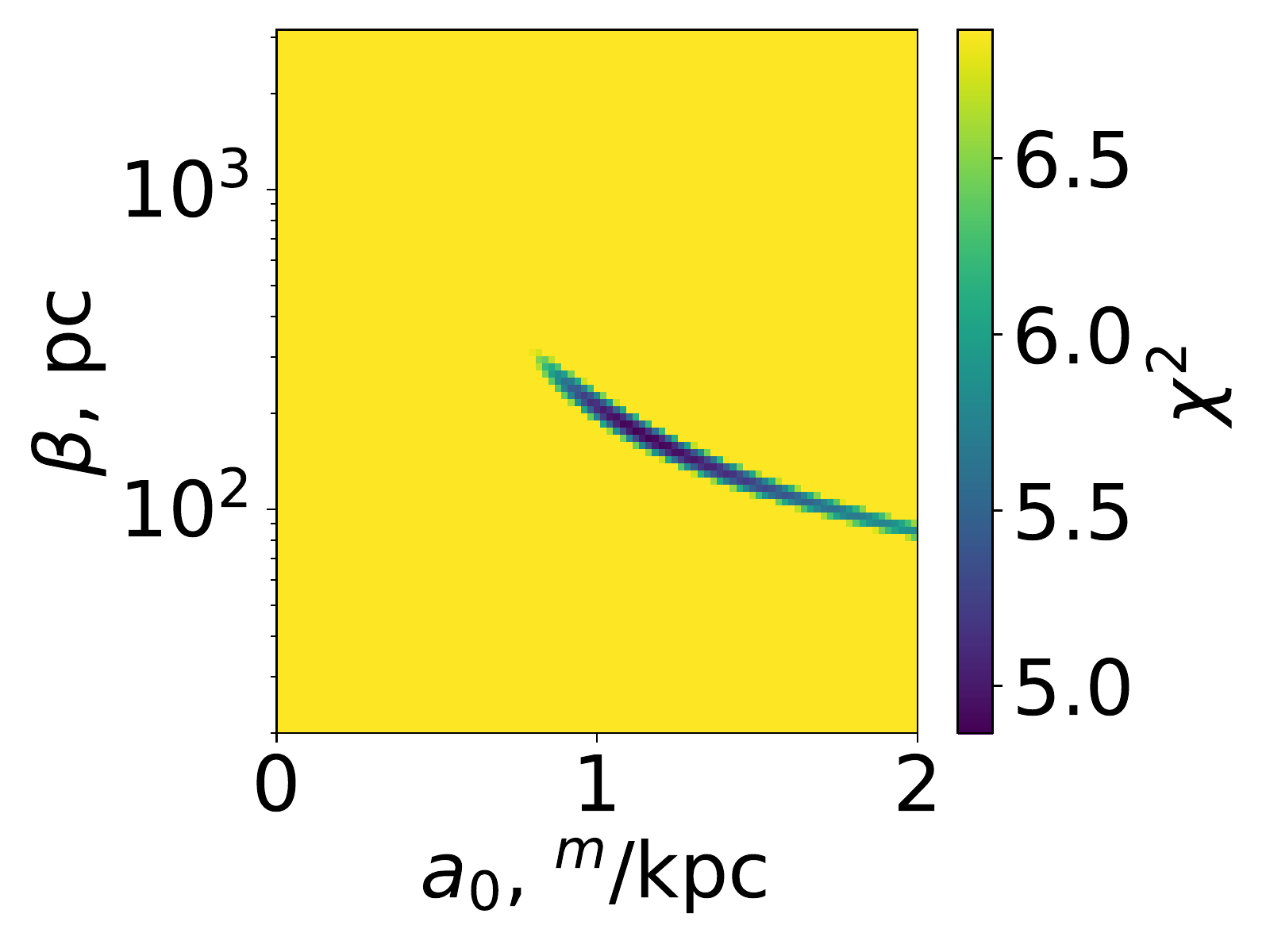}
  \includegraphics{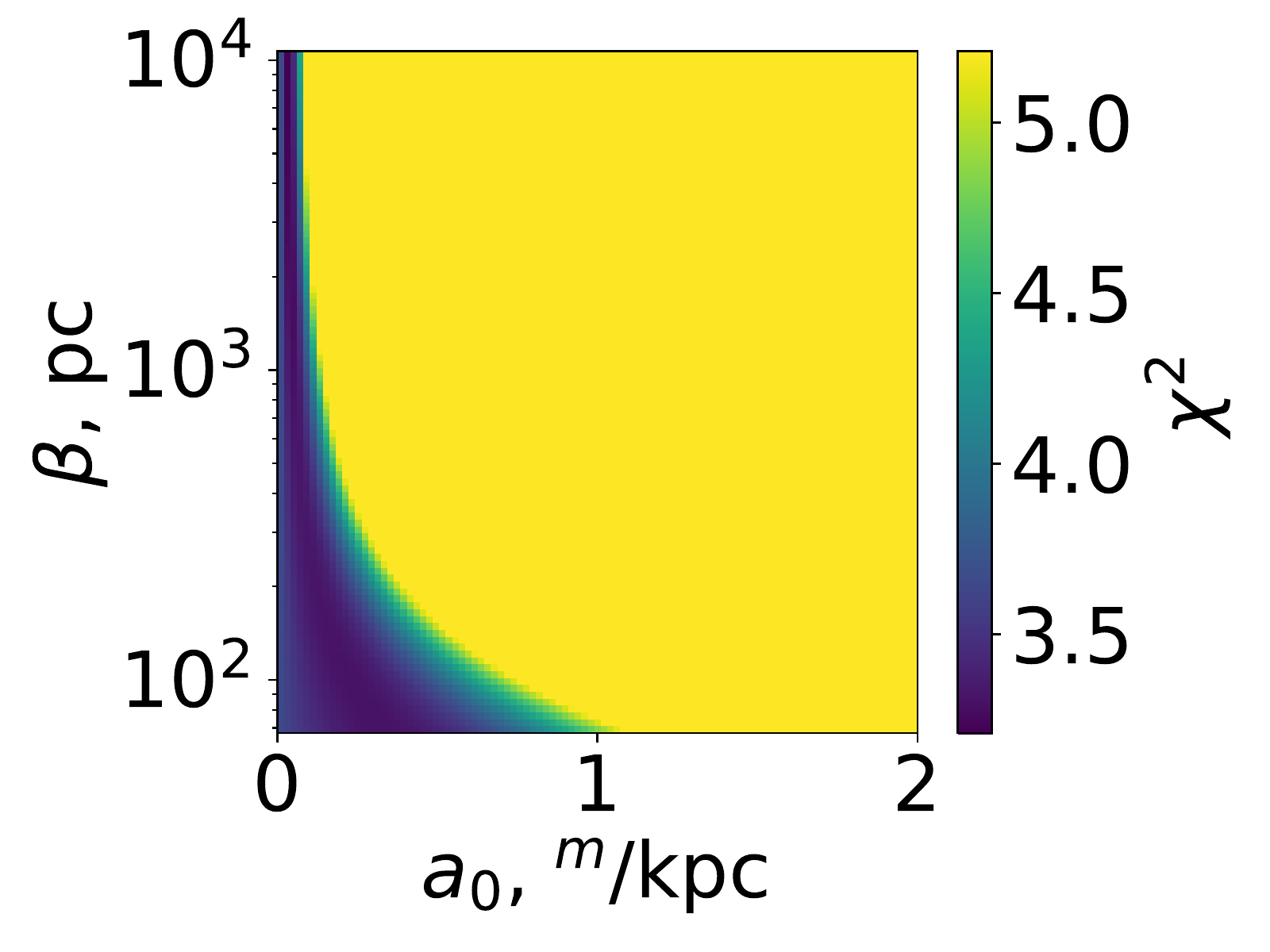}
  \includegraphics{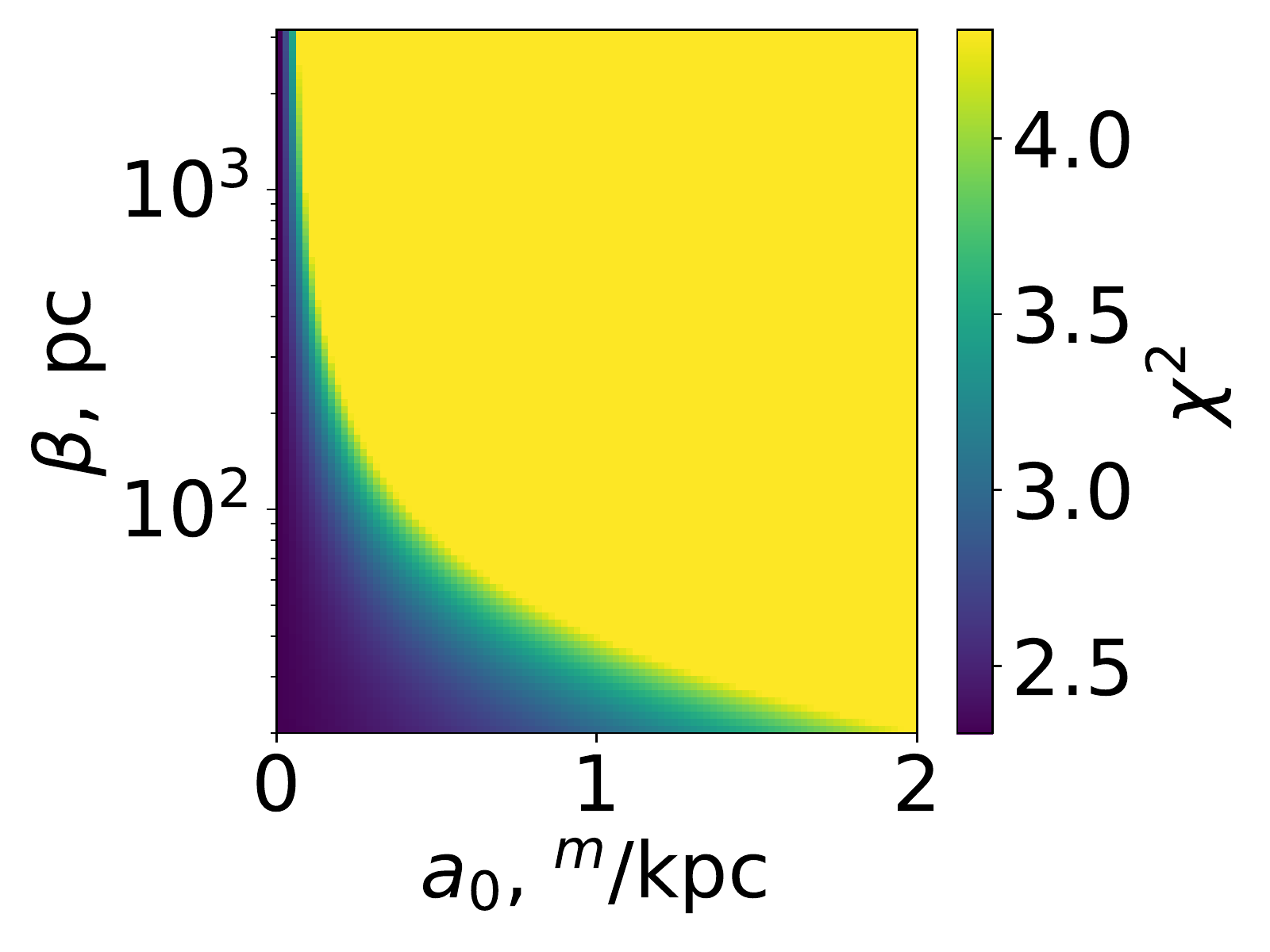}
 }
\caption{$\chi^2$ scan for 8789, 13305, 16342.}
\end{figure}

\begin{figure}[h]
\resizebox{1.0\columnwidth}{!}{%
  \includegraphics{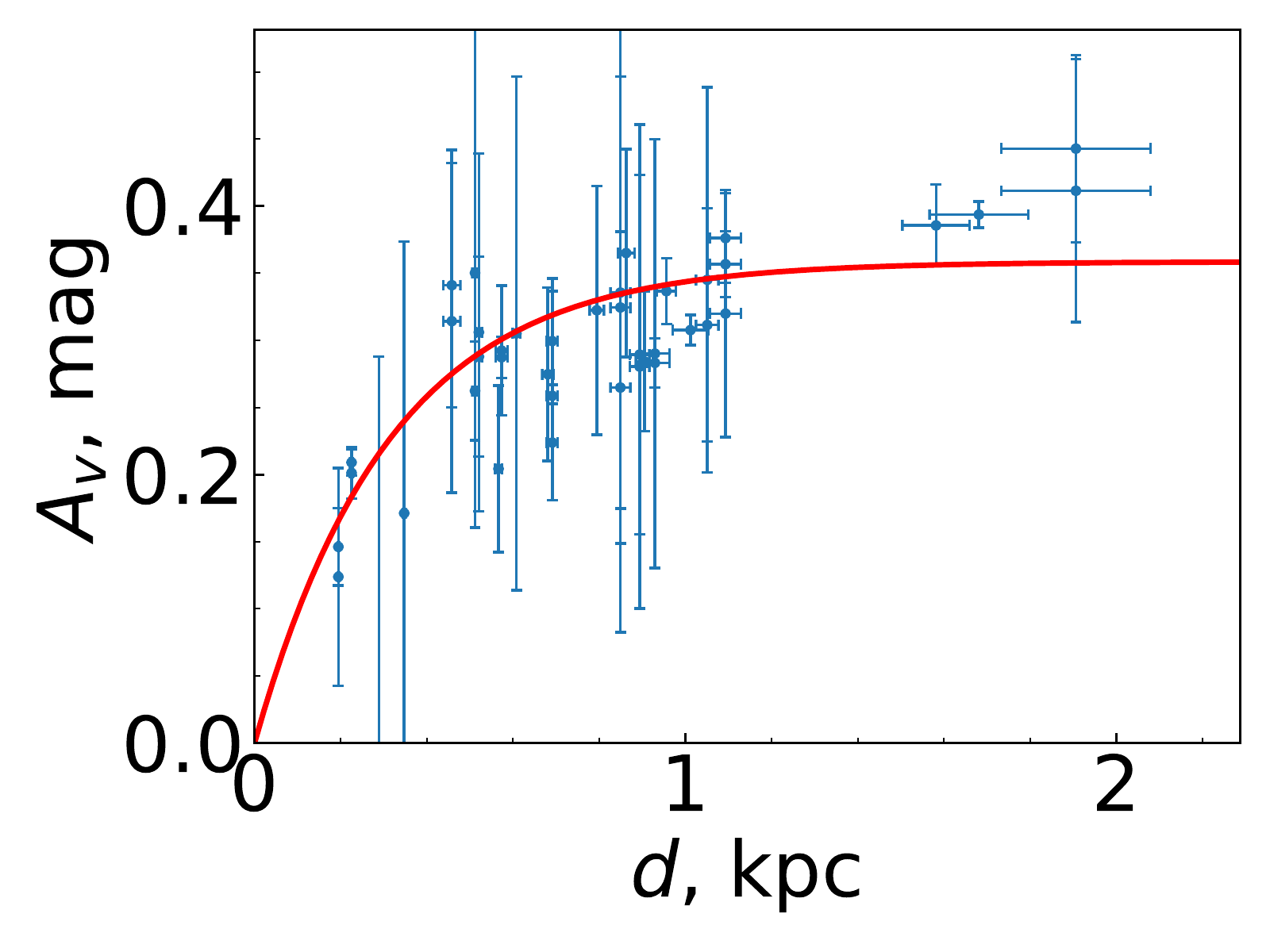}
  \includegraphics{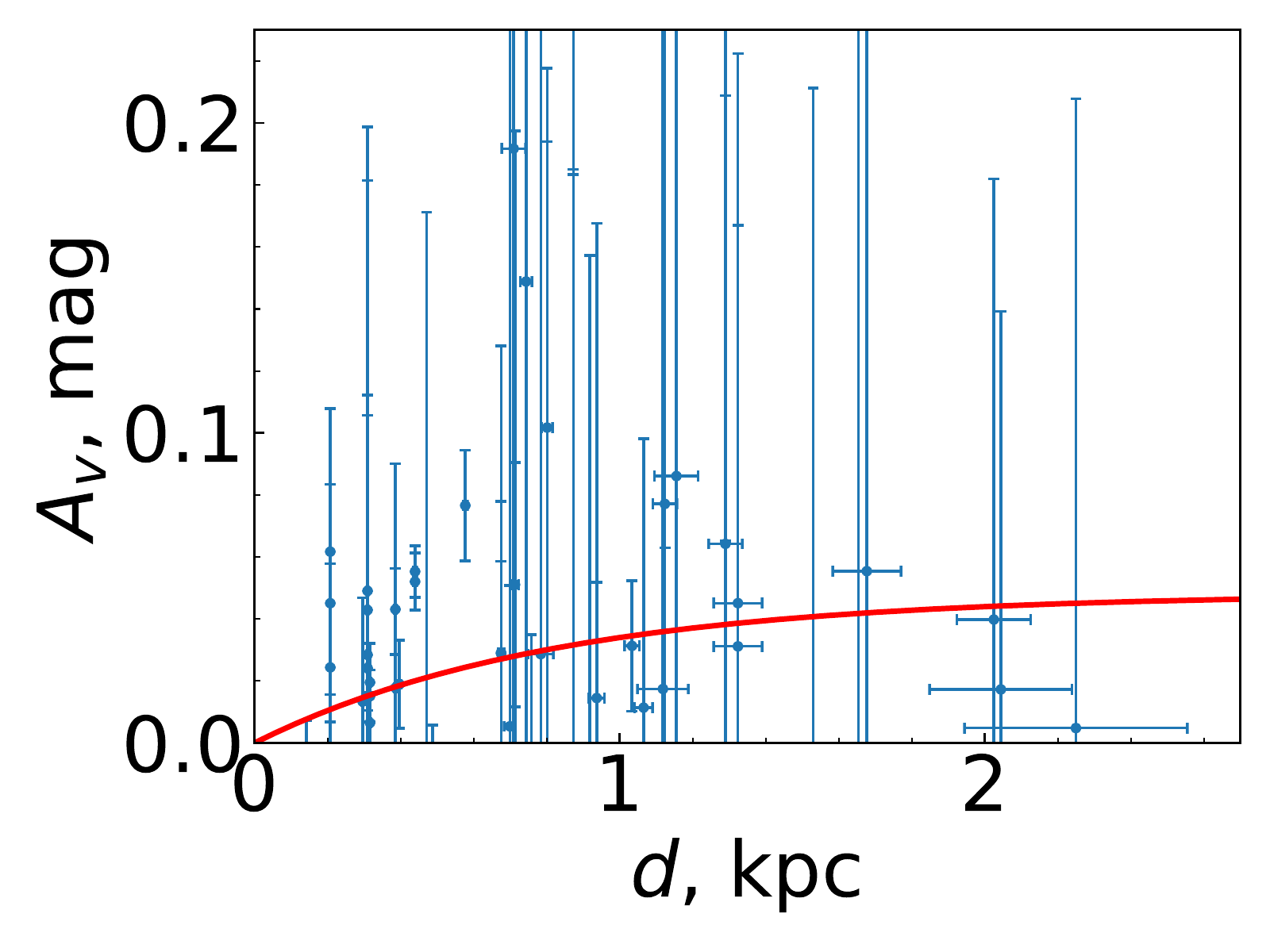}
  \includegraphics{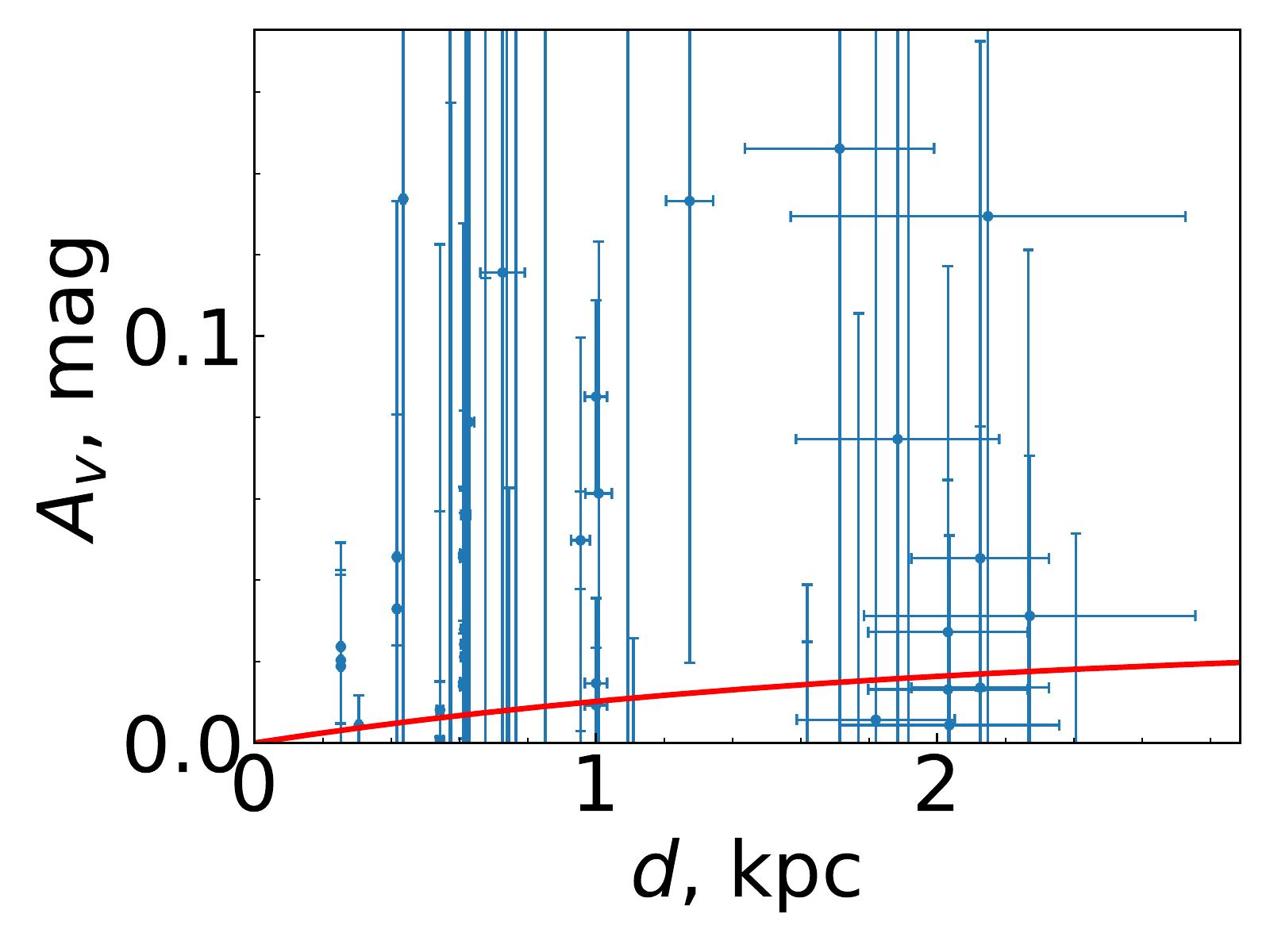}
 }
\caption{Best-fit for 17777, 28666, 29945.}
\end{figure}

\begin{figure}[h]
\resizebox{1.0\columnwidth}{!}{%
  \includegraphics{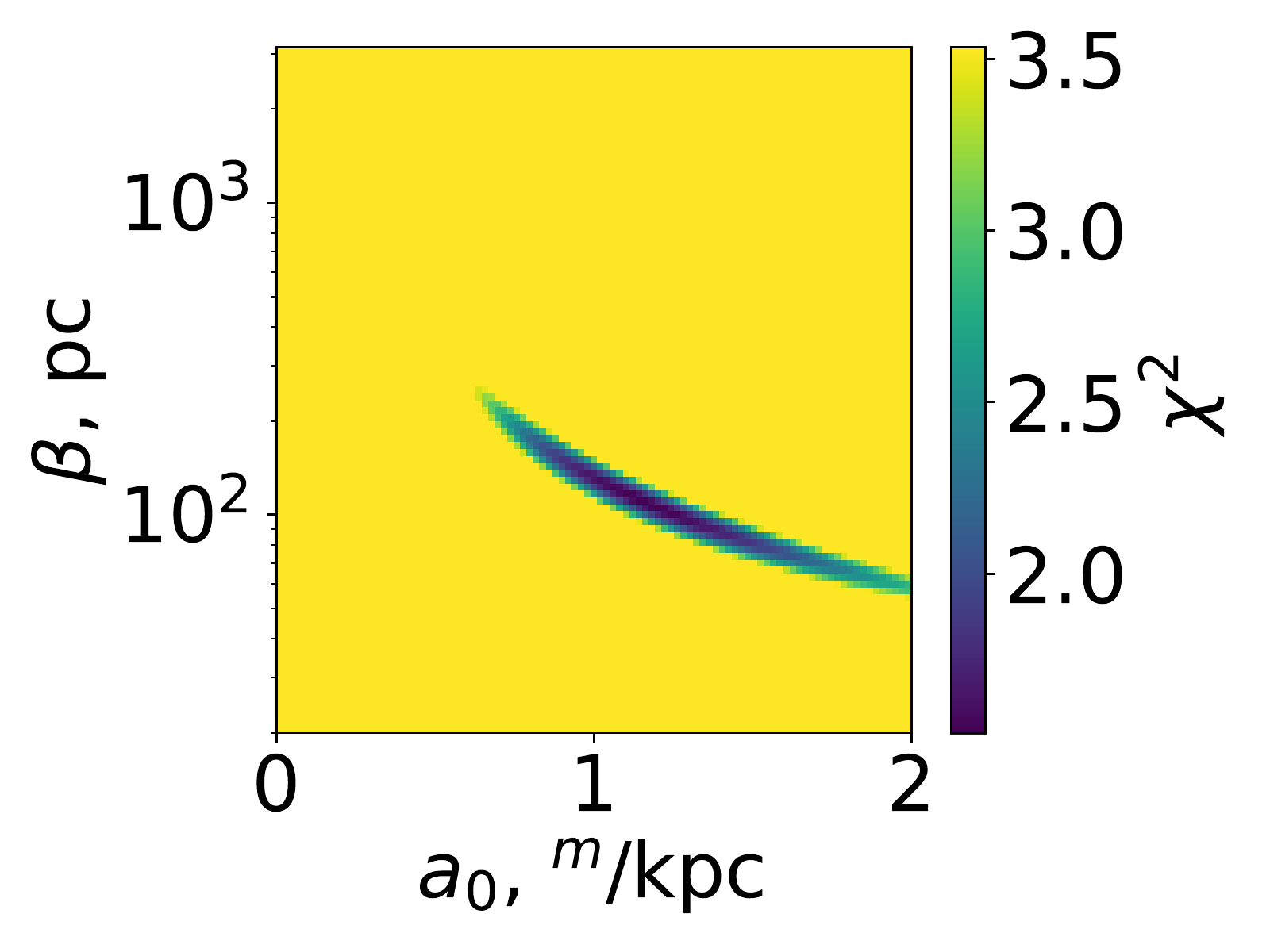}
  \includegraphics{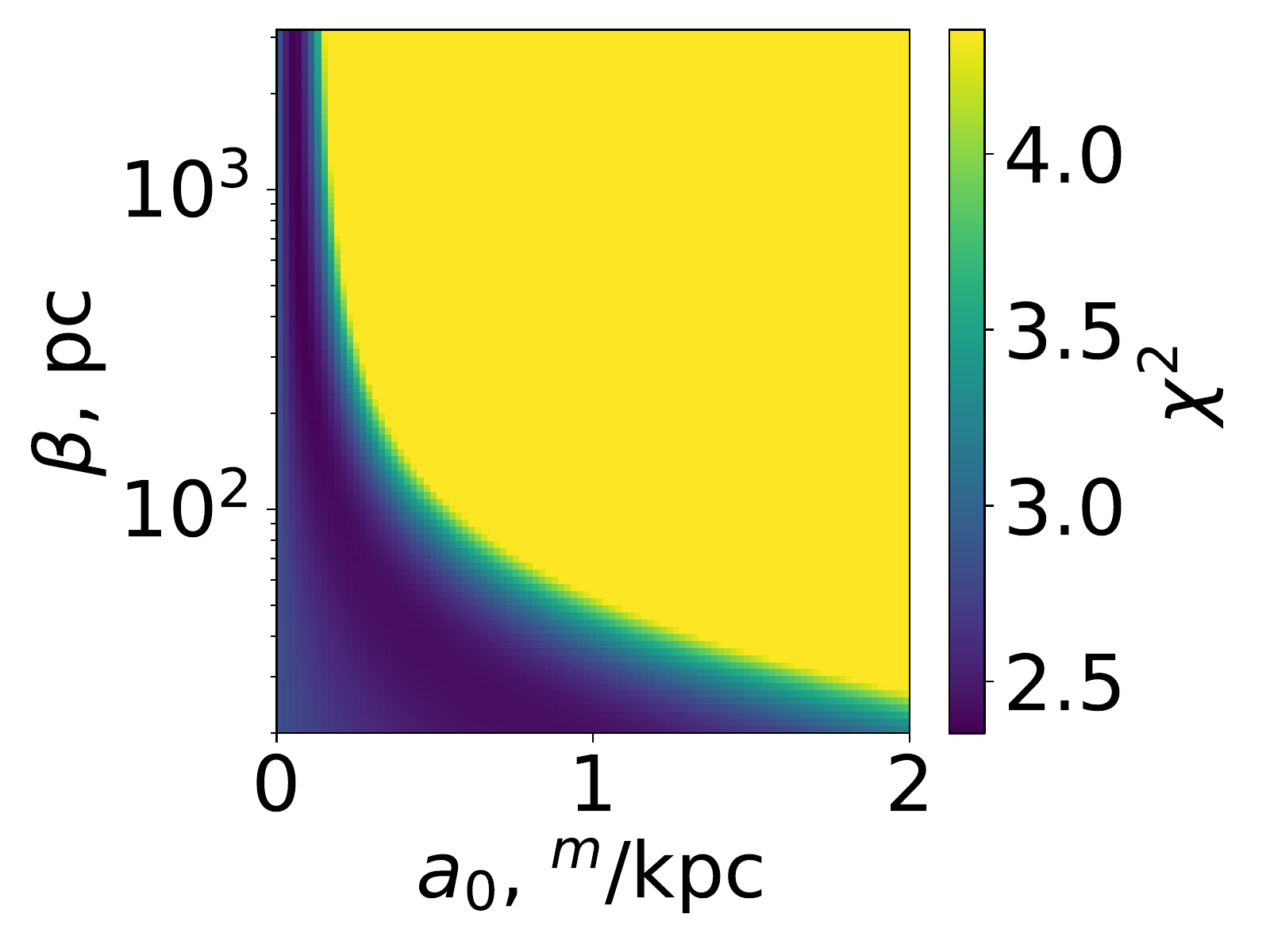}
  \includegraphics{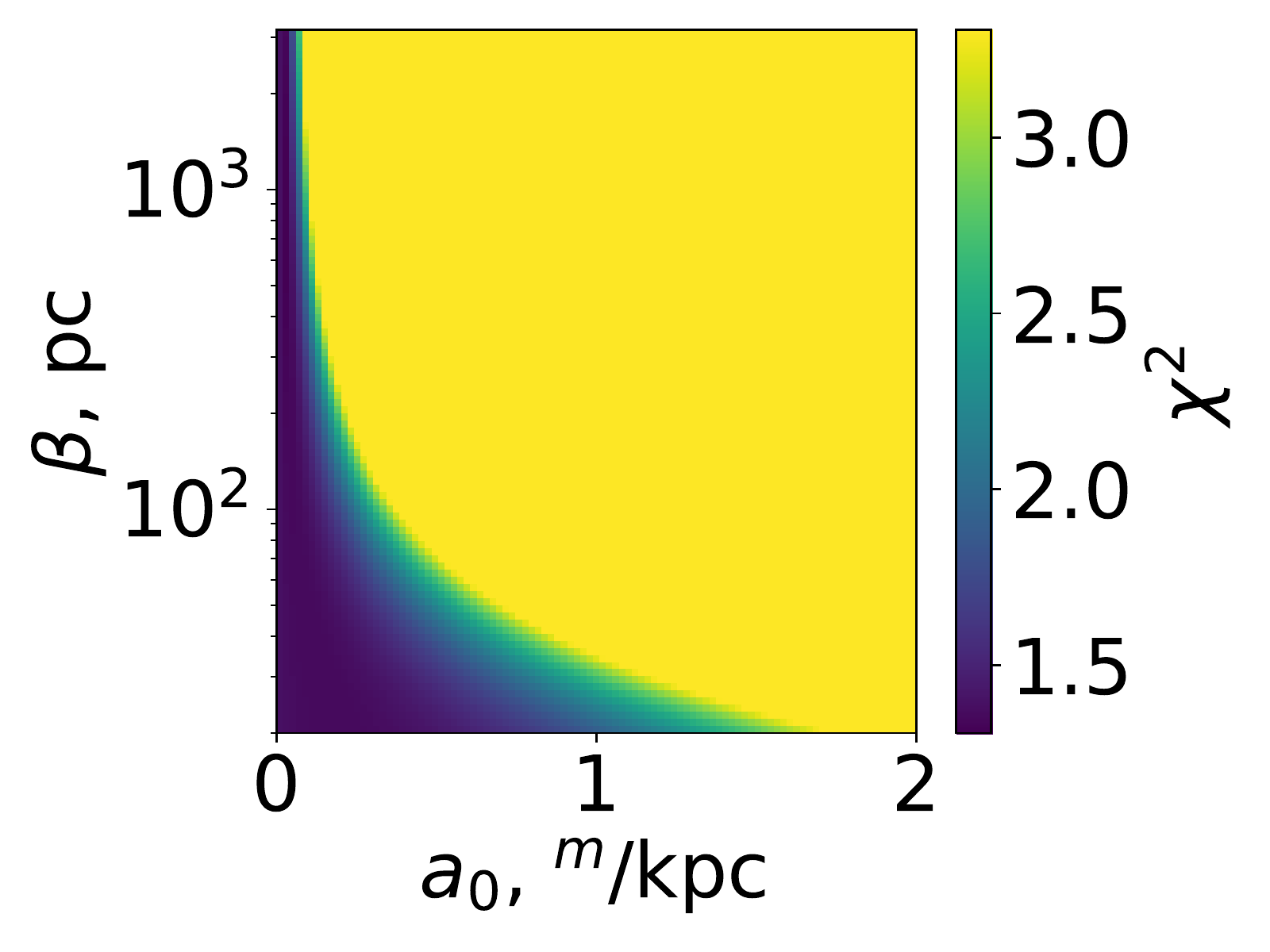}
 }
\caption{$\chi^2$ scan for 17777, 28666, 29945.}
\end{figure}

\begin{figure}[h]
\resizebox{1.0\columnwidth}{!}{%
  \includegraphics{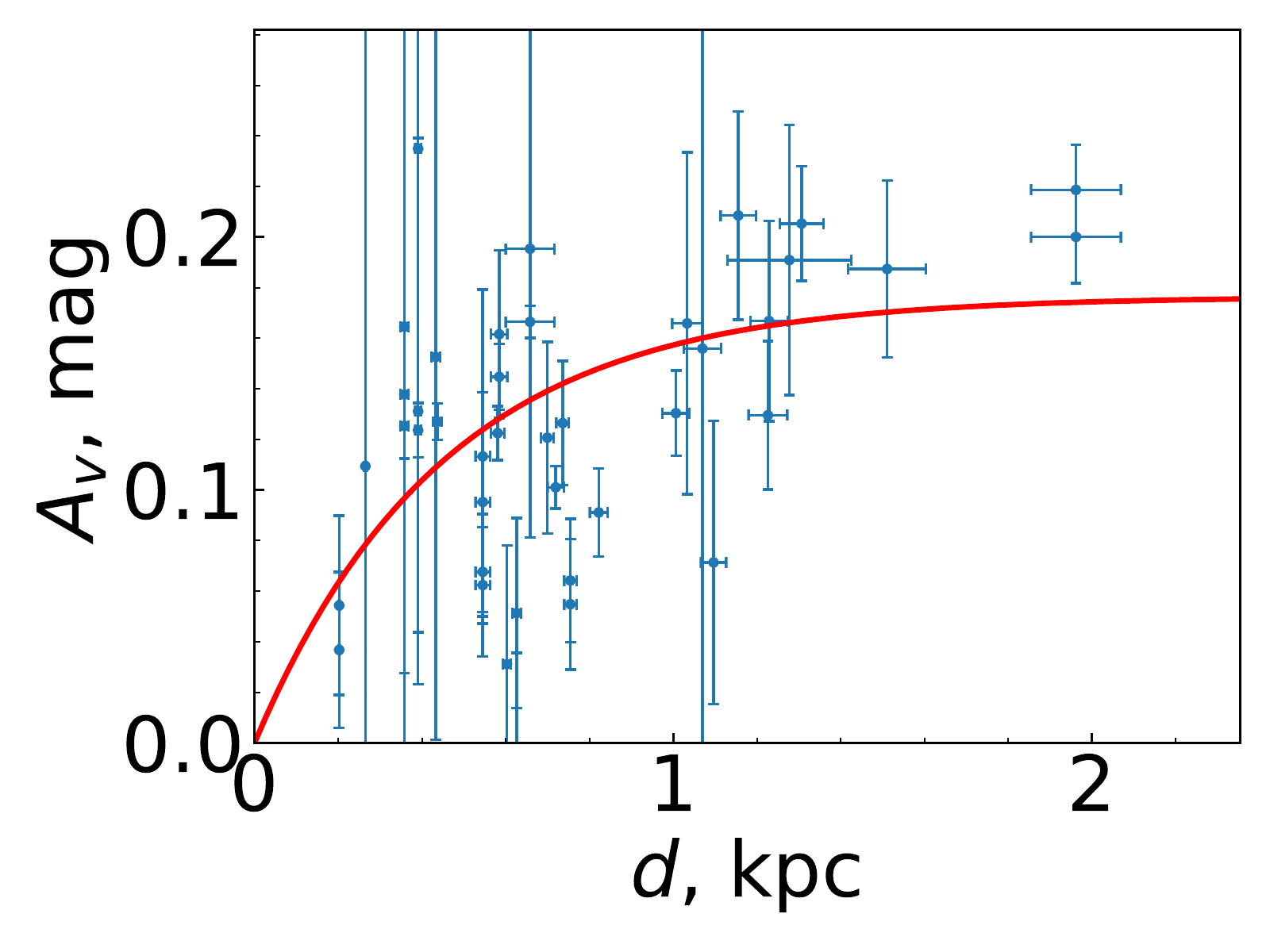}
  \includegraphics{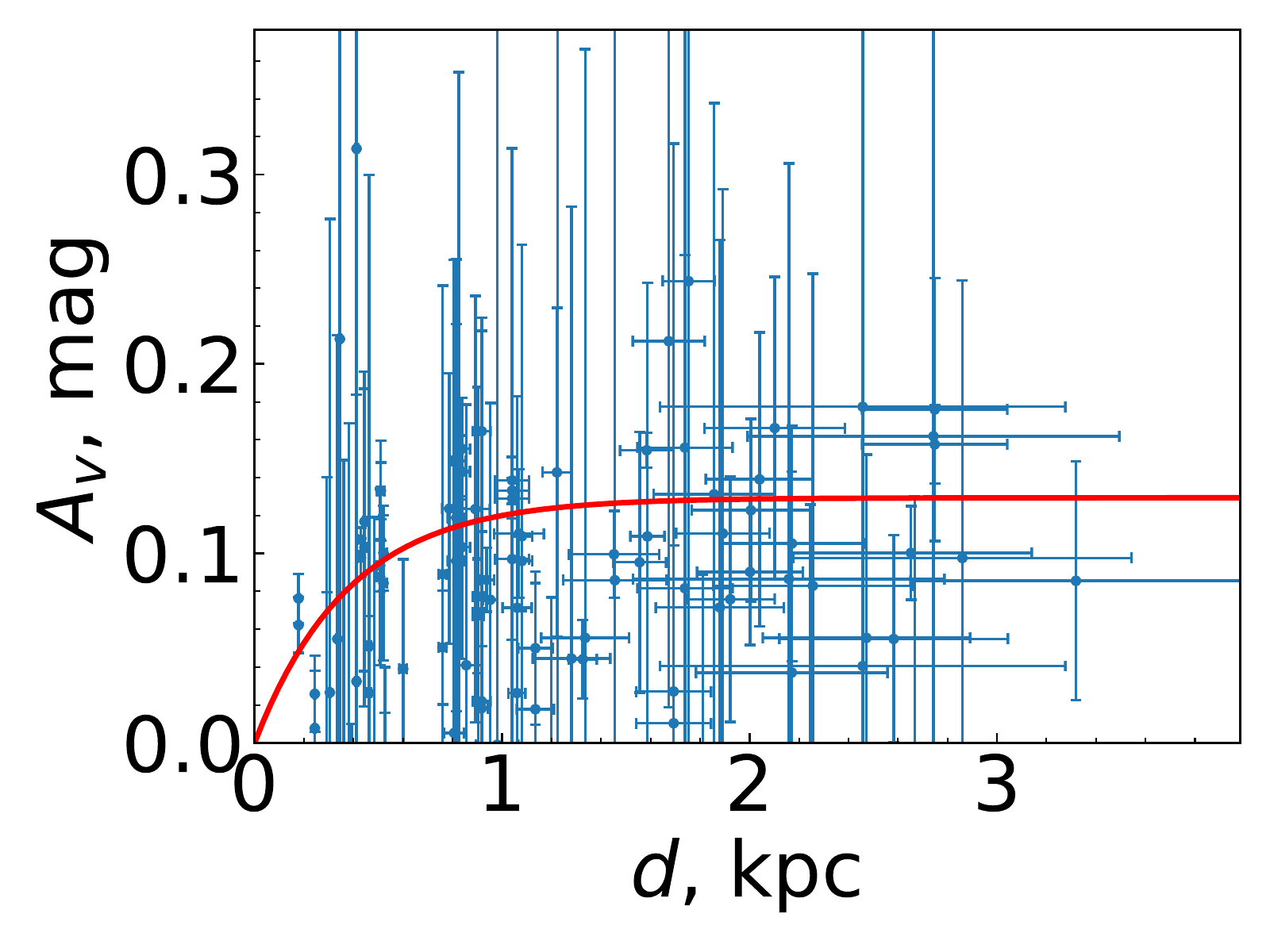}
  \includegraphics{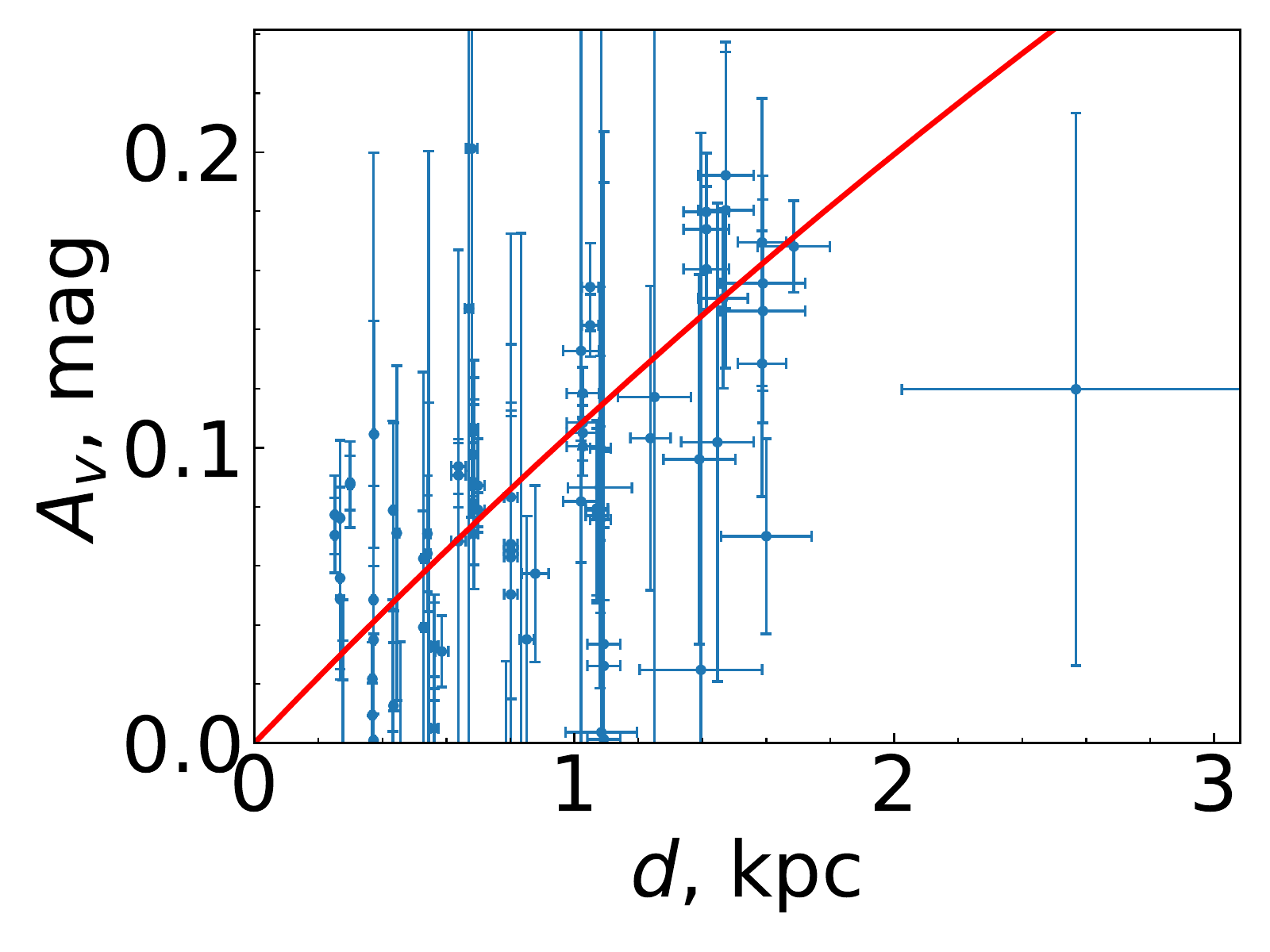}
 }
\caption{Best-fit for 34455, 35483, 36249.}
\end{figure}

\begin{figure}[h]
\resizebox{1.0\columnwidth}{!}{%
  \includegraphics{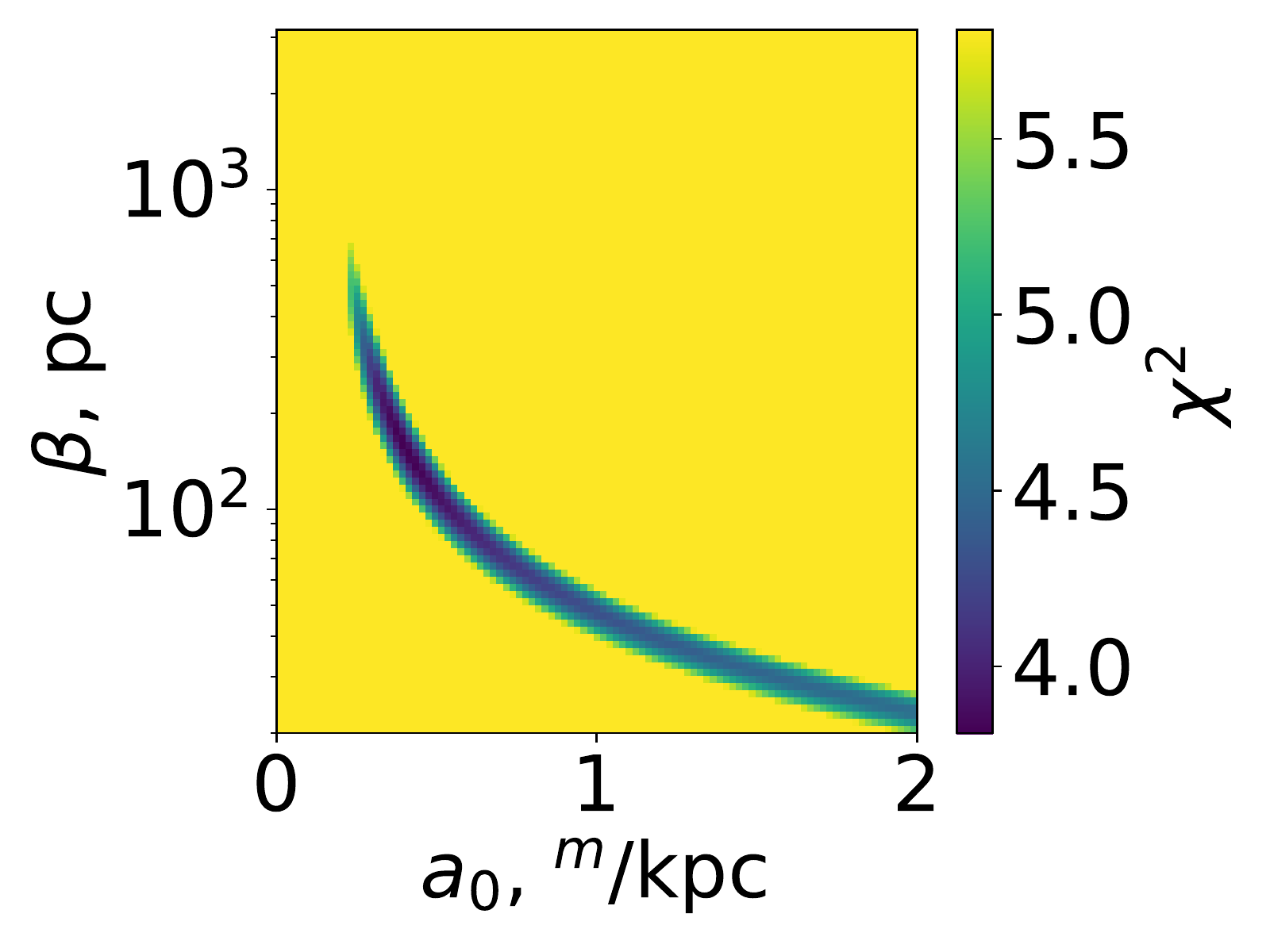}
  \includegraphics{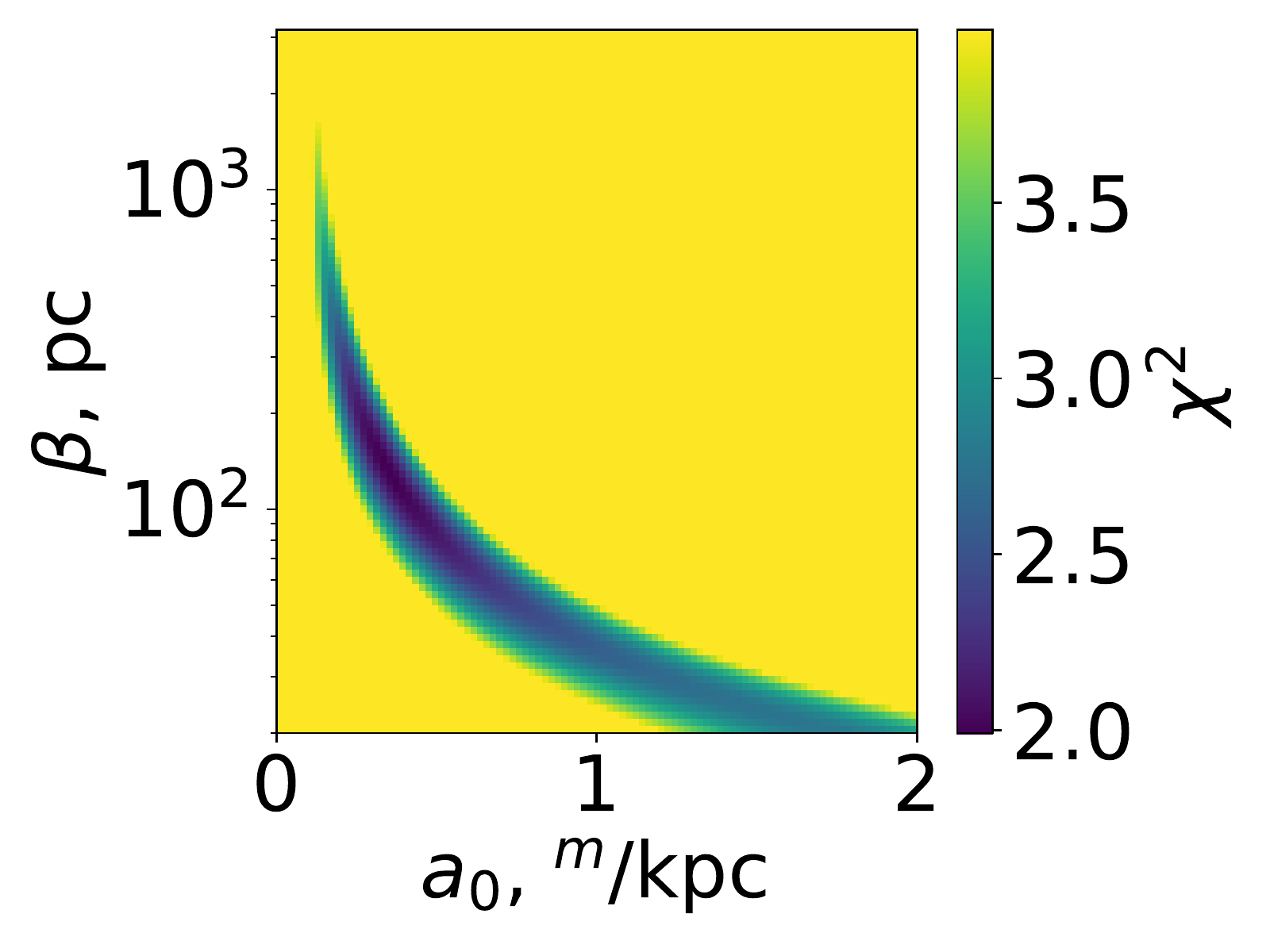}
  \includegraphics{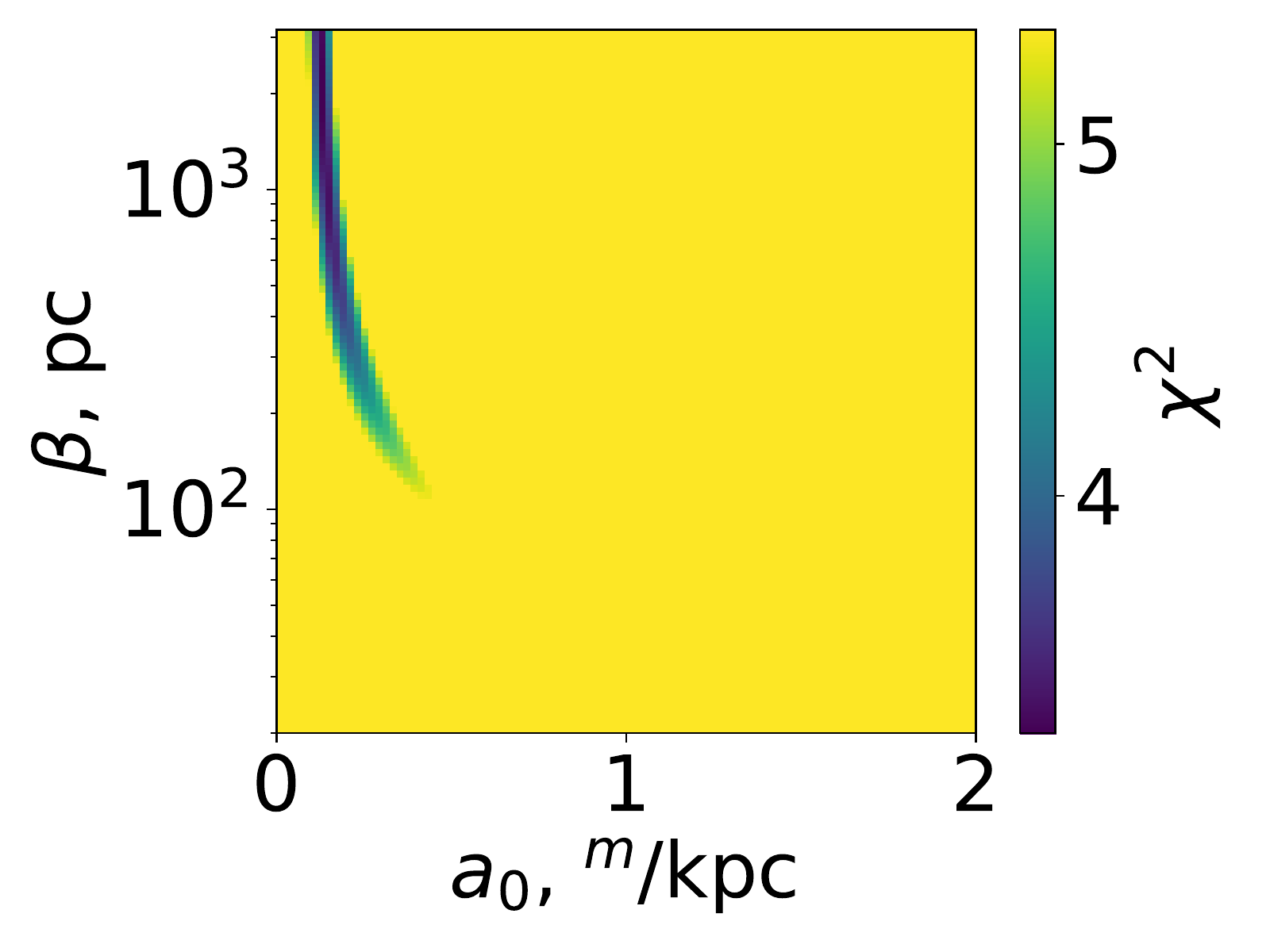}
 }
\caption{$\chi^2$ scan for 34455, 35483, 36249.}
\end{figure}

\begin{figure}[h]
\resizebox{1.0\columnwidth}{!}{%
  \includegraphics{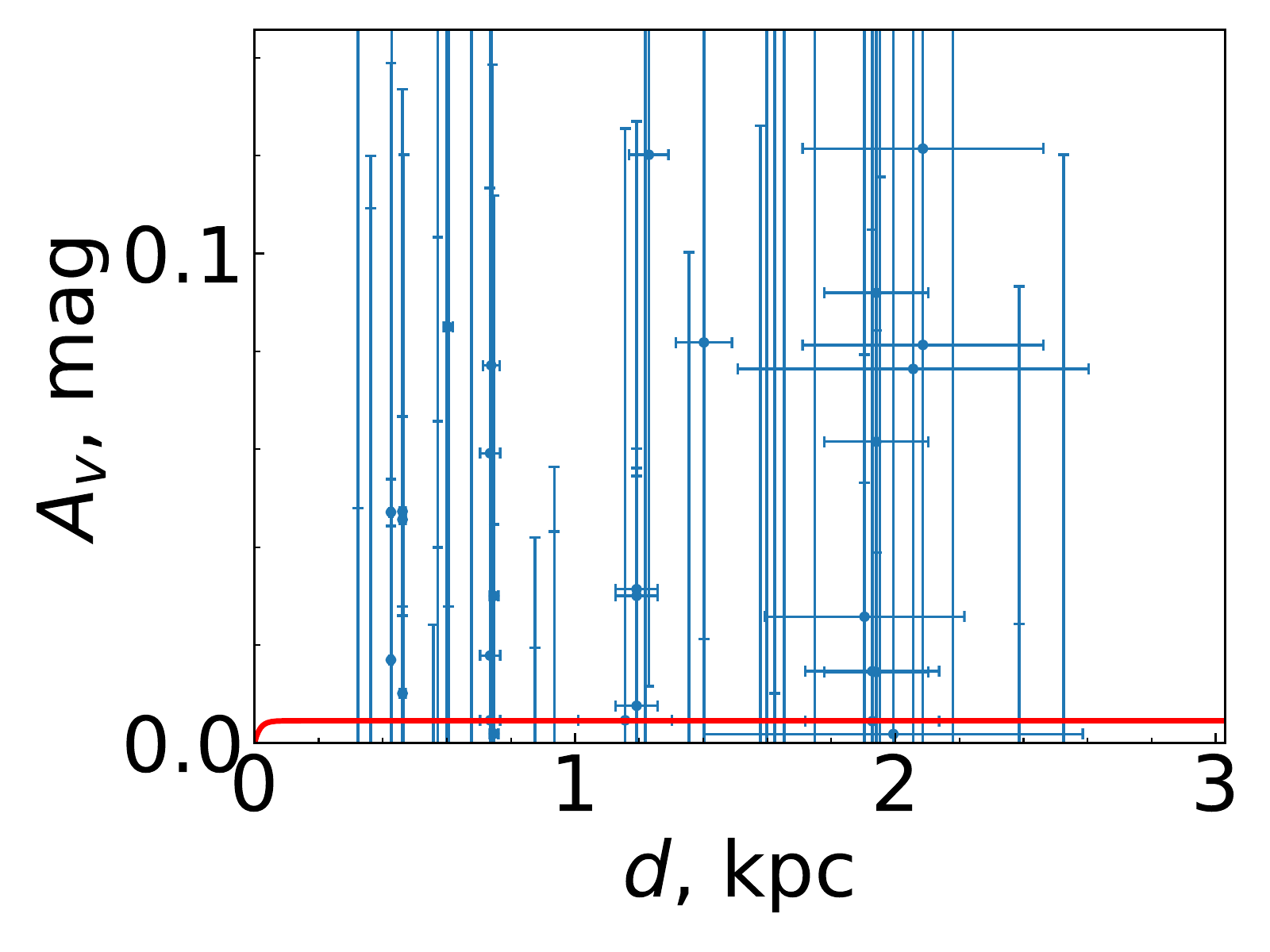}
  \includegraphics{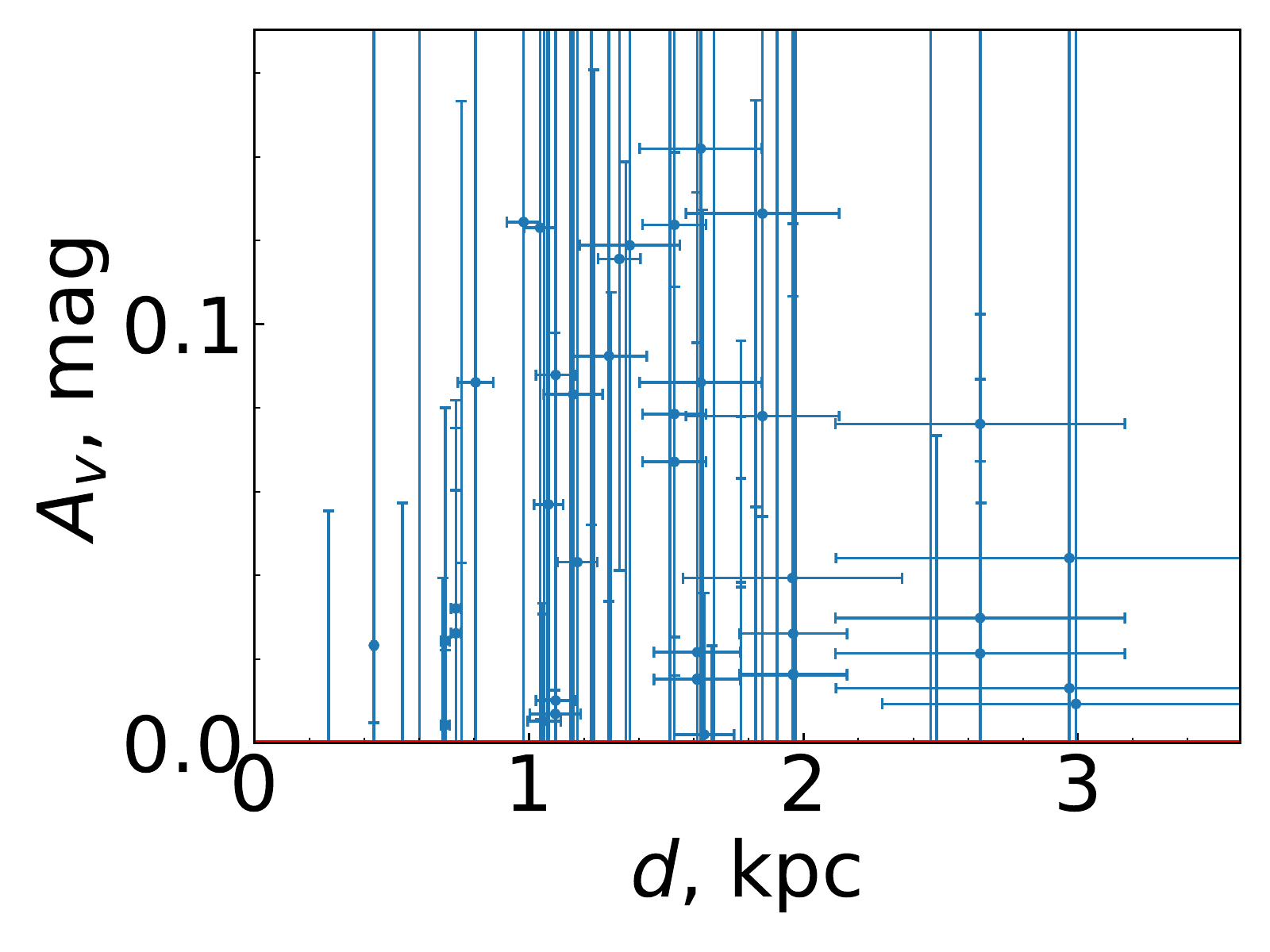}
  \includegraphics{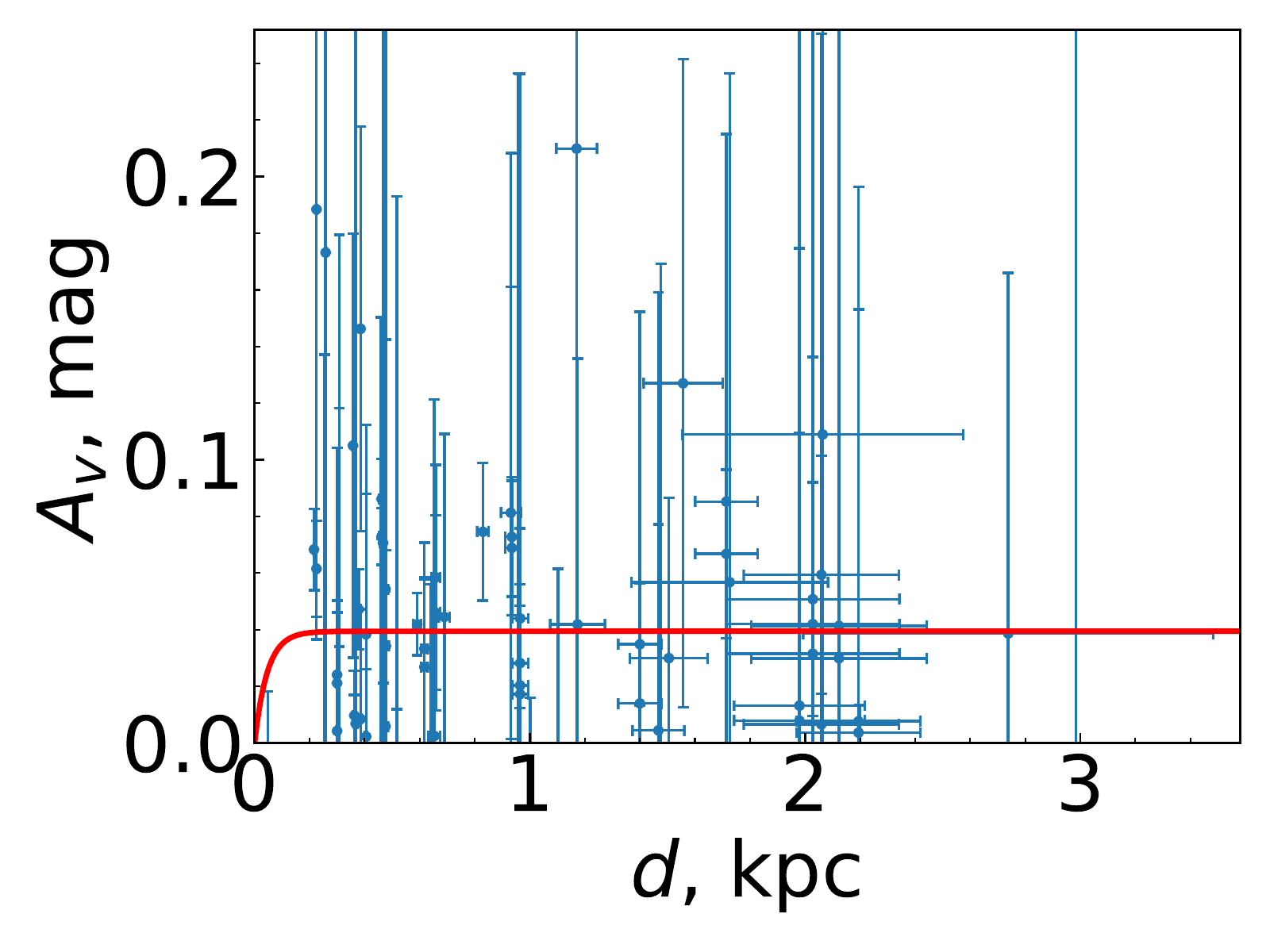}
 }
\caption{Best-fit for 47350, 49084, 62853.}
\end{figure}

\begin{figure}[h]
\resizebox{1.0\columnwidth}{!}{%
  \includegraphics{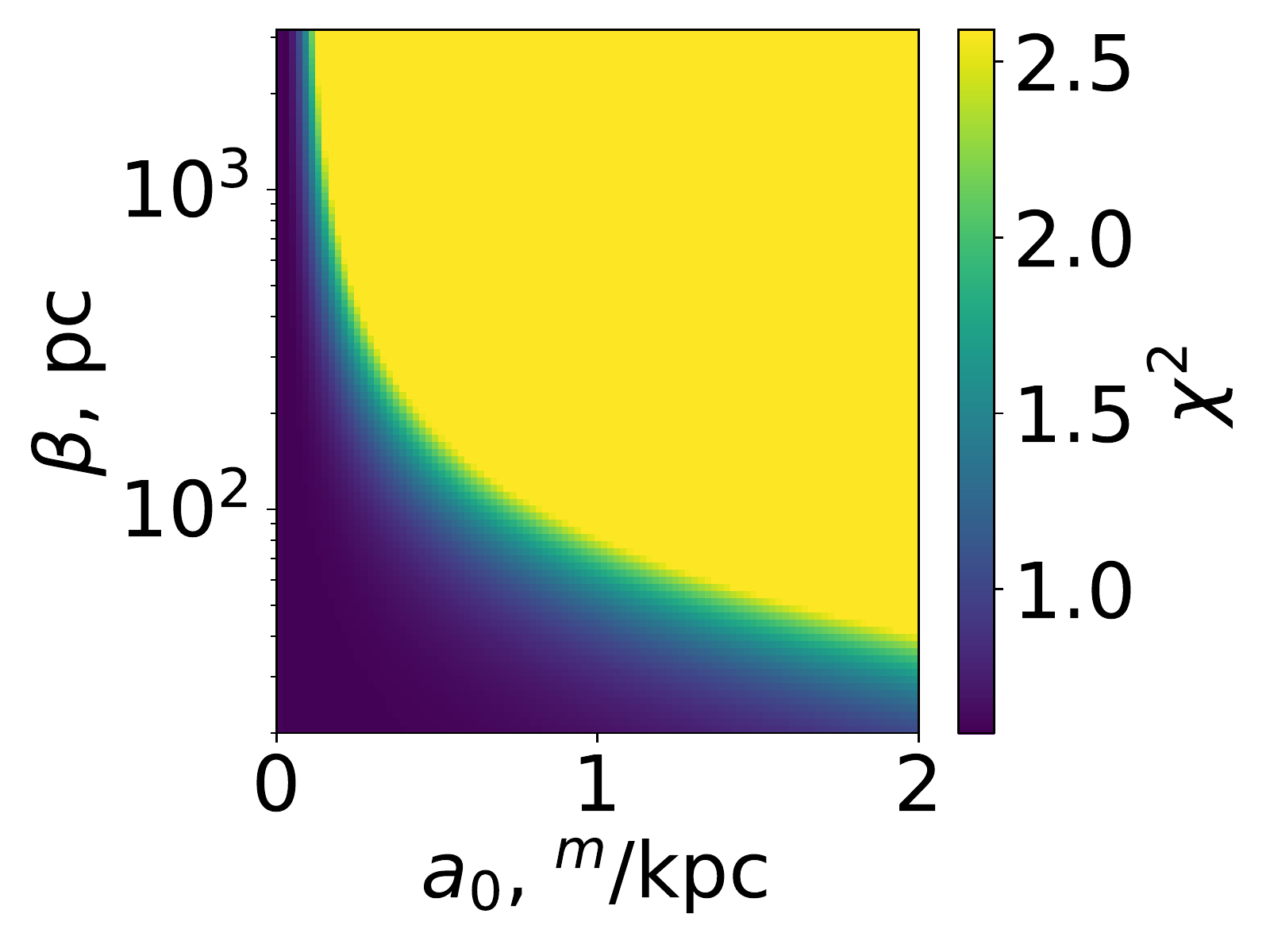}
  \includegraphics{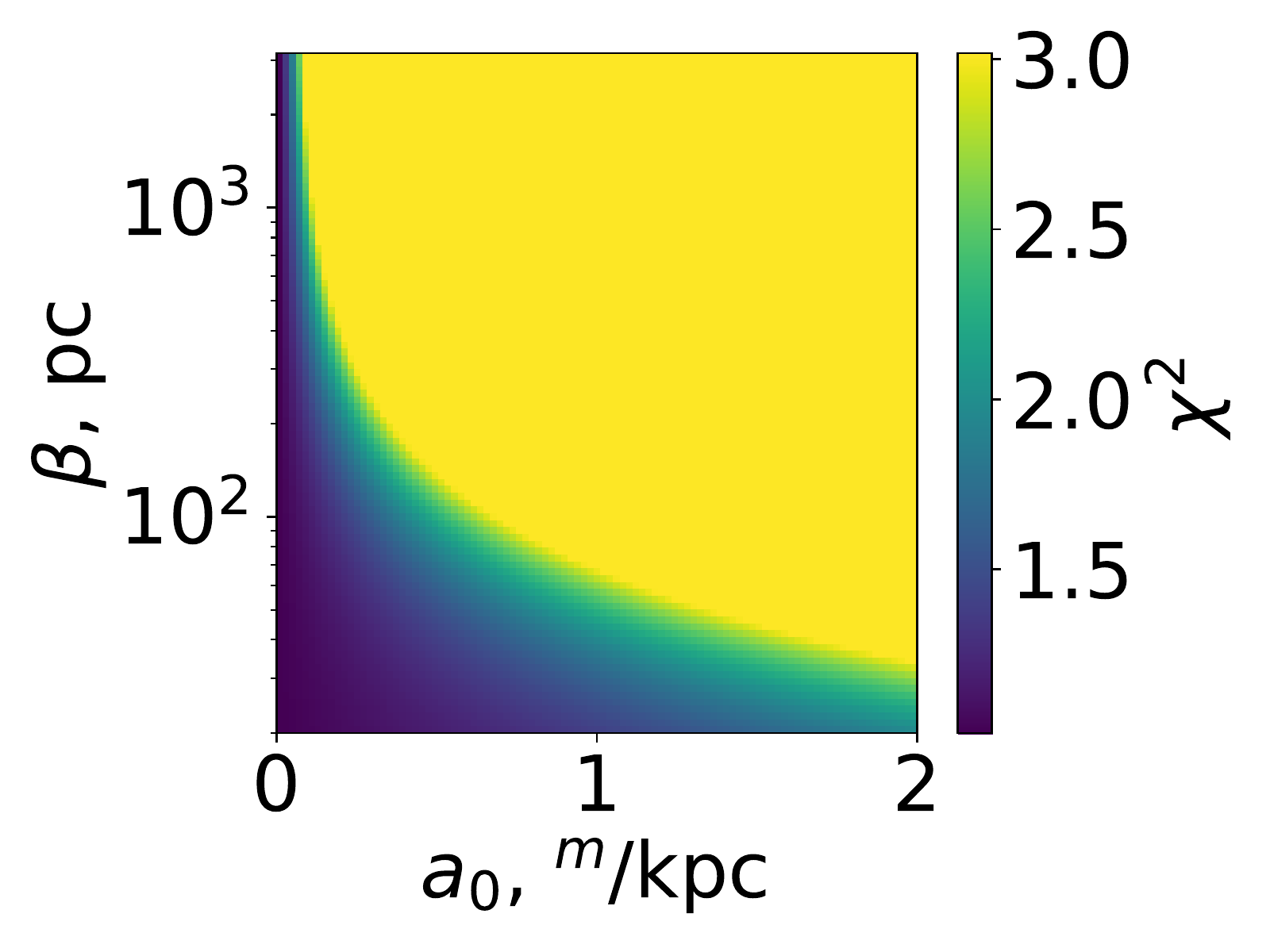}
  \includegraphics{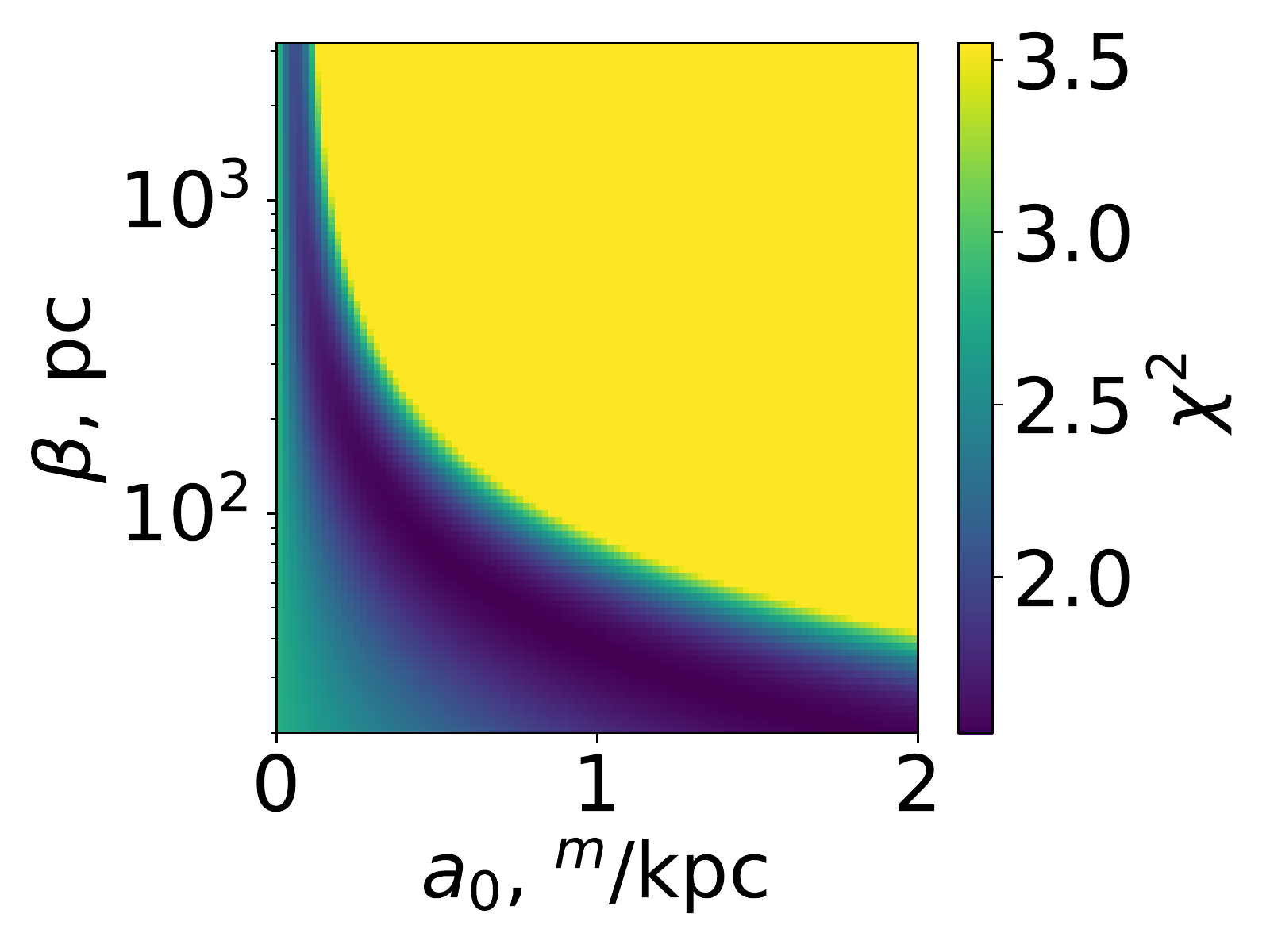}
 }
\caption{$\chi^2$ scan for 47350, 49084, 62853.}
\end{figure}

\begin{figure}[h]
\resizebox{1.0\columnwidth}{!}{%
  \includegraphics{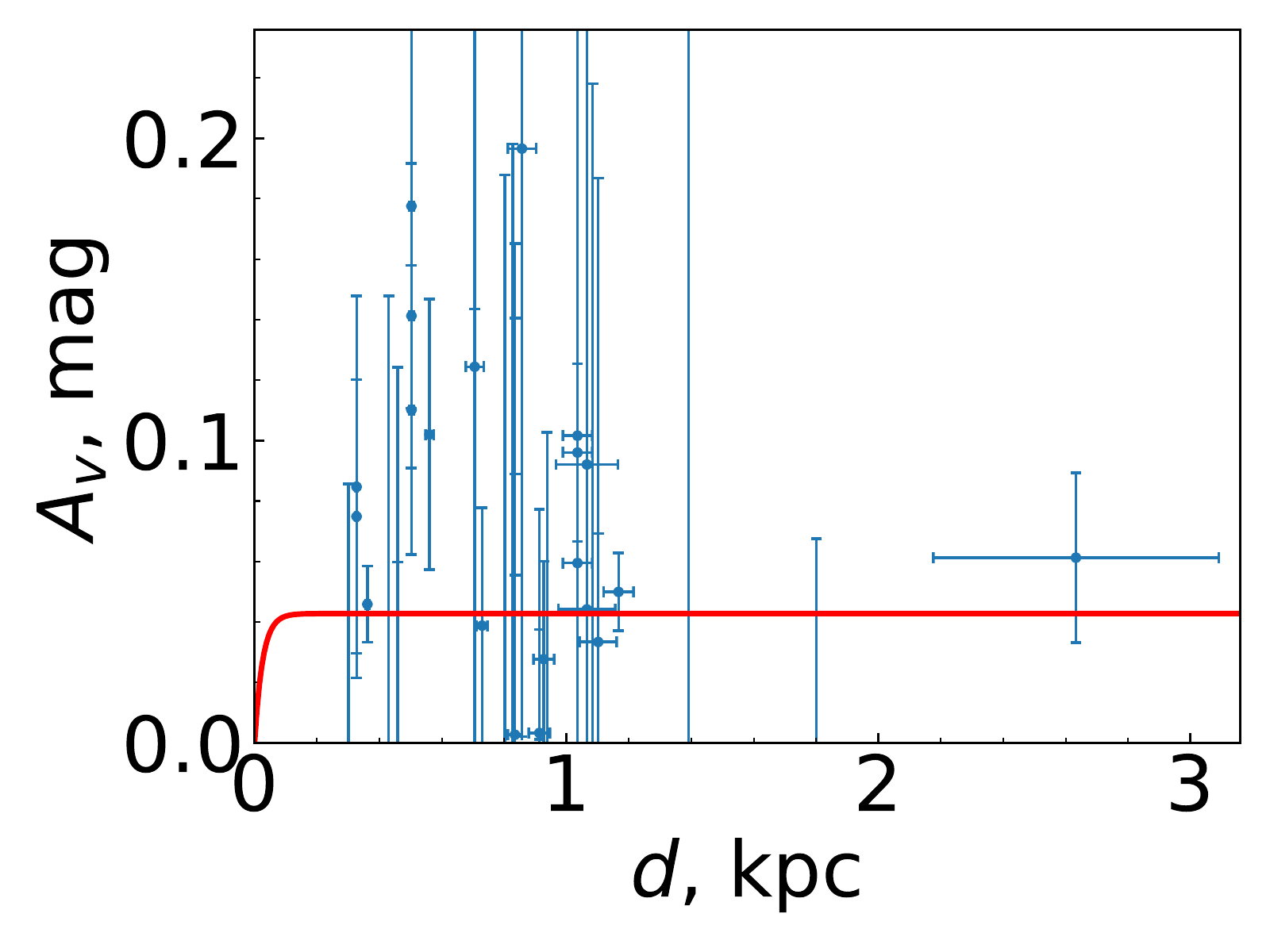}
  \includegraphics{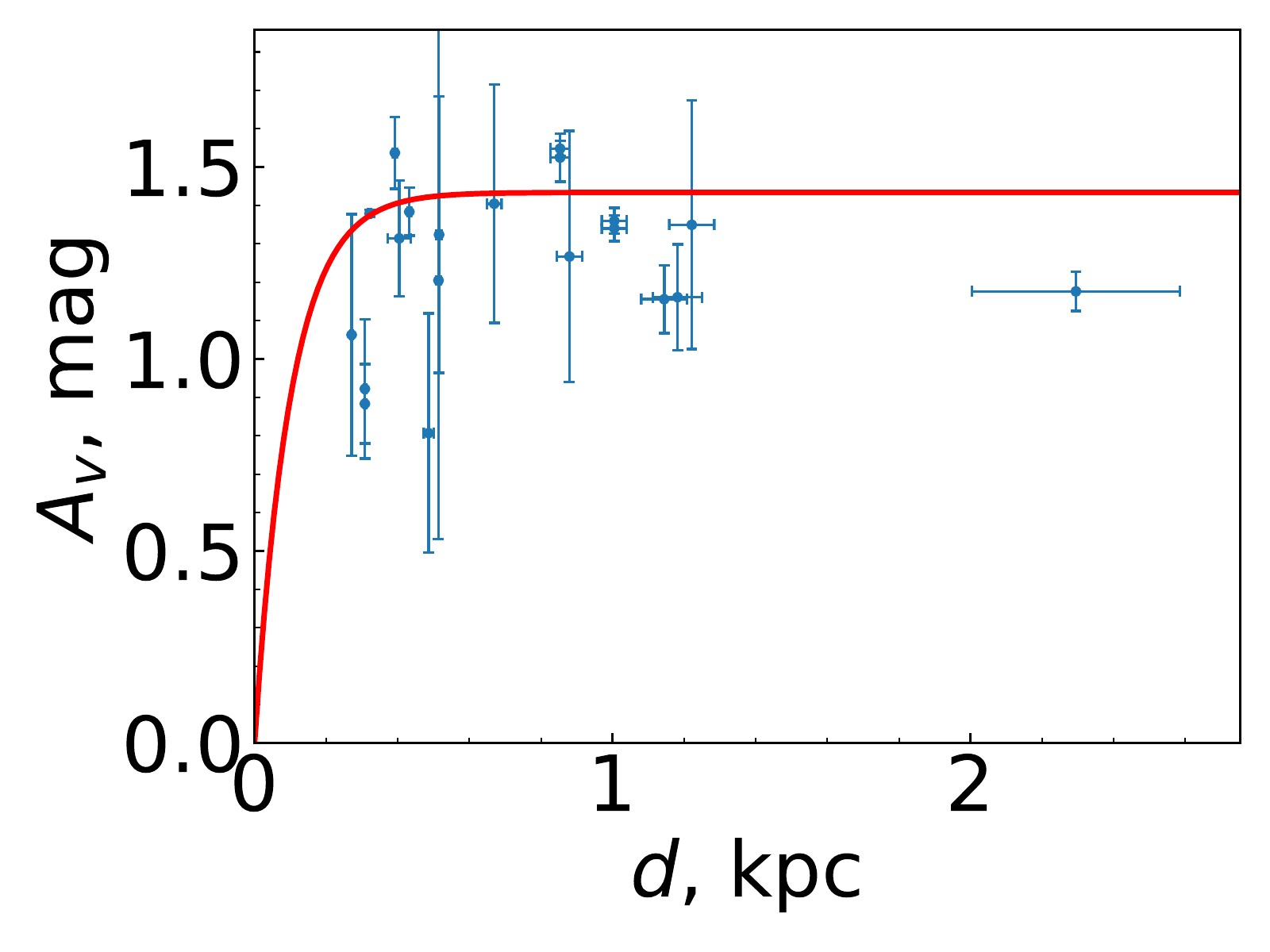}
  \includegraphics{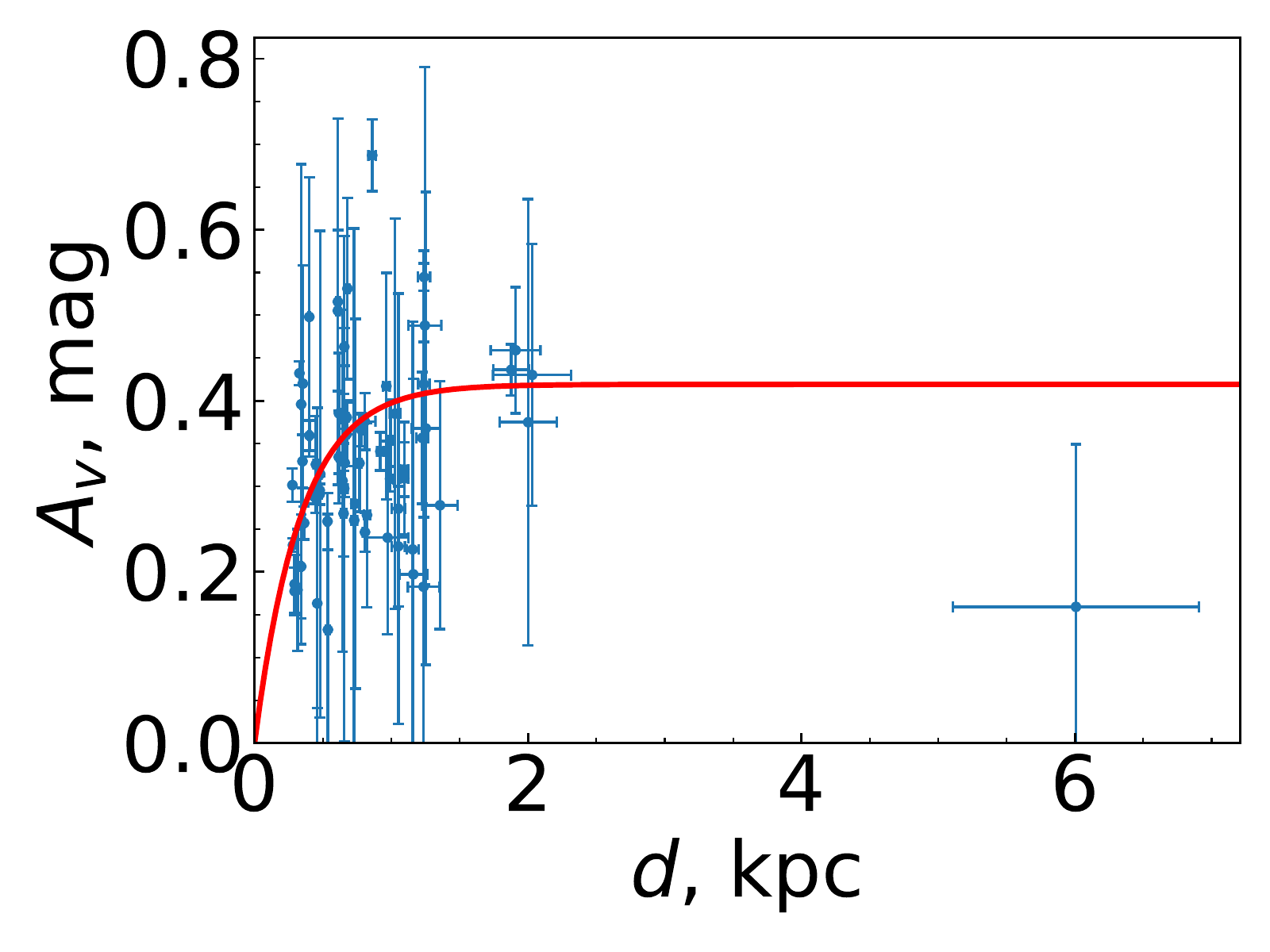}
 }
\caption{Best-fit for 64053, 79217, 82277.}
\end{figure}

\begin{figure}[h]
\resizebox{1.0\columnwidth}{!}{%
  \includegraphics{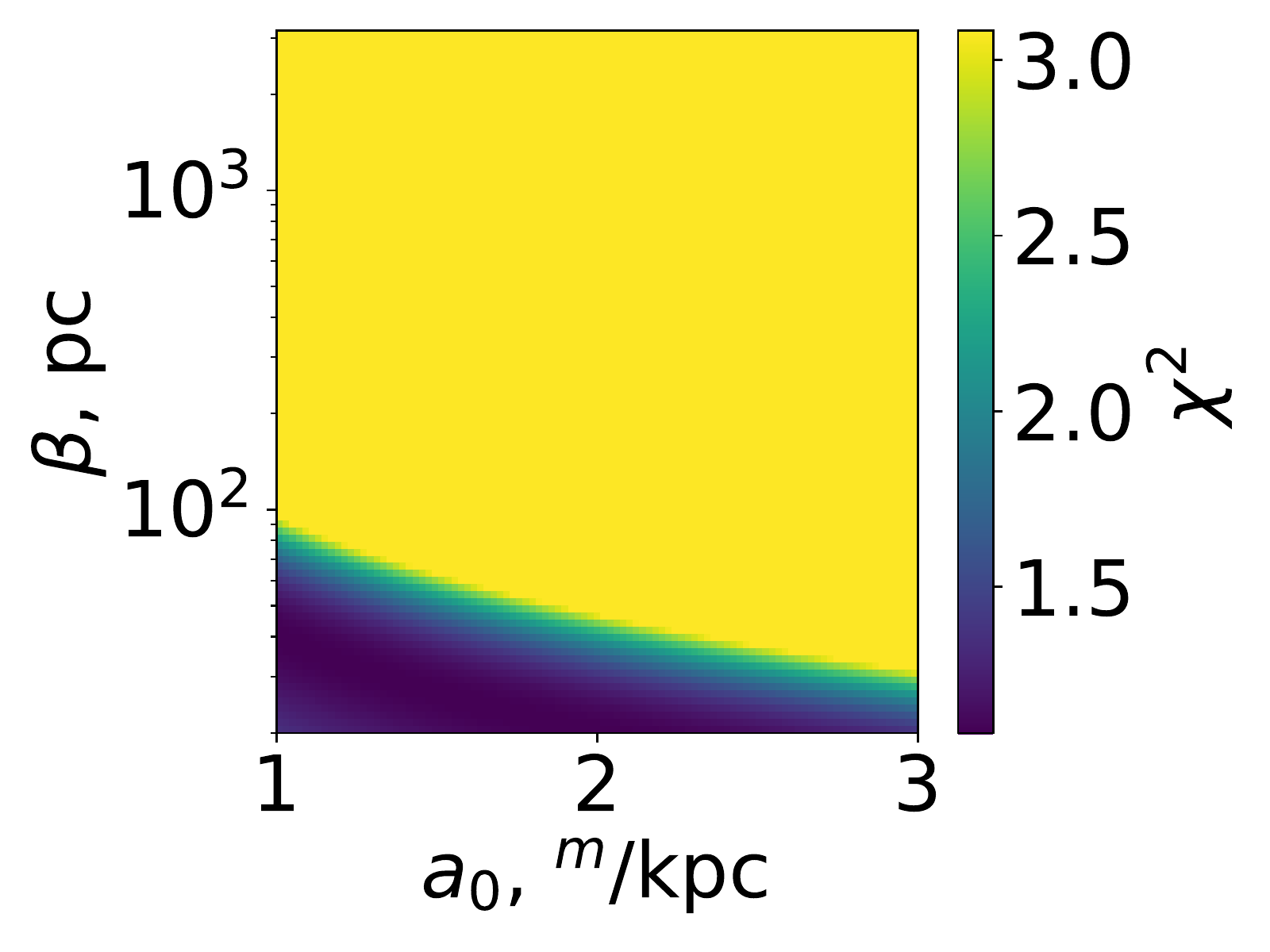}
  \includegraphics{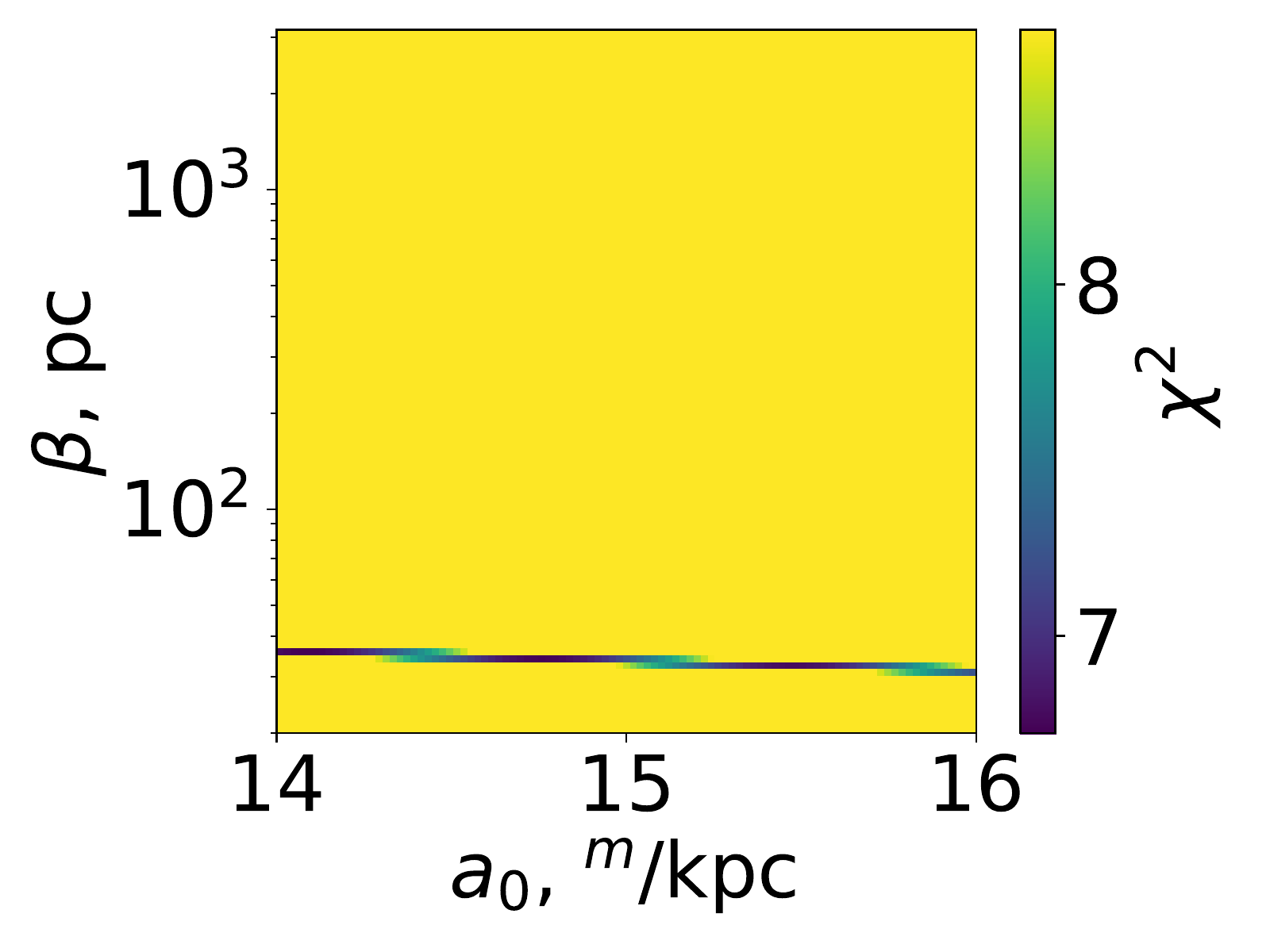}
  \includegraphics{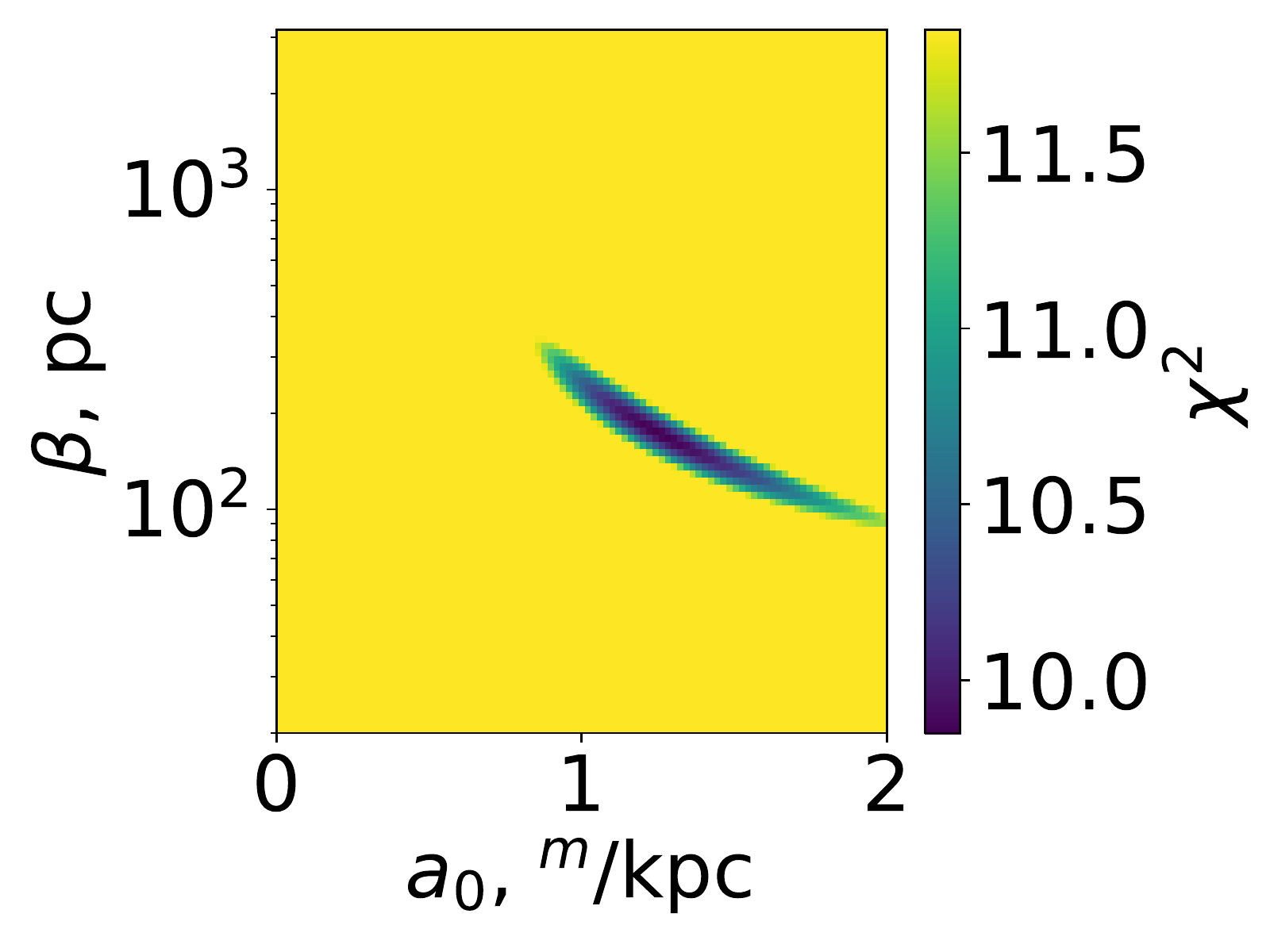}
 }
\caption{$\chi^2$ scan for 64053, 79217, 82277.}
\end{figure}

\begin{figure}[h]
\resizebox{1.0\columnwidth}{!}{%
  \includegraphics{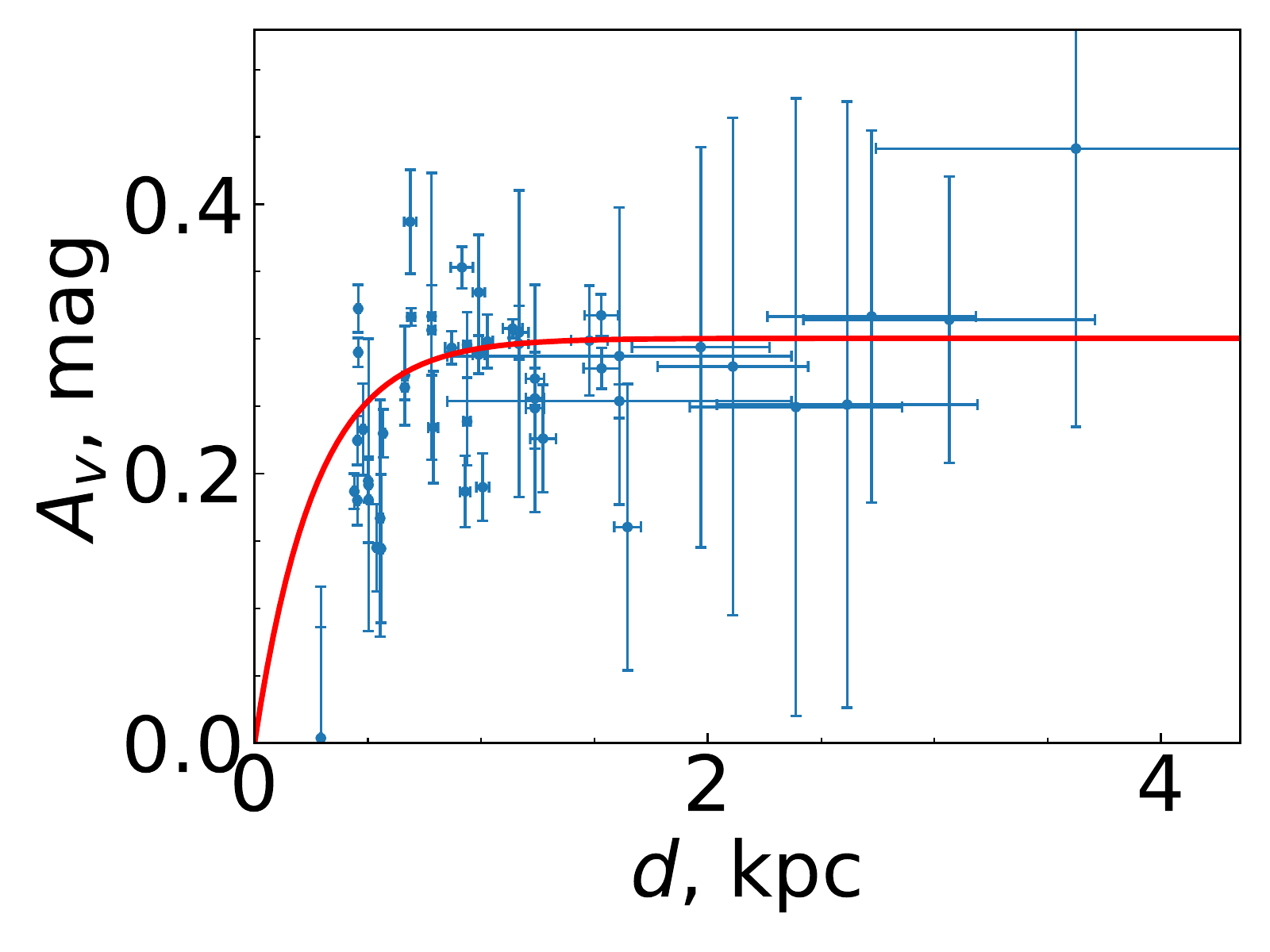}
  \includegraphics{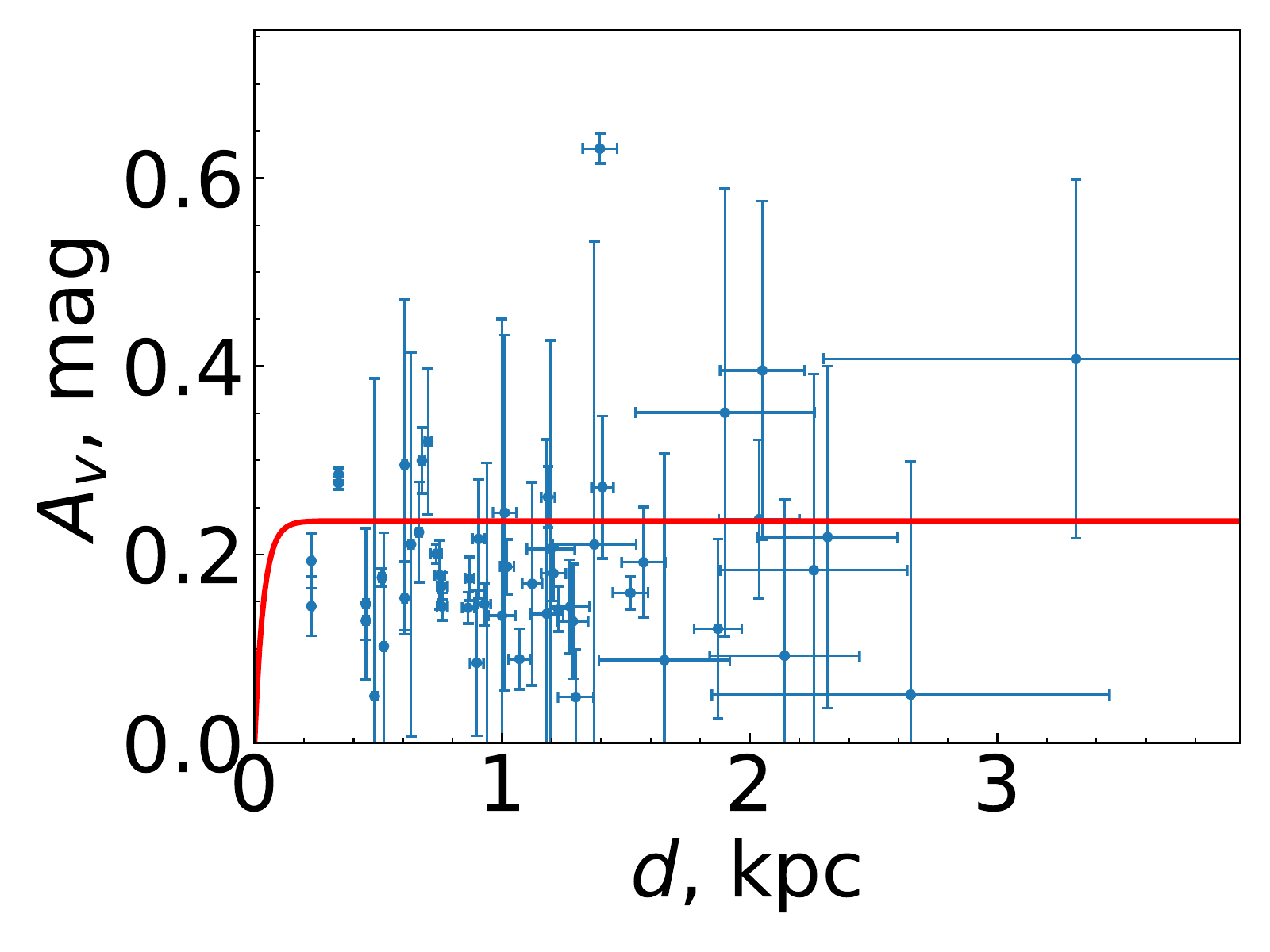}
  \includegraphics{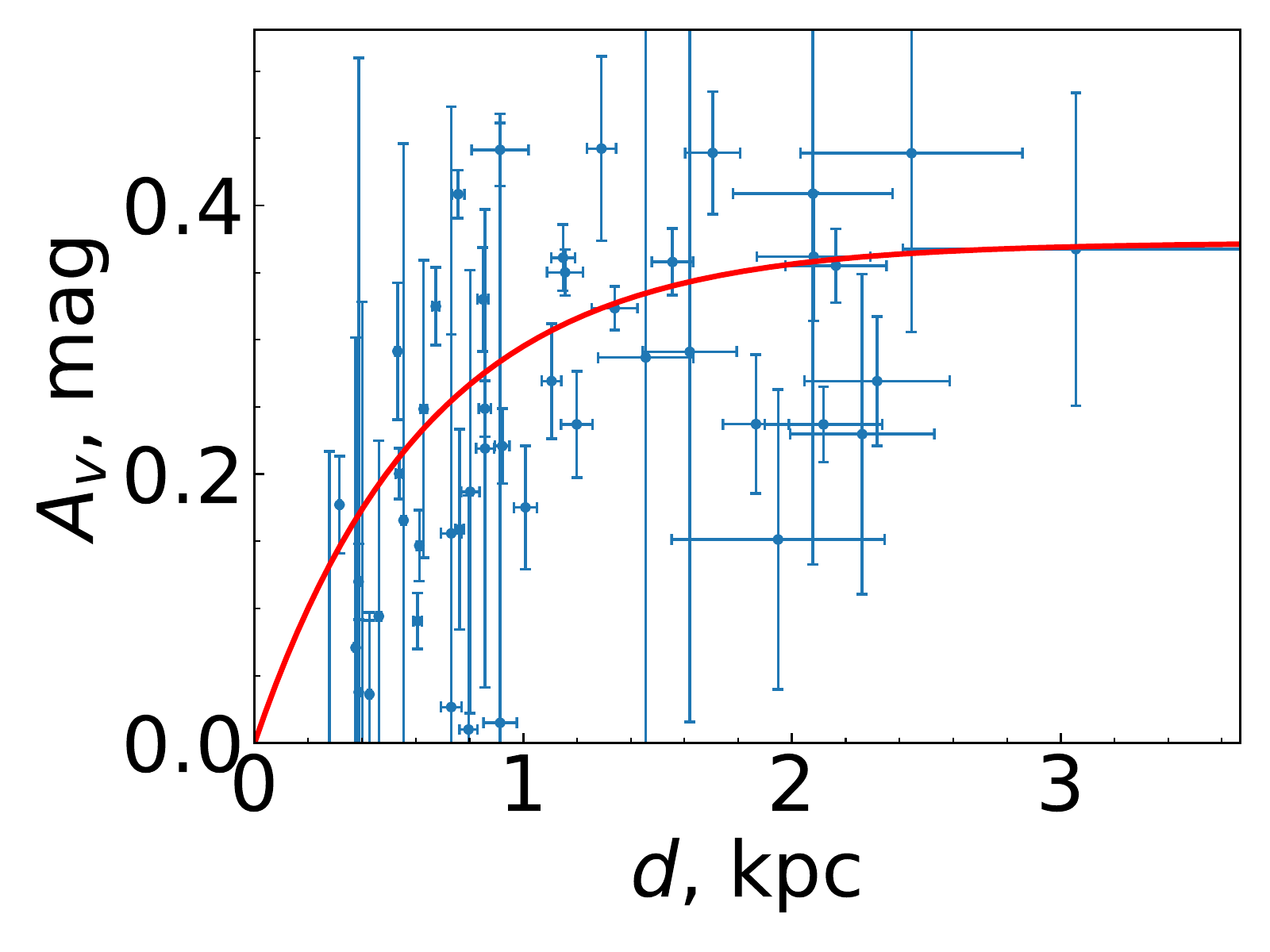}
 }
\caption{Best-fit for 83604, 84386, 84622.}
\end{figure}

\begin{figure}[h]
\resizebox{1.0\columnwidth}{!}{%
  \includegraphics{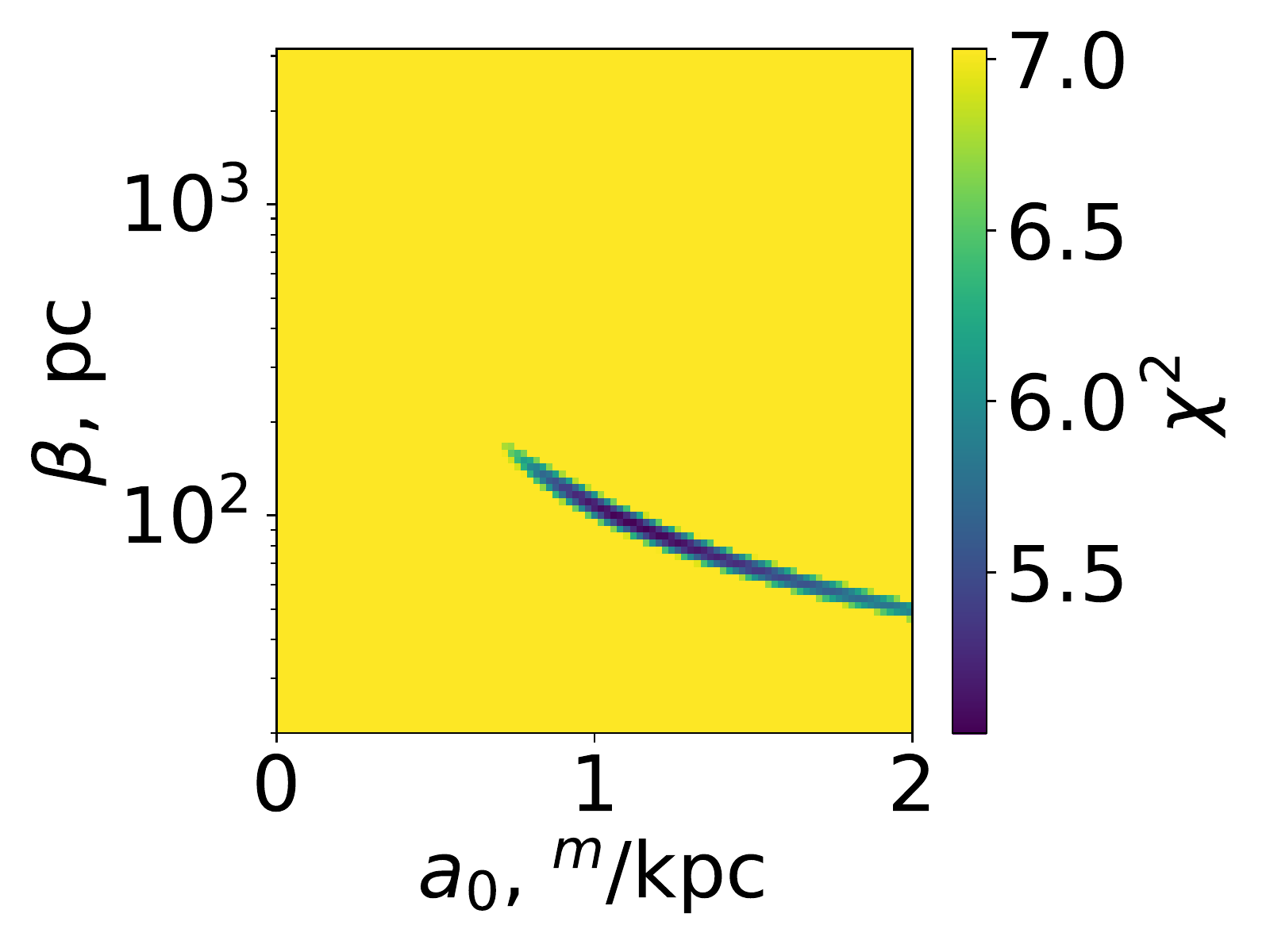}
  \includegraphics{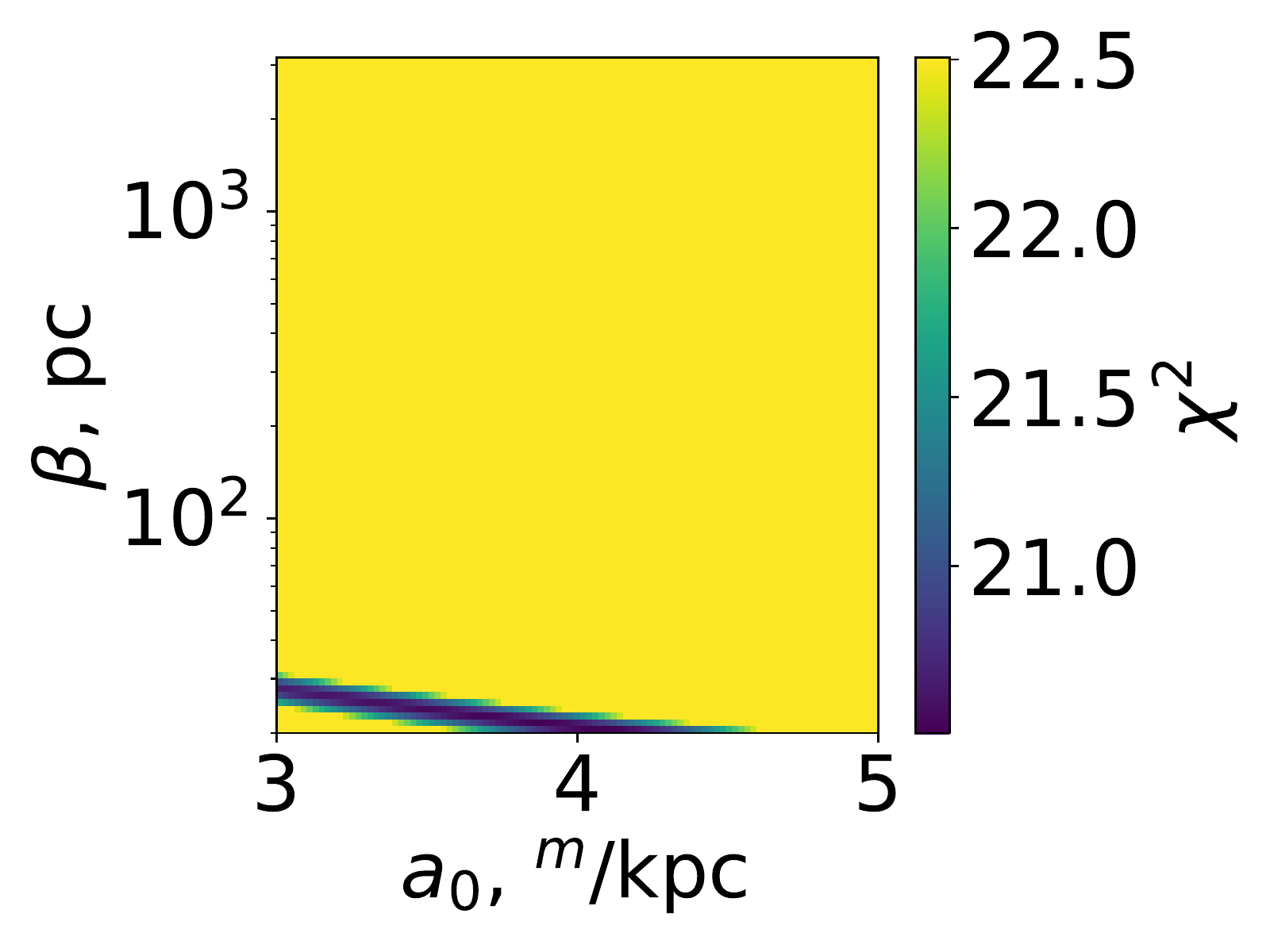}
  \includegraphics{84622_chi2_map.pdf}
 }
\caption{$\chi^2$ scan for 83604, 84386, 84622.}
\end{figure}

\begin{figure}[h]
\resizebox{1.0\columnwidth}{!}{%
  \includegraphics{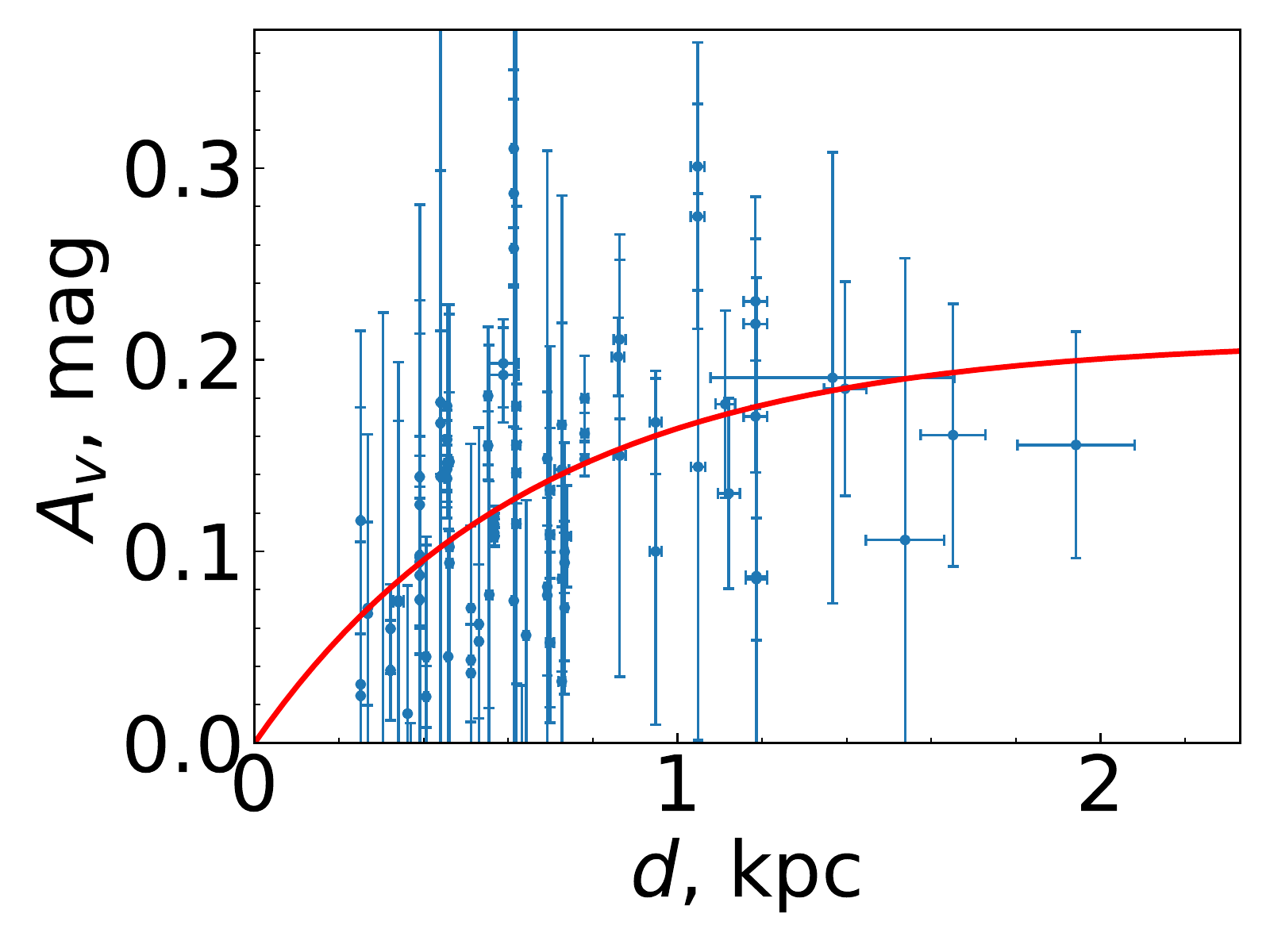}
  \includegraphics{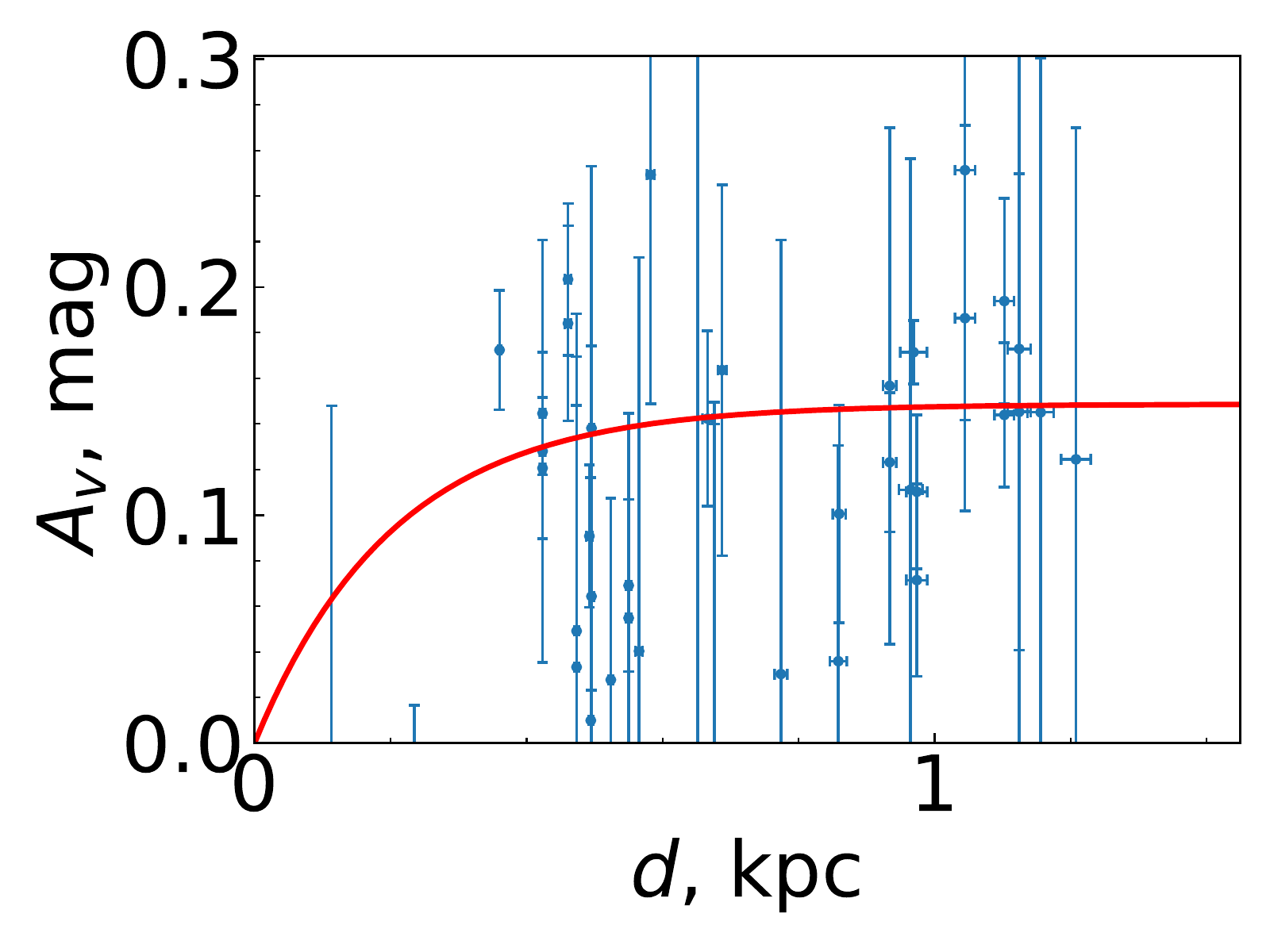}
  \includegraphics{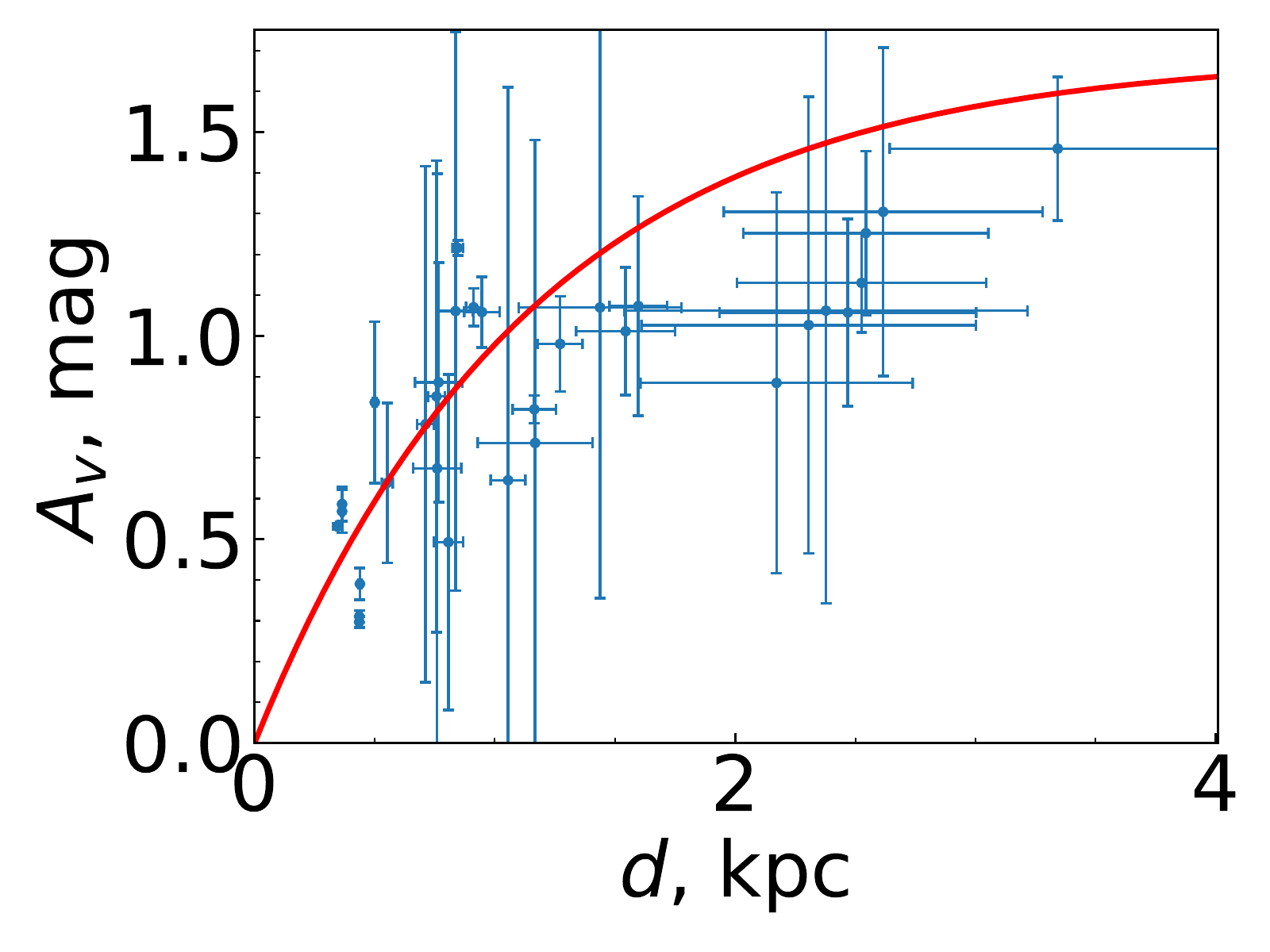}
 }
\caption{Best-fit for 96855, 97942, 98650.}
\end{figure}

\begin{figure}[h]
\resizebox{1.0\columnwidth}{!}{%
  \includegraphics{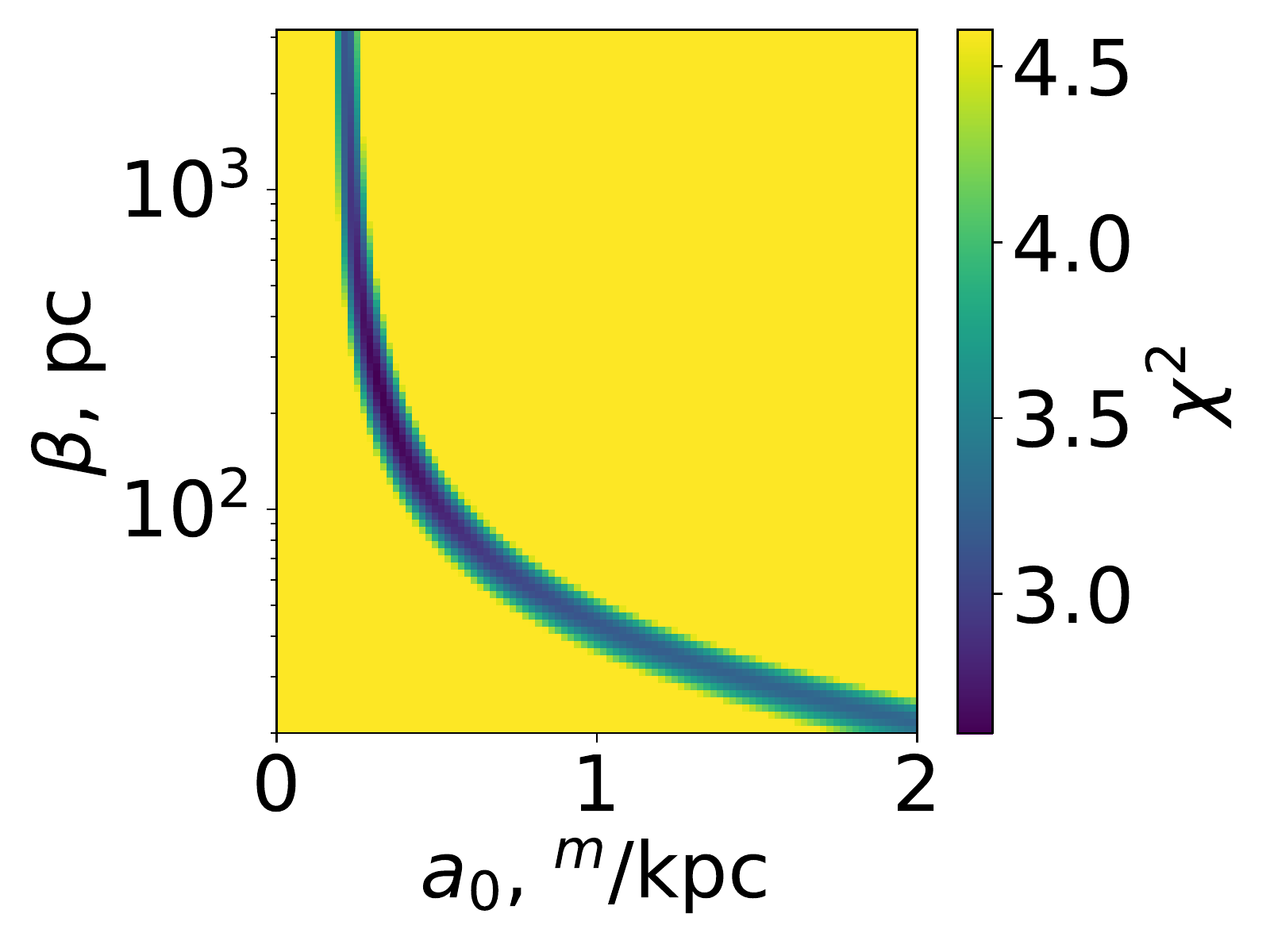}
  \includegraphics{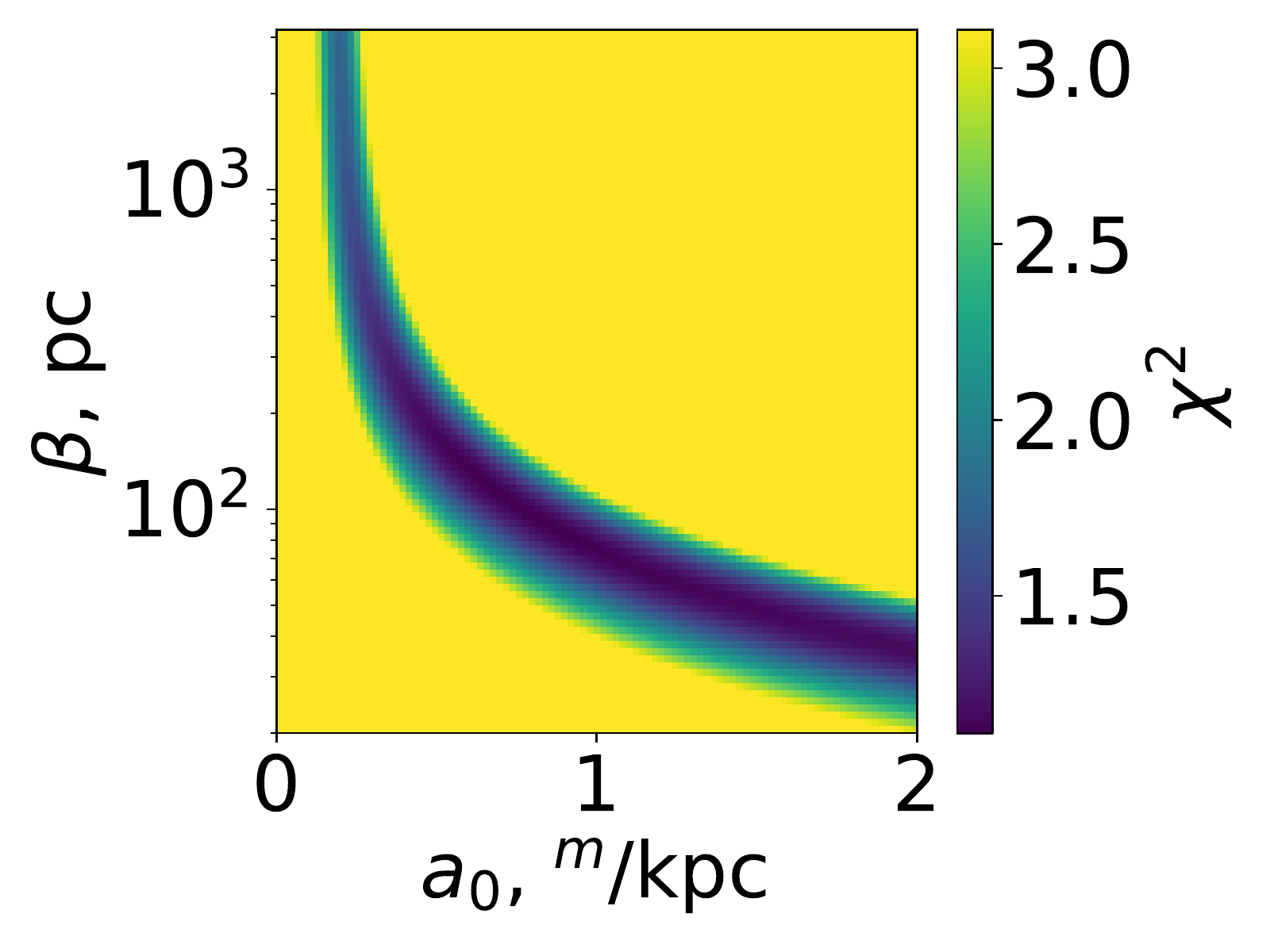}
  \includegraphics{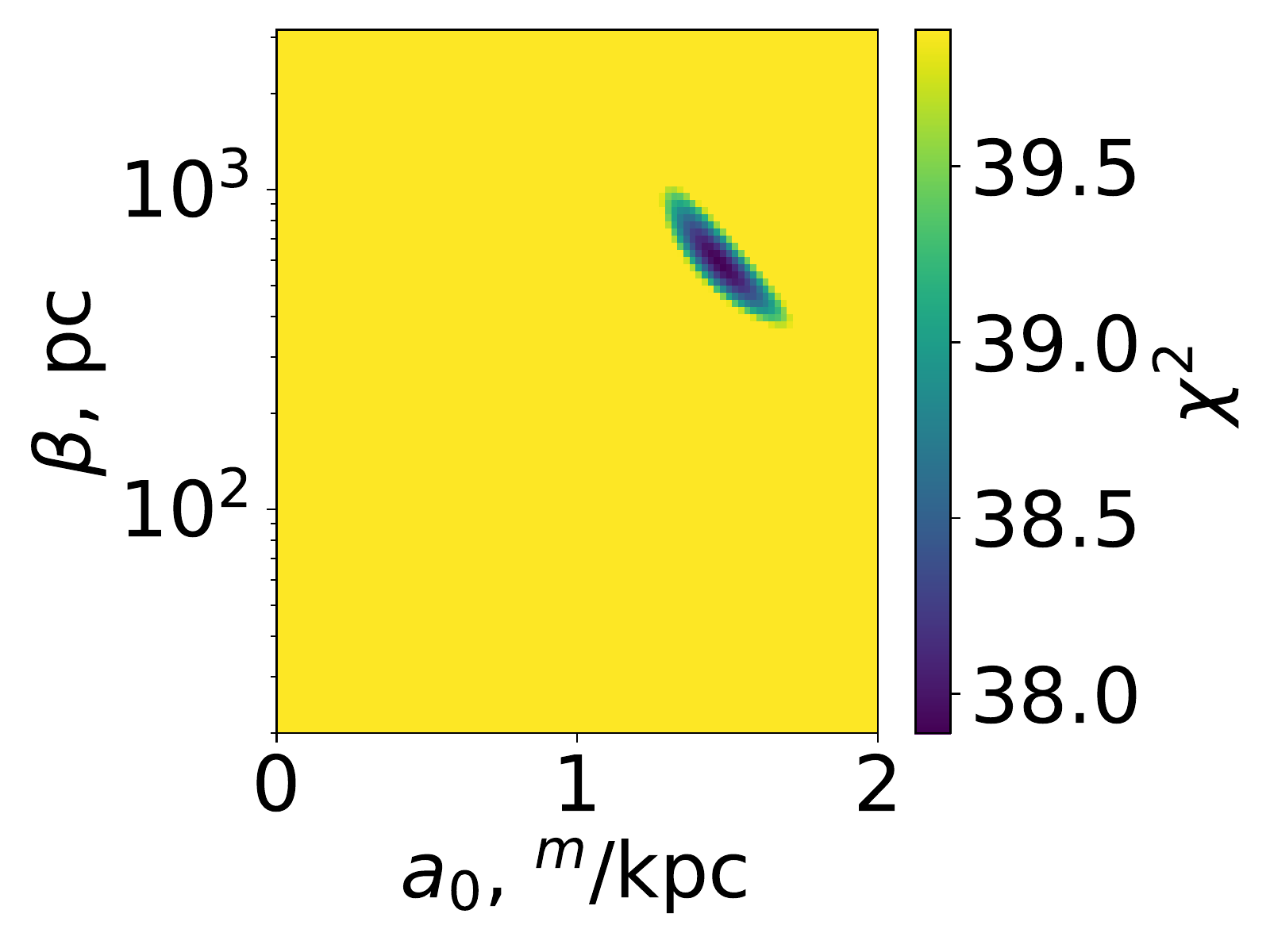}
 }
\caption{$\chi^2$ scan for 96855, 97942, 98650.}
\end{figure}

\begin{figure}[h]
\resizebox{1.0\columnwidth}{!}{%
  \includegraphics{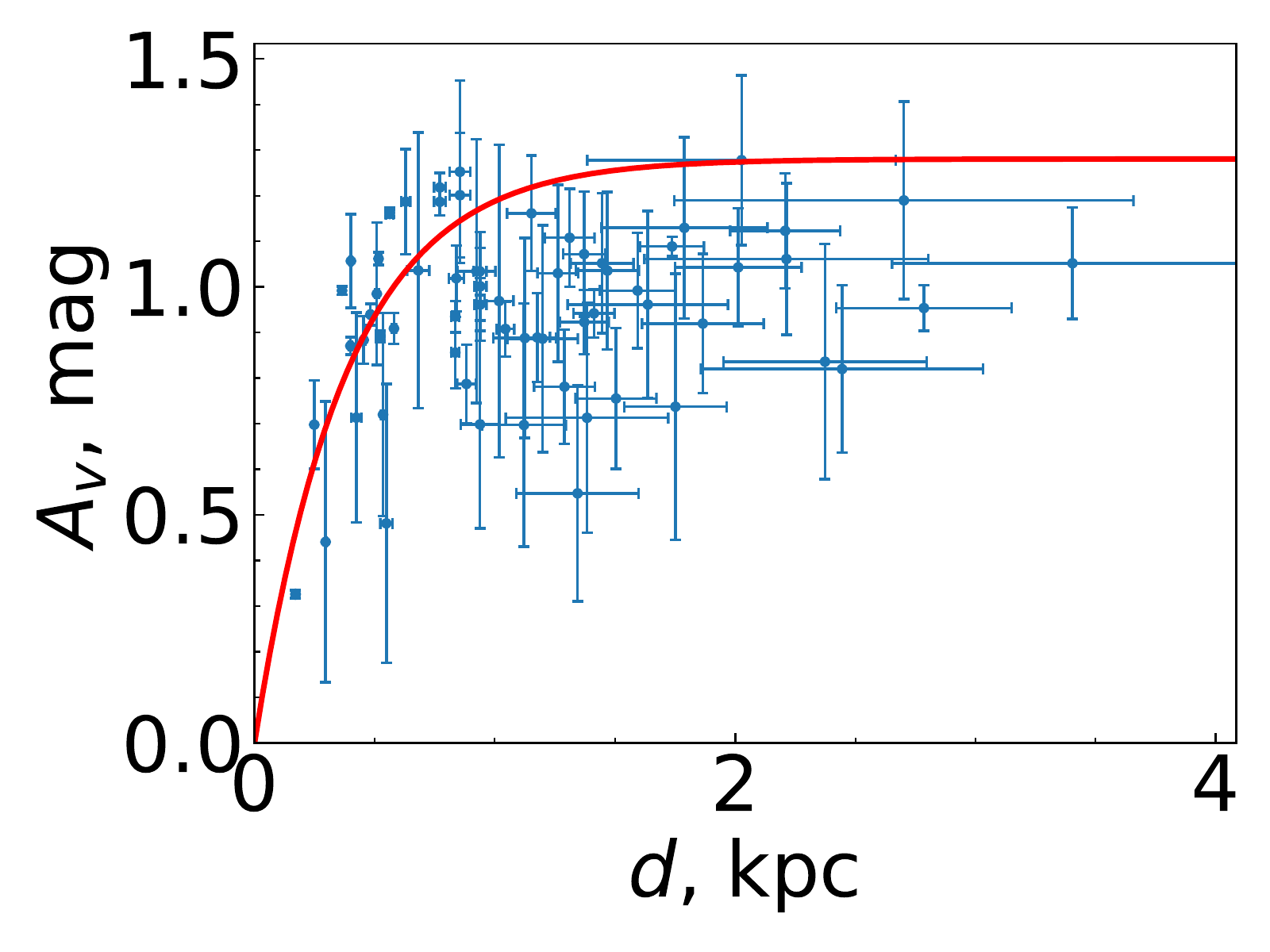}
  \includegraphics{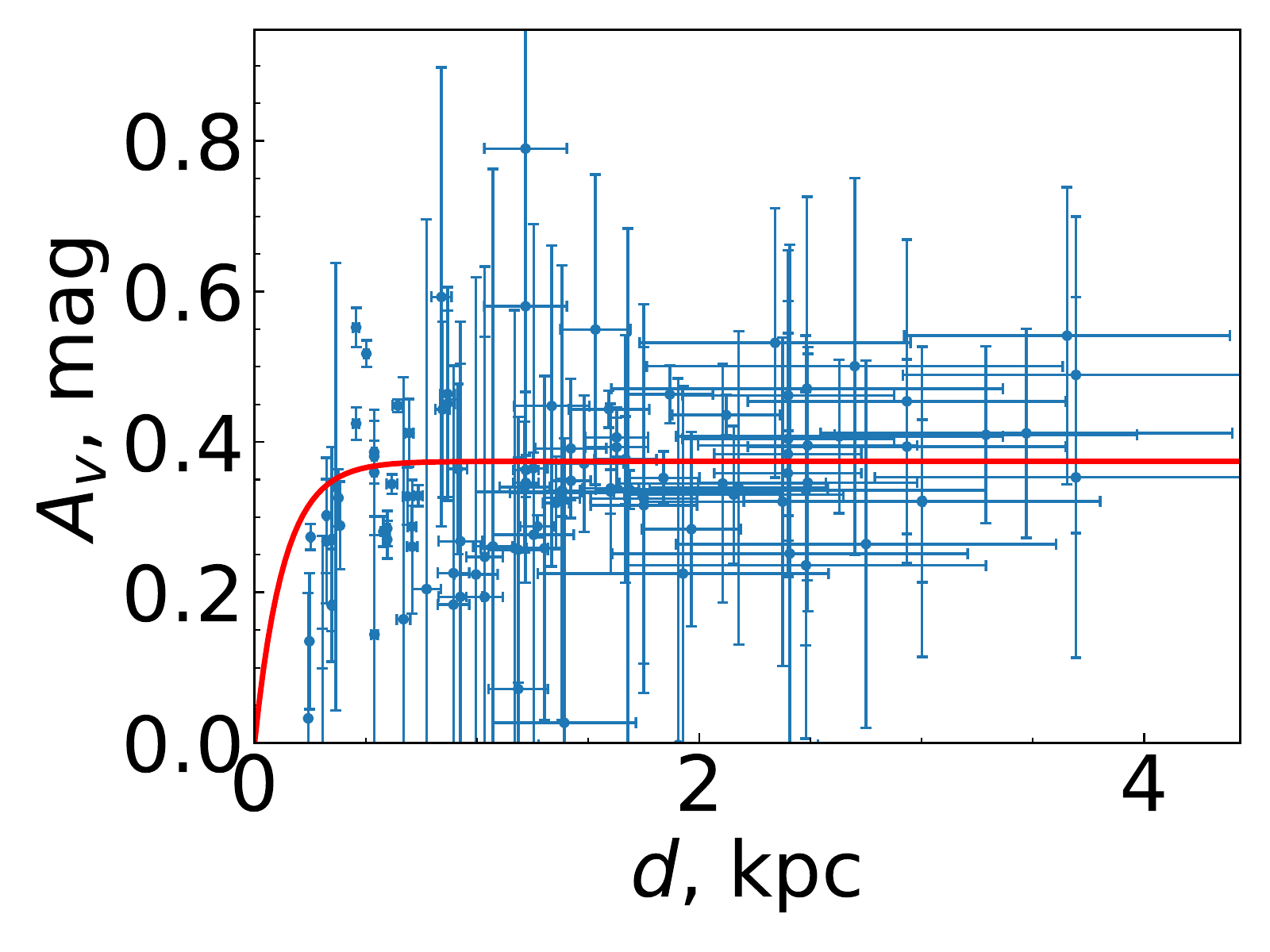}
  \includegraphics{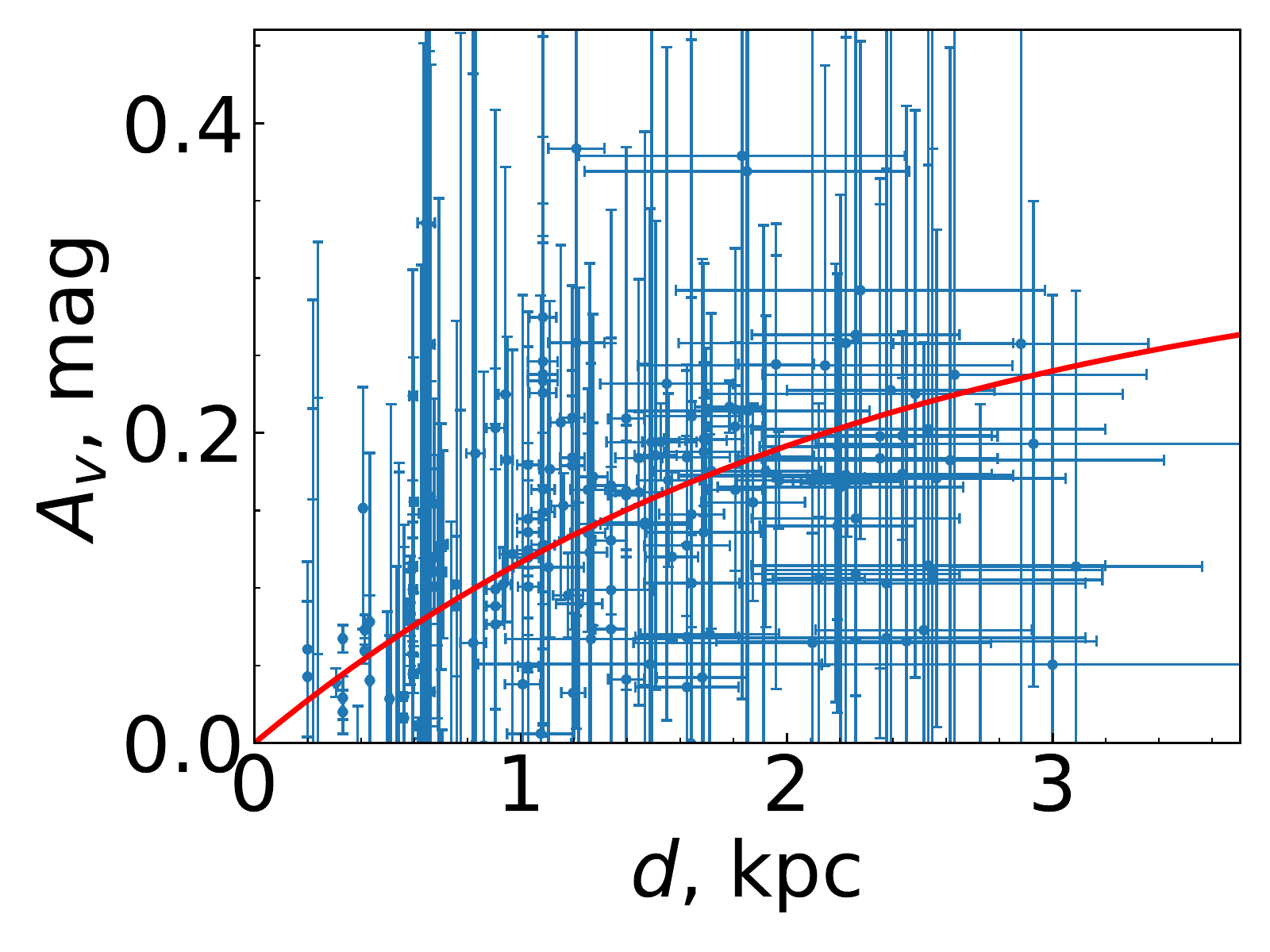}
 }
\caption{Best-fit for 99901, 100961, 112048.}
\end{figure}

\begin{figure}[h]
\resizebox{1.0\columnwidth}{!}{%
  \includegraphics{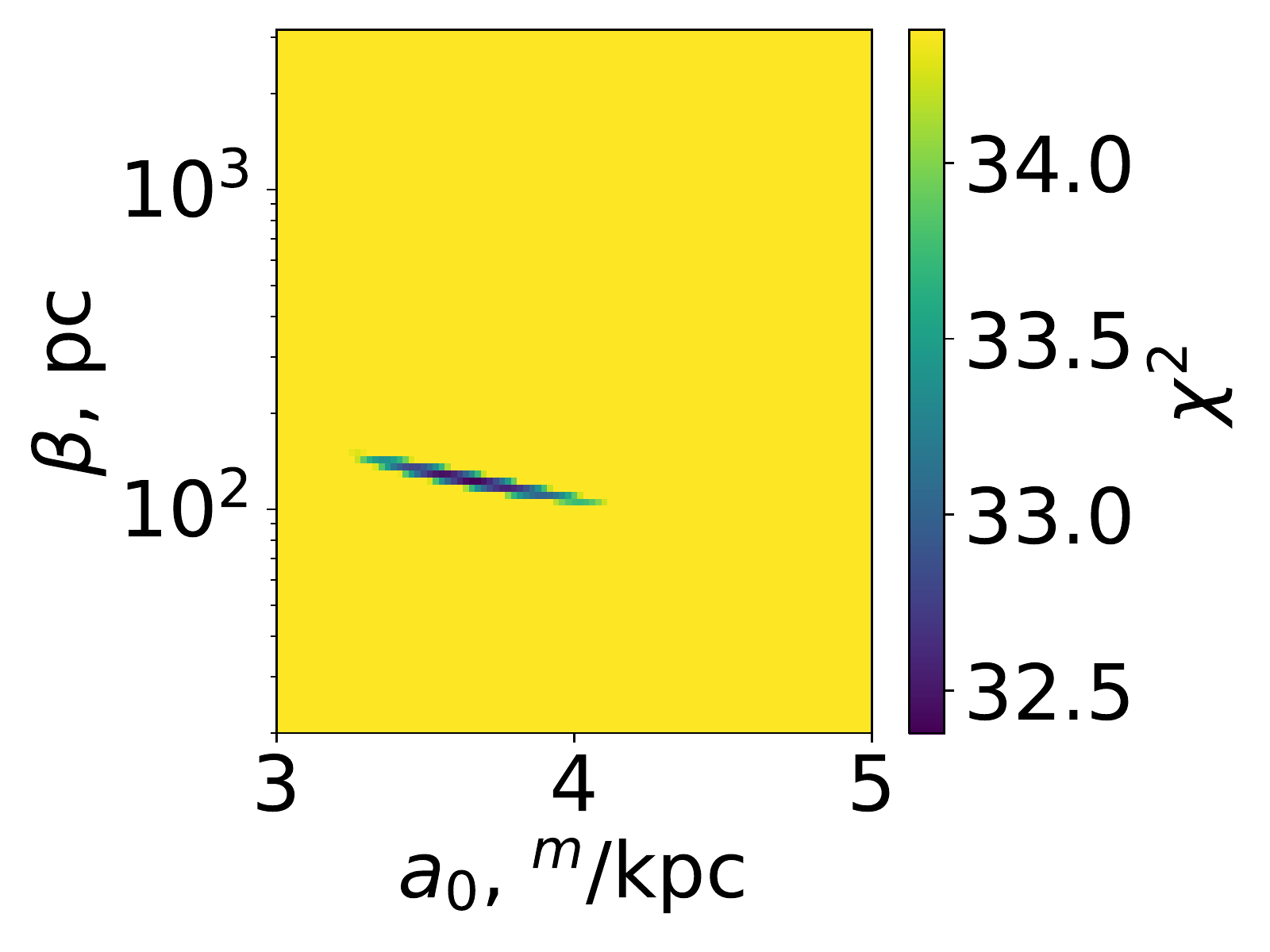}
  \includegraphics{100961_chi2_map.pdf}
  \includegraphics{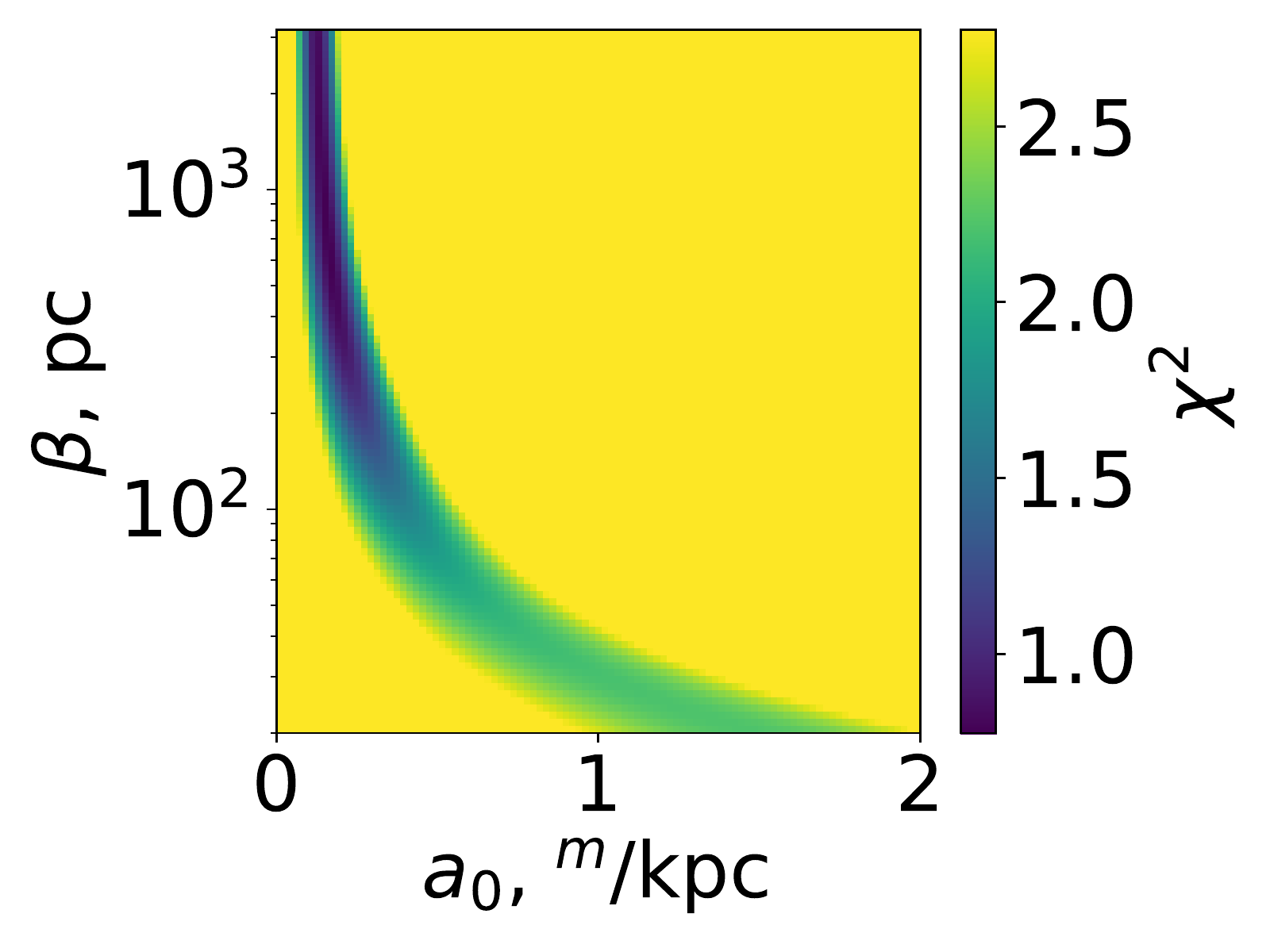}
 }
\caption{$\chi^2$ scan for 99901, 100961, 112048.}
\end{figure}

\begin{figure}[h]
\resizebox{1.0\columnwidth}{!}{%
  \includegraphics{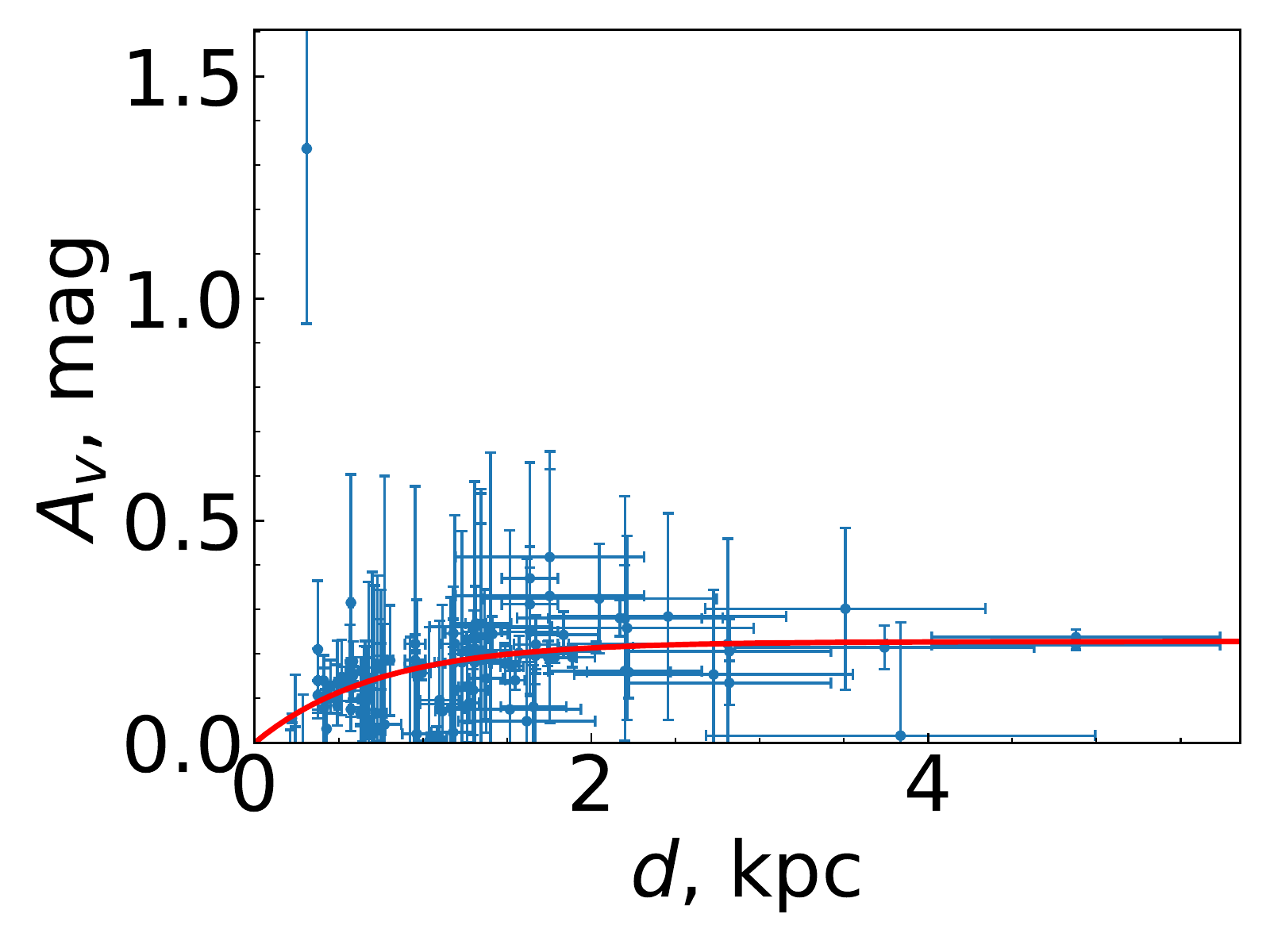}
  \includegraphics{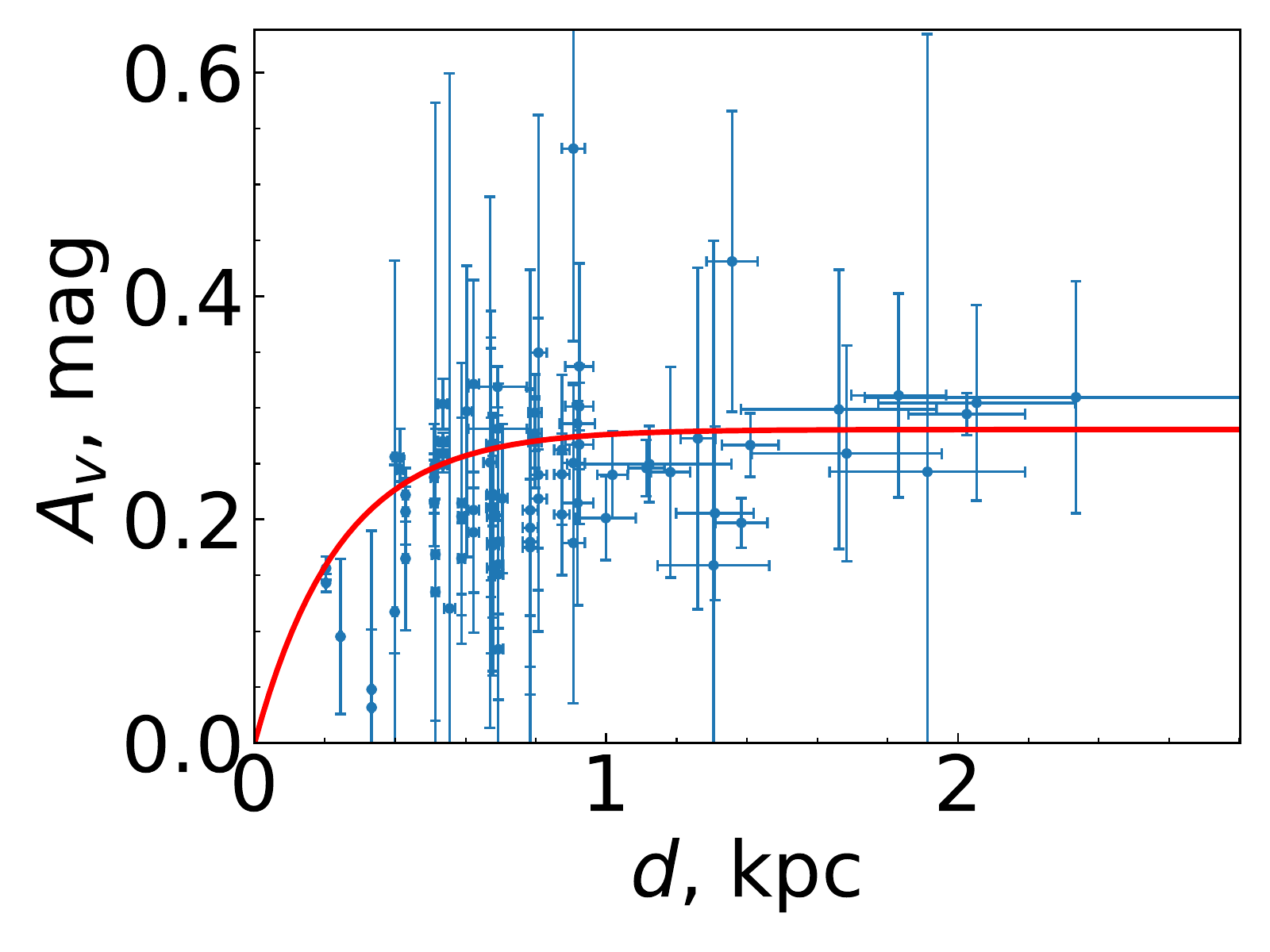}
  \includegraphics{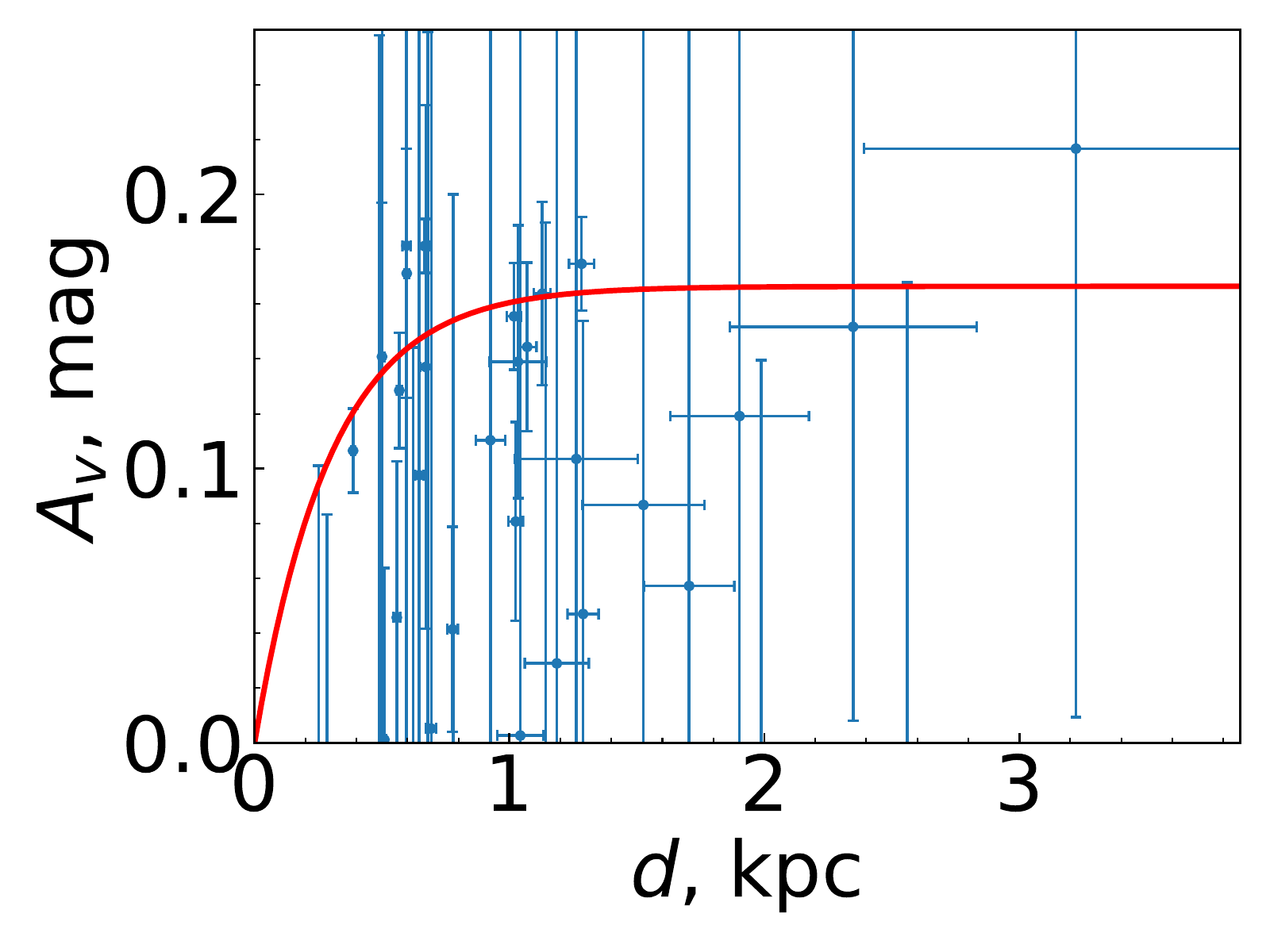}
 }
\caption{Best-fit for 112305, 113239, 114025.}
\end{figure}

\begin{figure}[h]
\resizebox{1.0\columnwidth}{!}{%
  \includegraphics{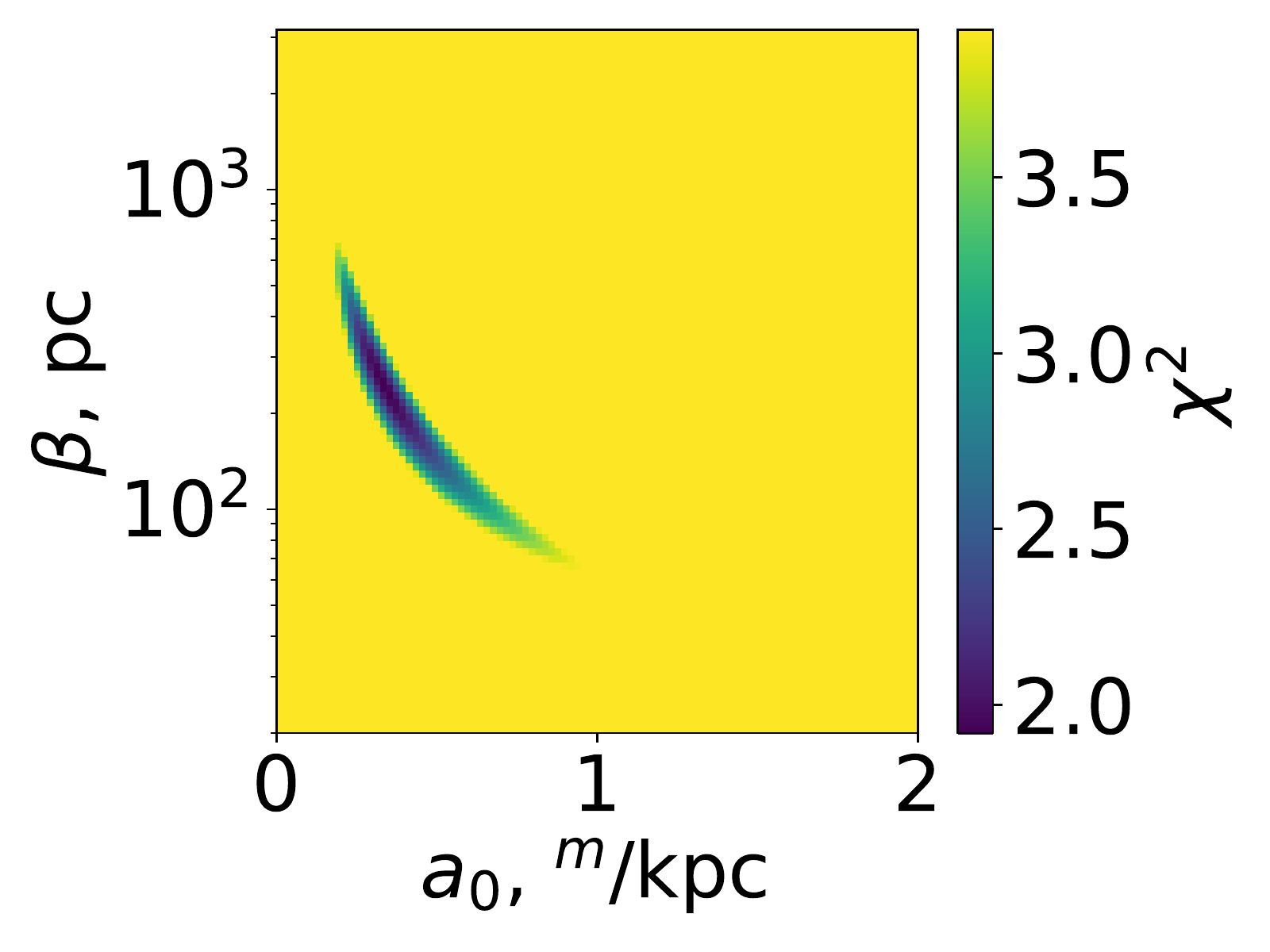}
  \includegraphics{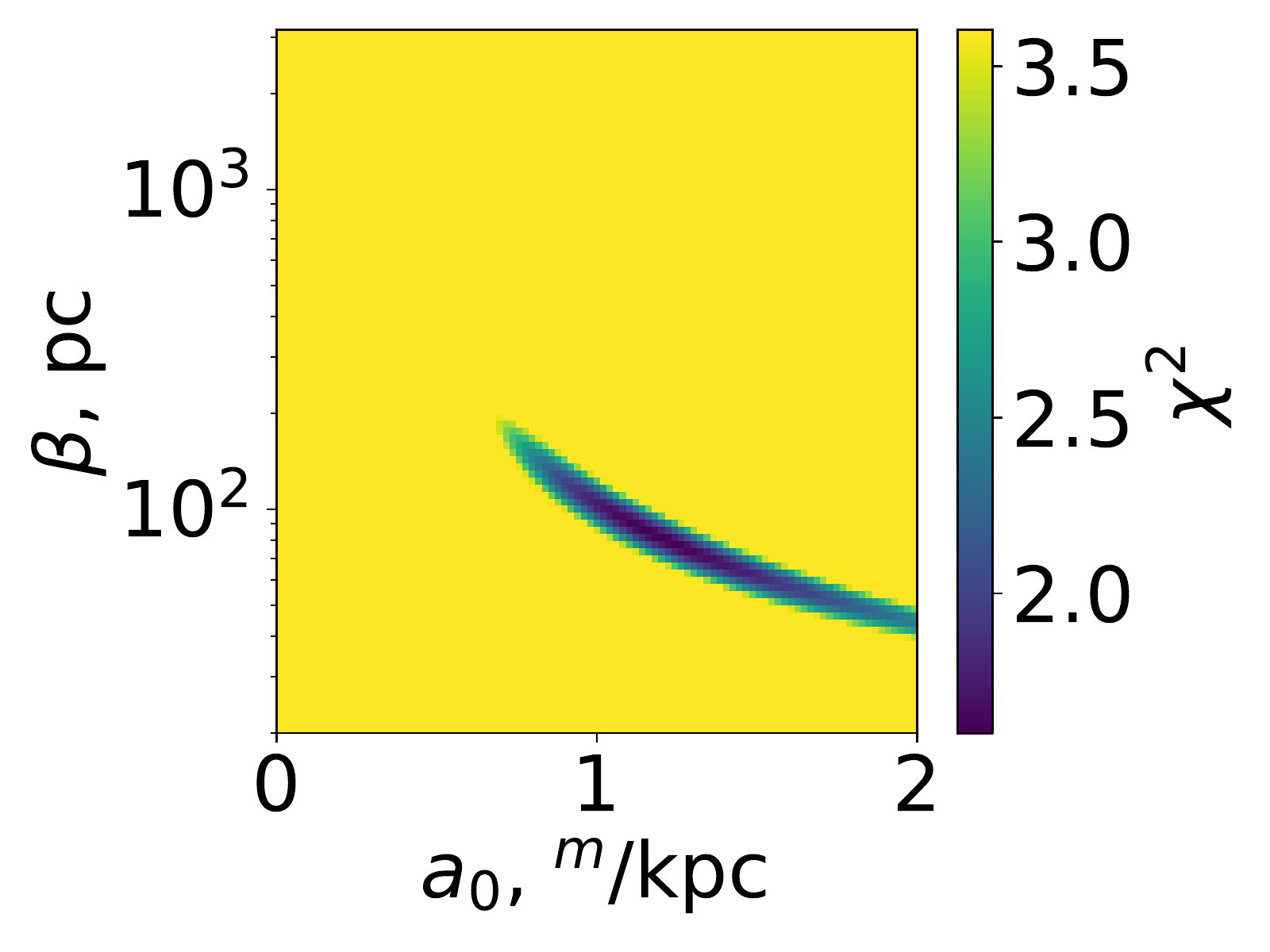}
  \includegraphics{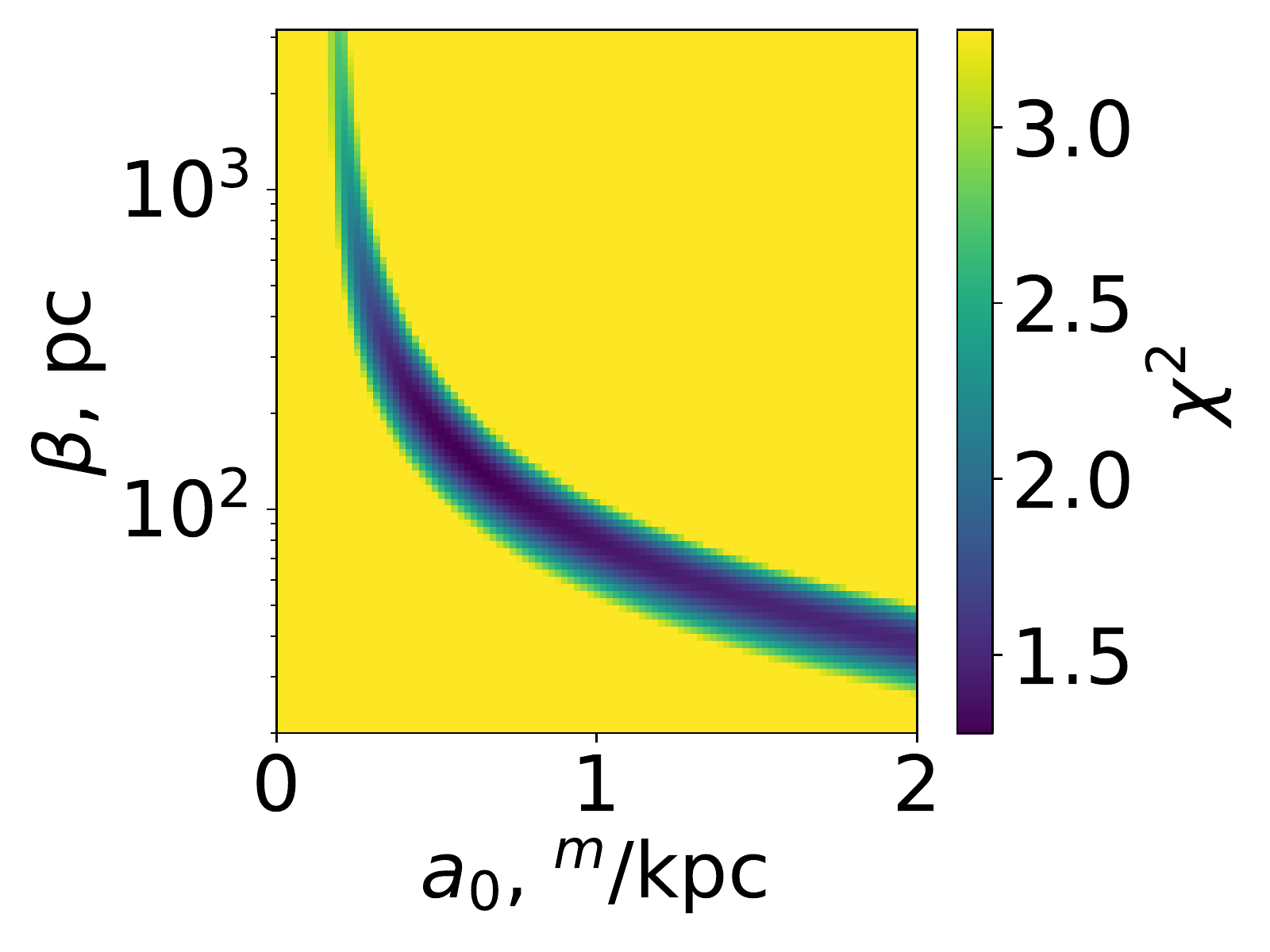}
 }
\caption{$\chi^2$ scan for 112305, 113239, 114025.}
\end{figure}

\begin{figure}[h]
\resizebox{1.0\columnwidth}{!}{%
  \includegraphics{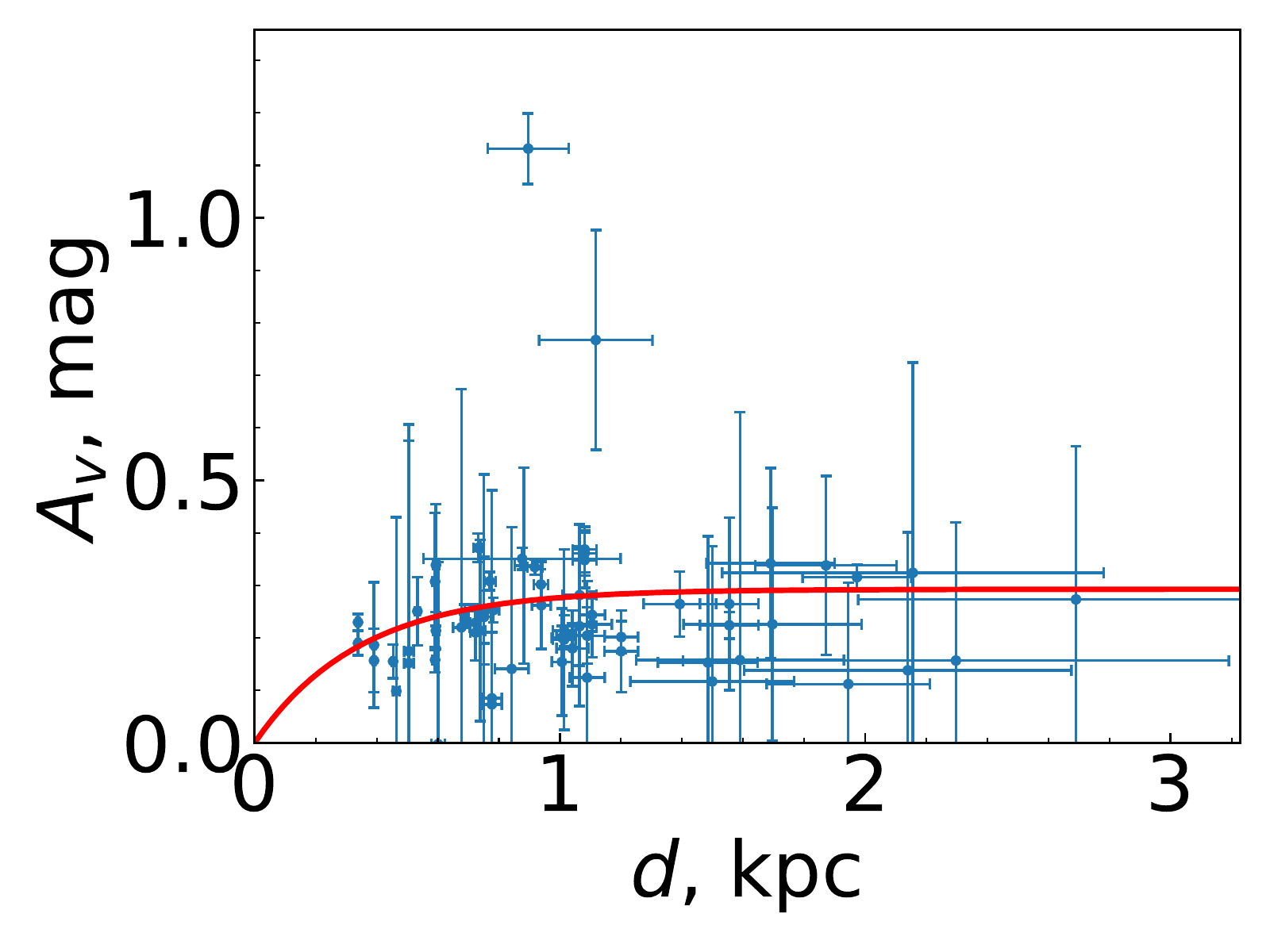}
  \includegraphics{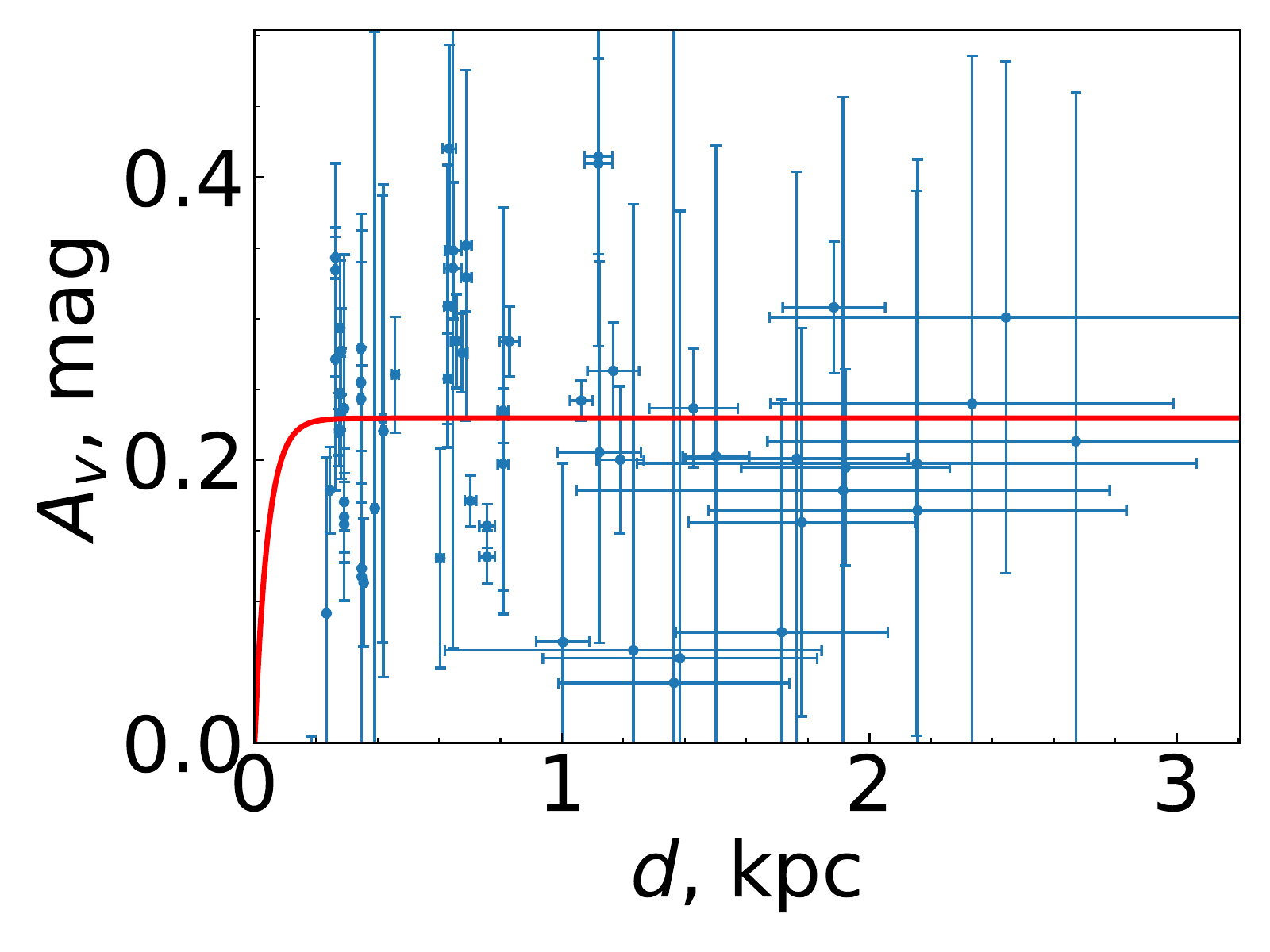}
  \includegraphics{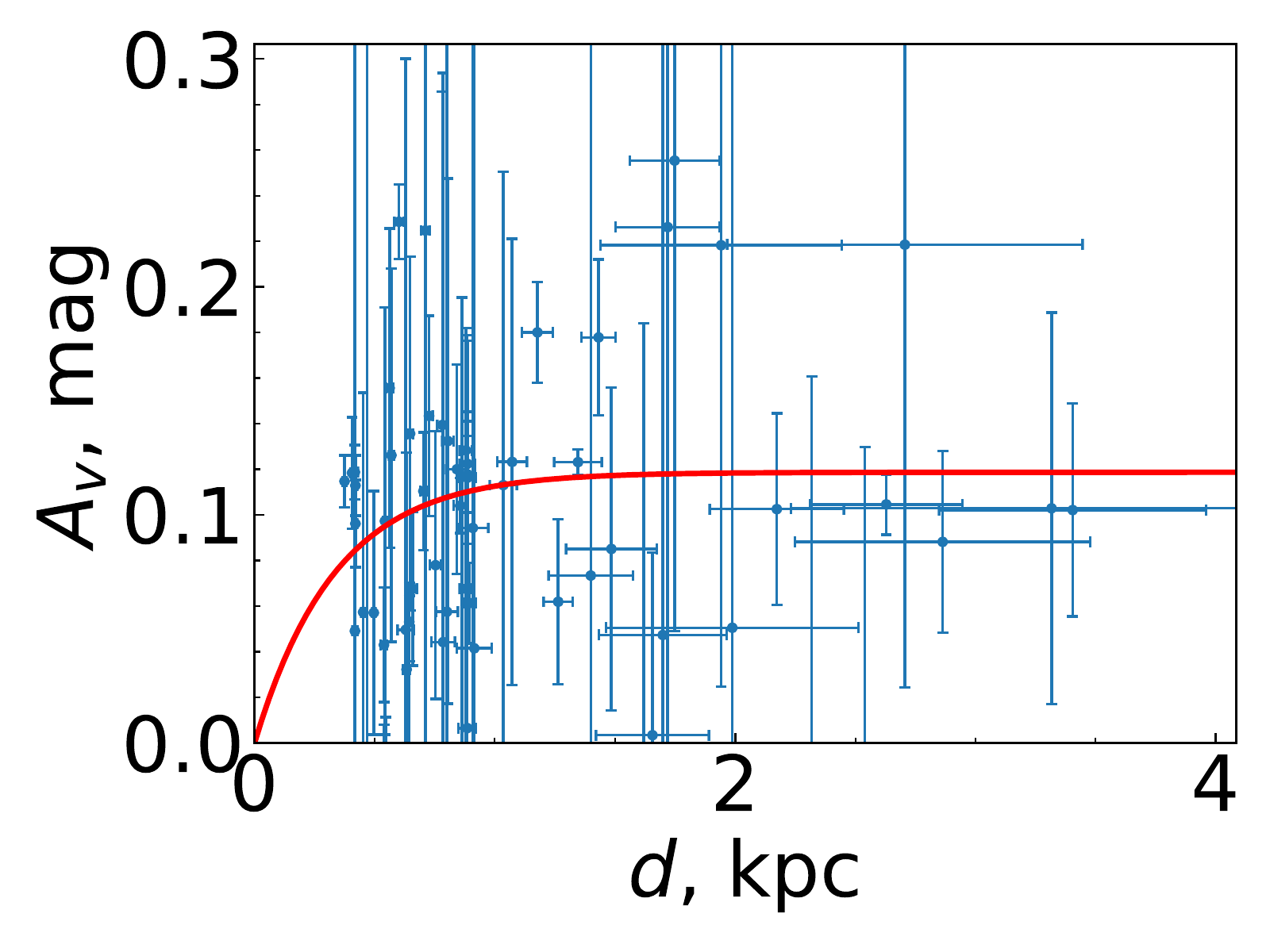}
 }
\caption{Best-fit for 136719, 137487, 138857.}
\end{figure}

\begin{figure}[h]
\resizebox{1.0\columnwidth}{!}{%
  \includegraphics{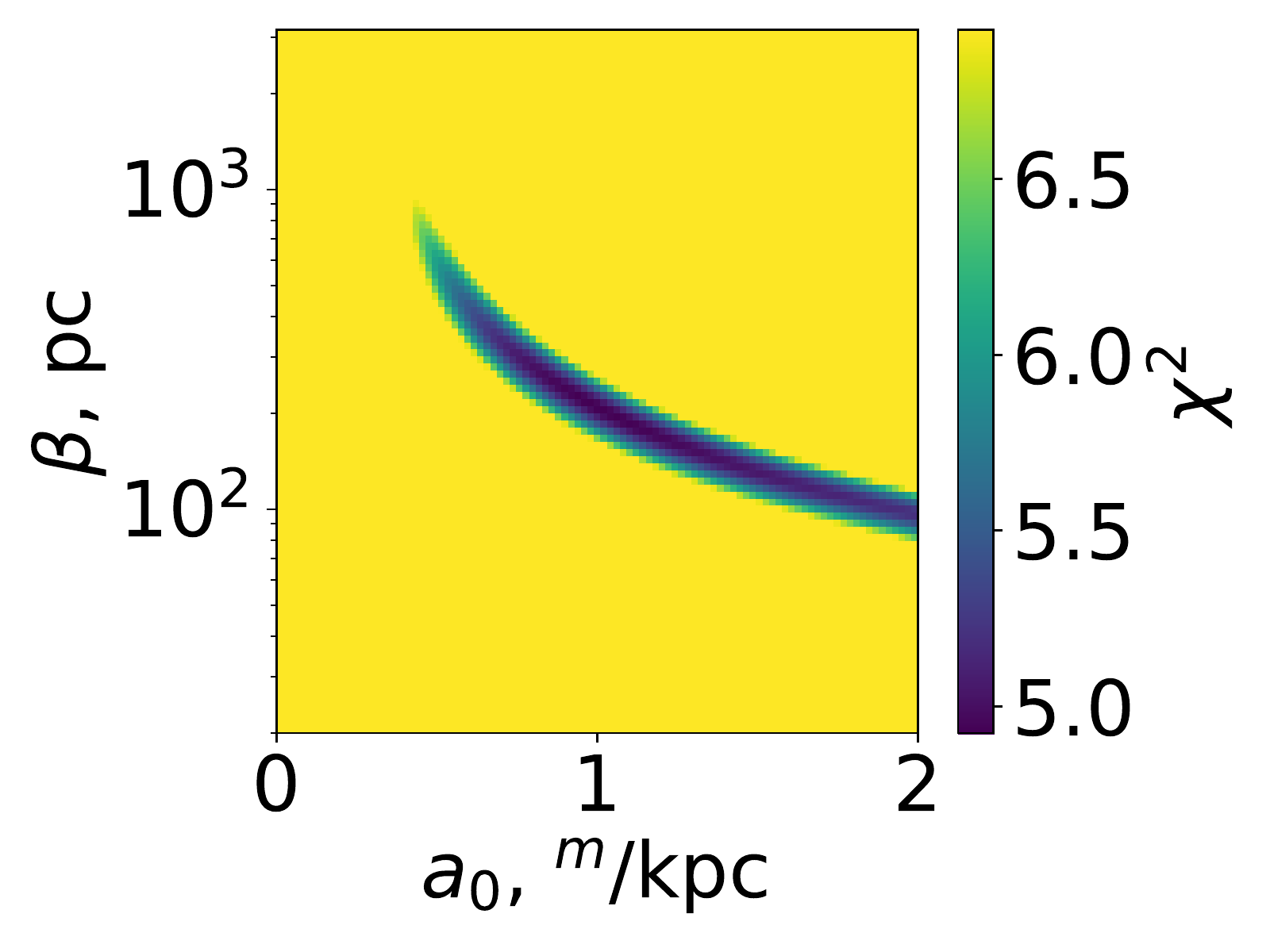}
  \includegraphics{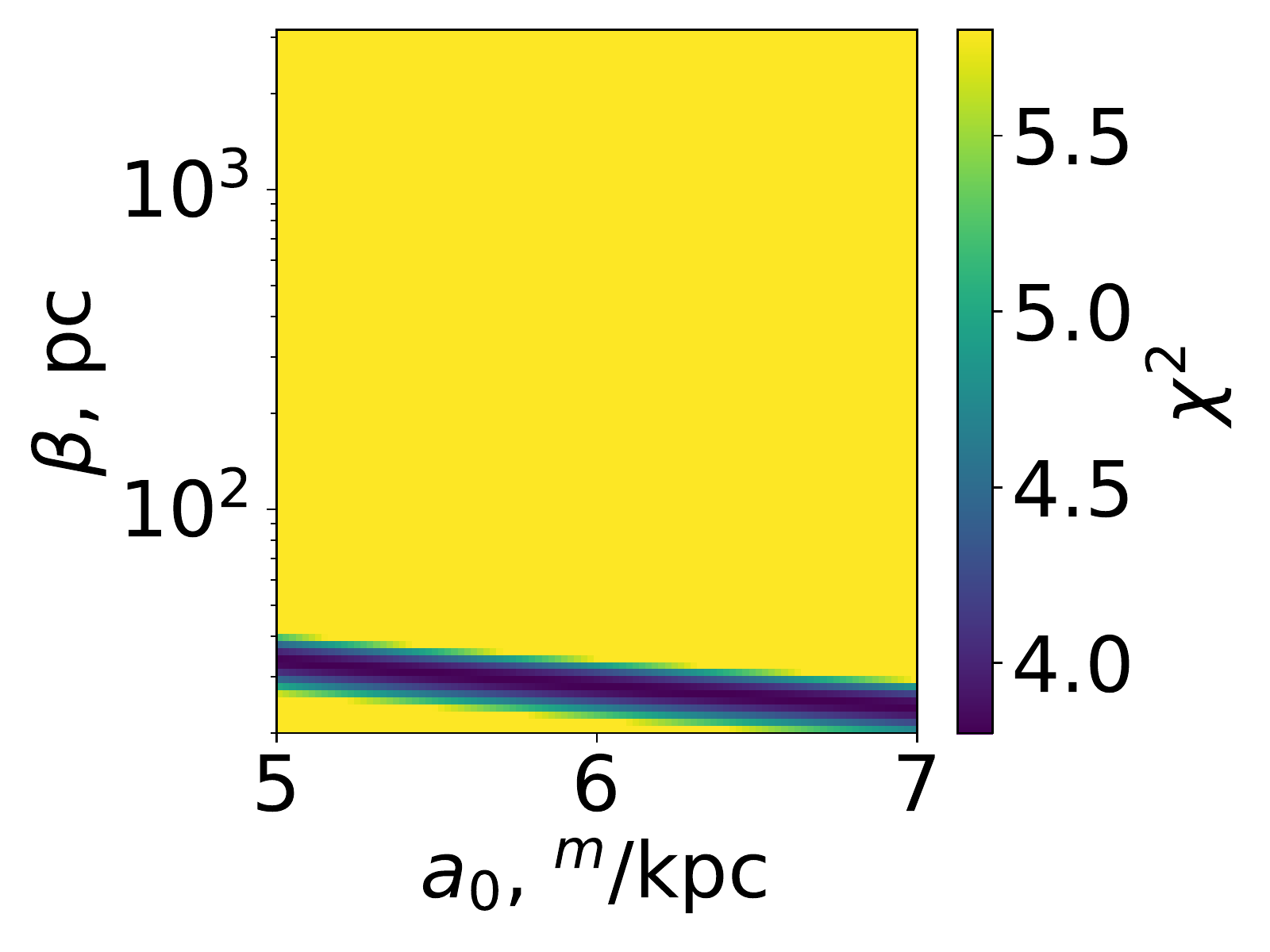}
  \includegraphics{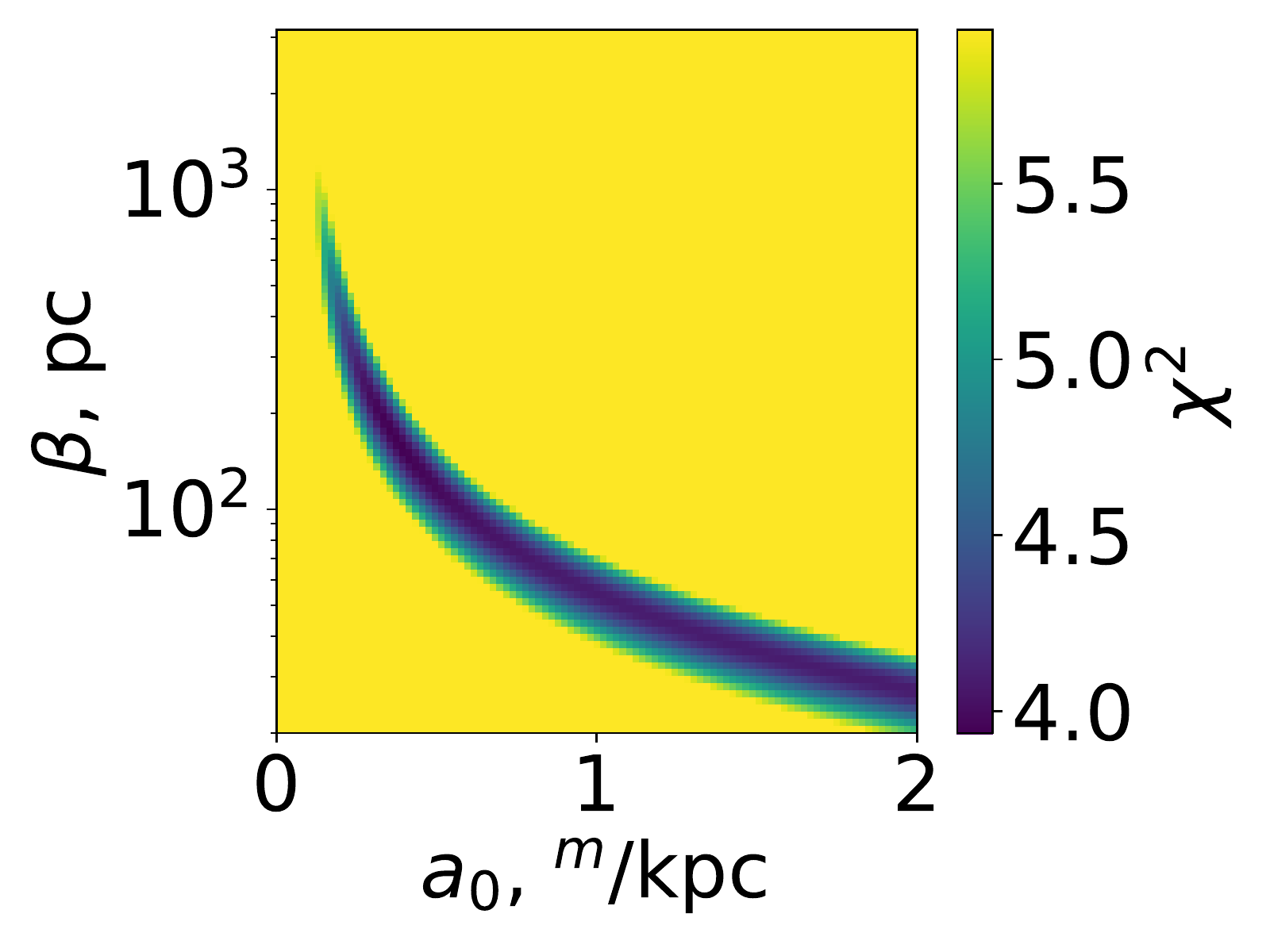}
 }
\caption{$\chi^2$ scan for 136719, 137487, 138857.}
\end{figure}

\begin{figure}[h]
\resizebox{1.0\columnwidth}{!}{%
  \includegraphics{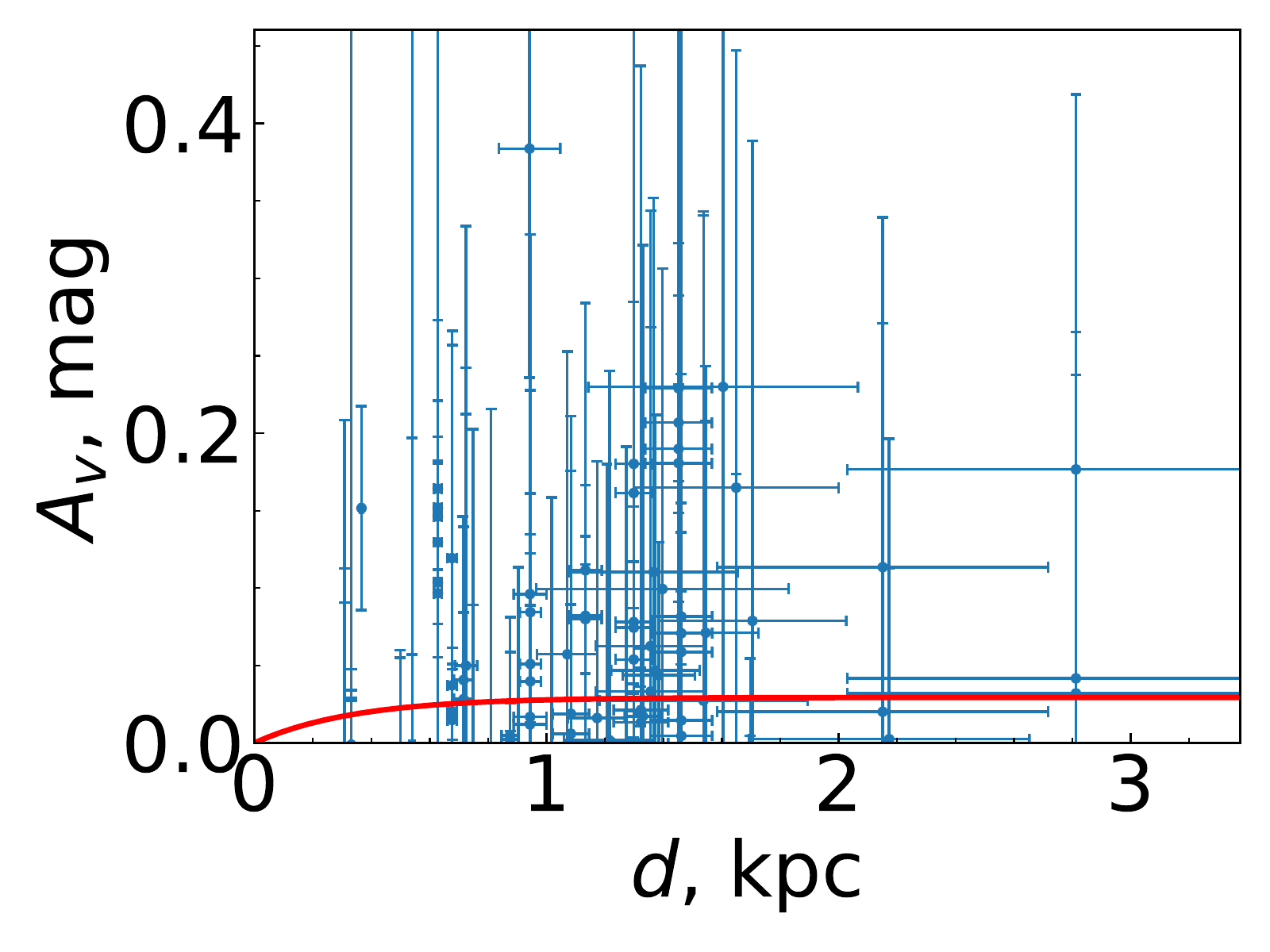}
  \includegraphics{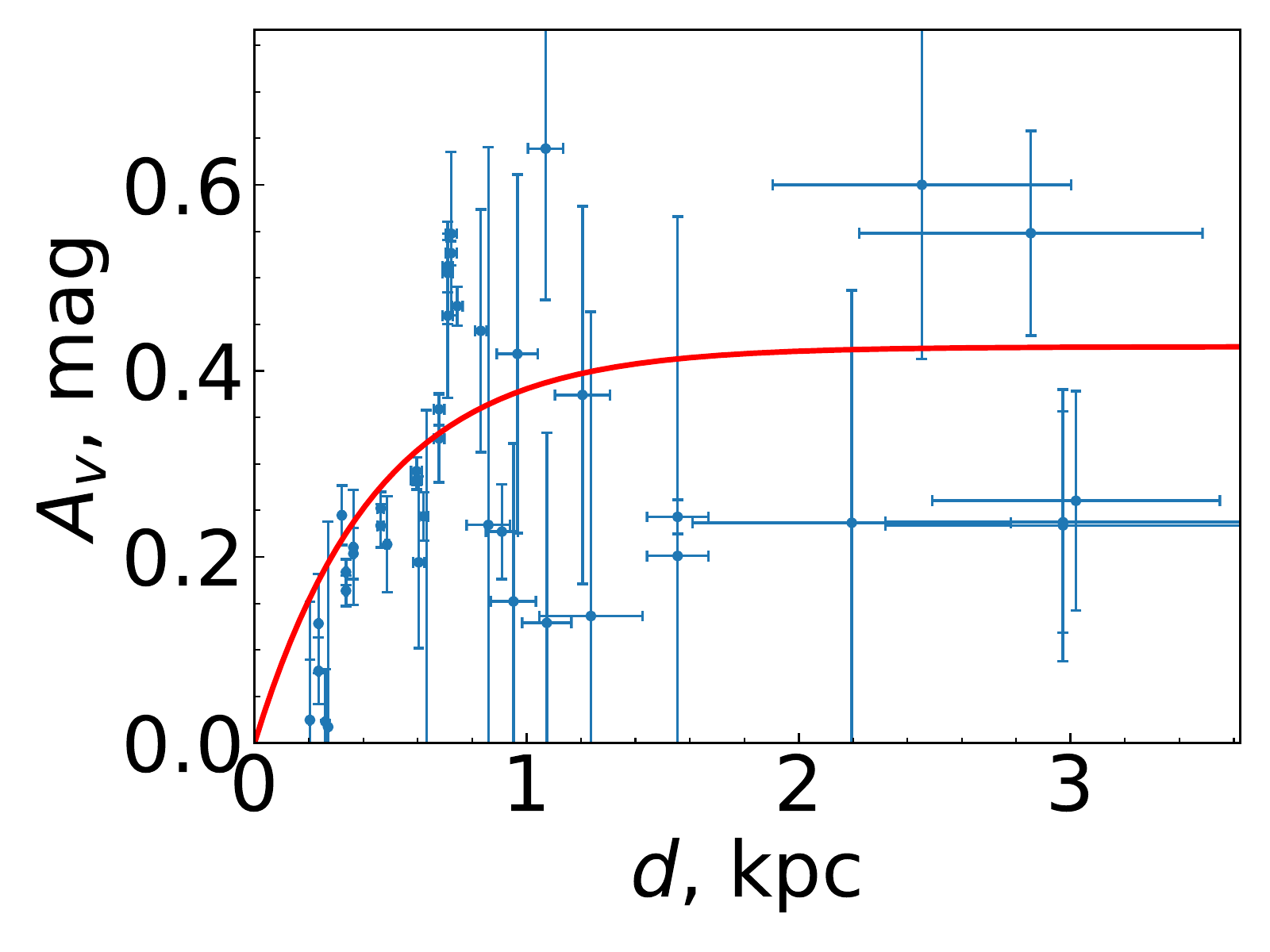}
  \includegraphics{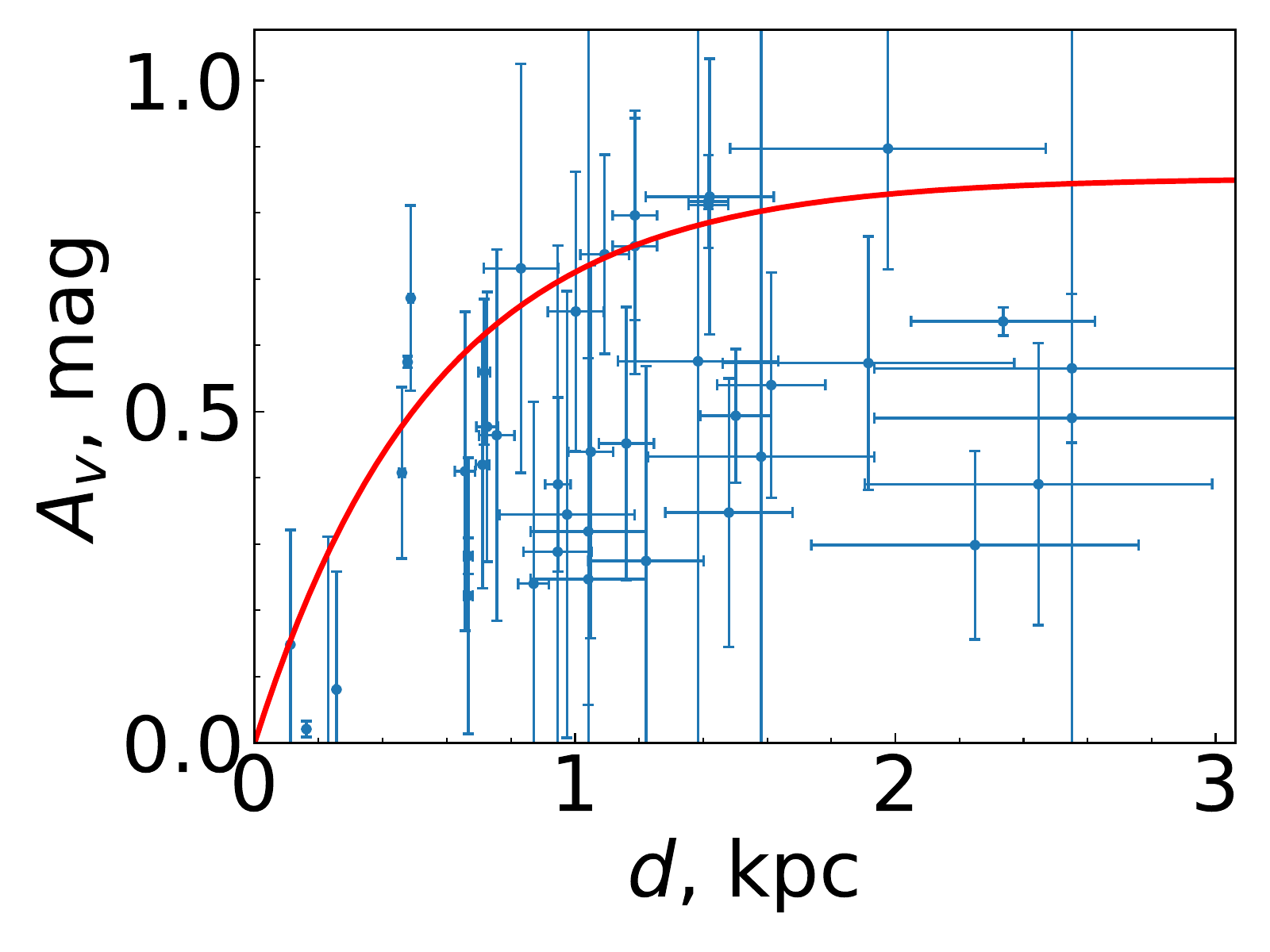}
 }
\caption{Best-fit for 149151, 151503, 152806.}
\end{figure}

\begin{figure}[h]
\resizebox{1.0\columnwidth}{!}{%
  \includegraphics{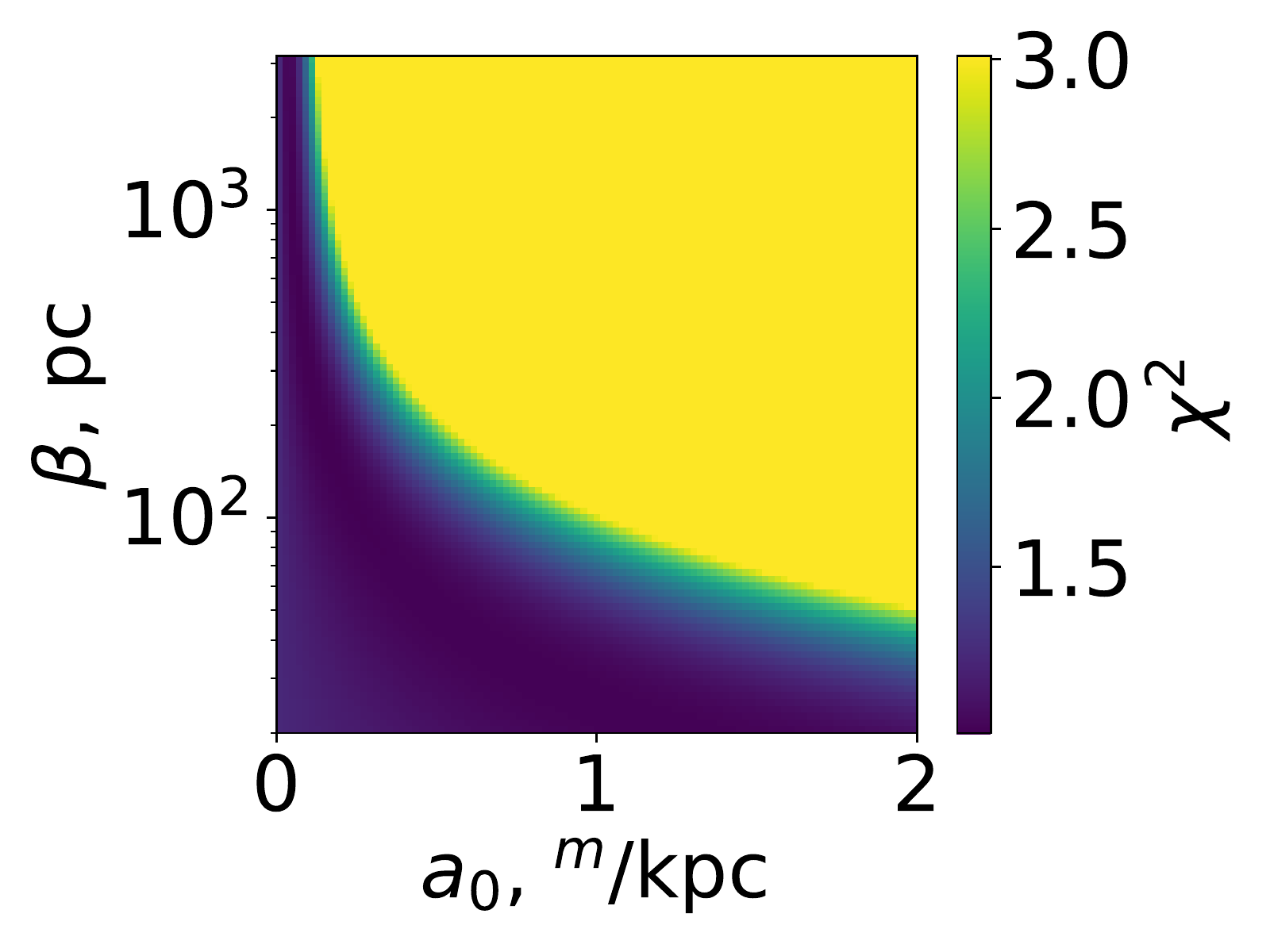}
  \includegraphics{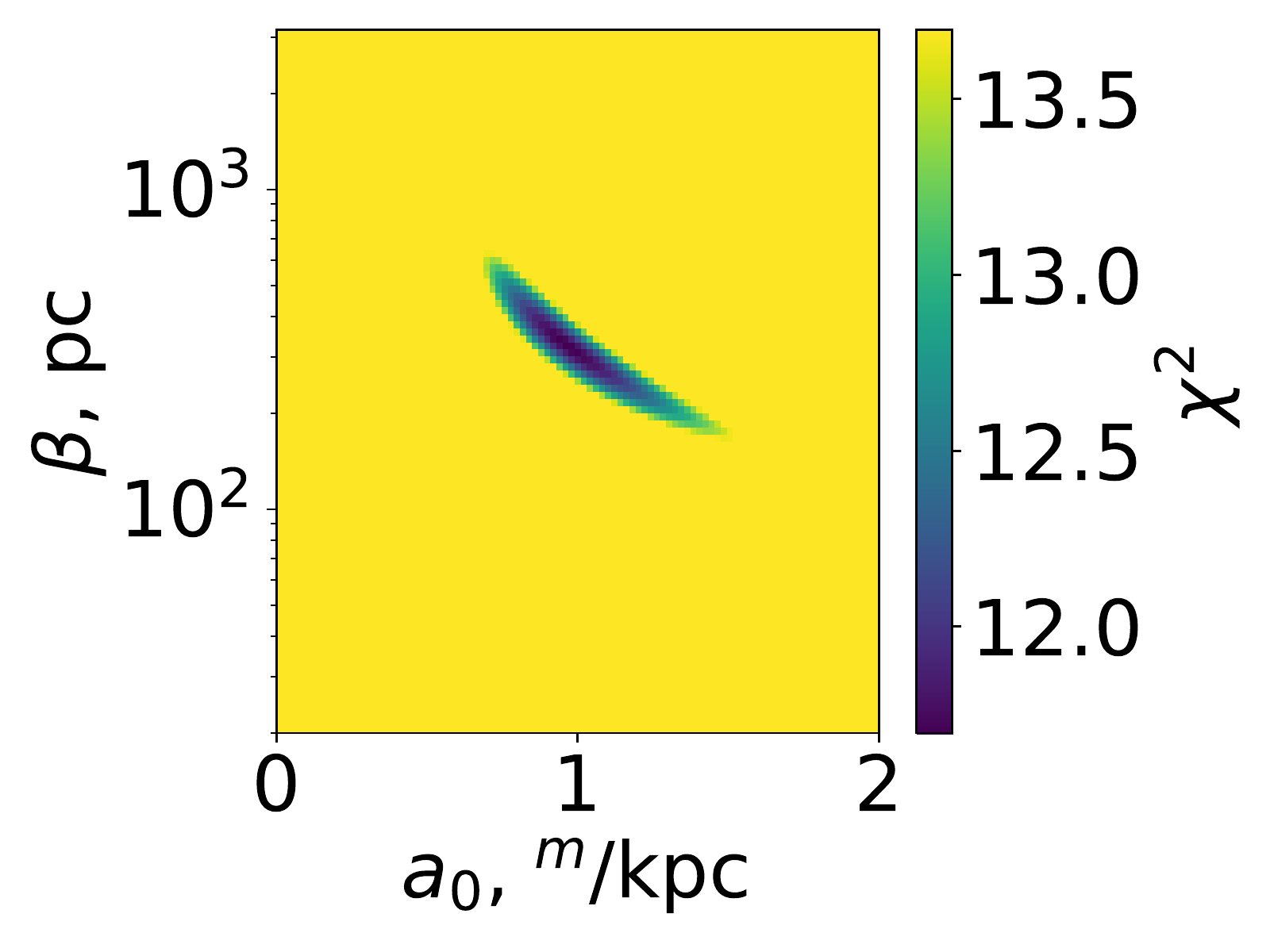}
  \includegraphics{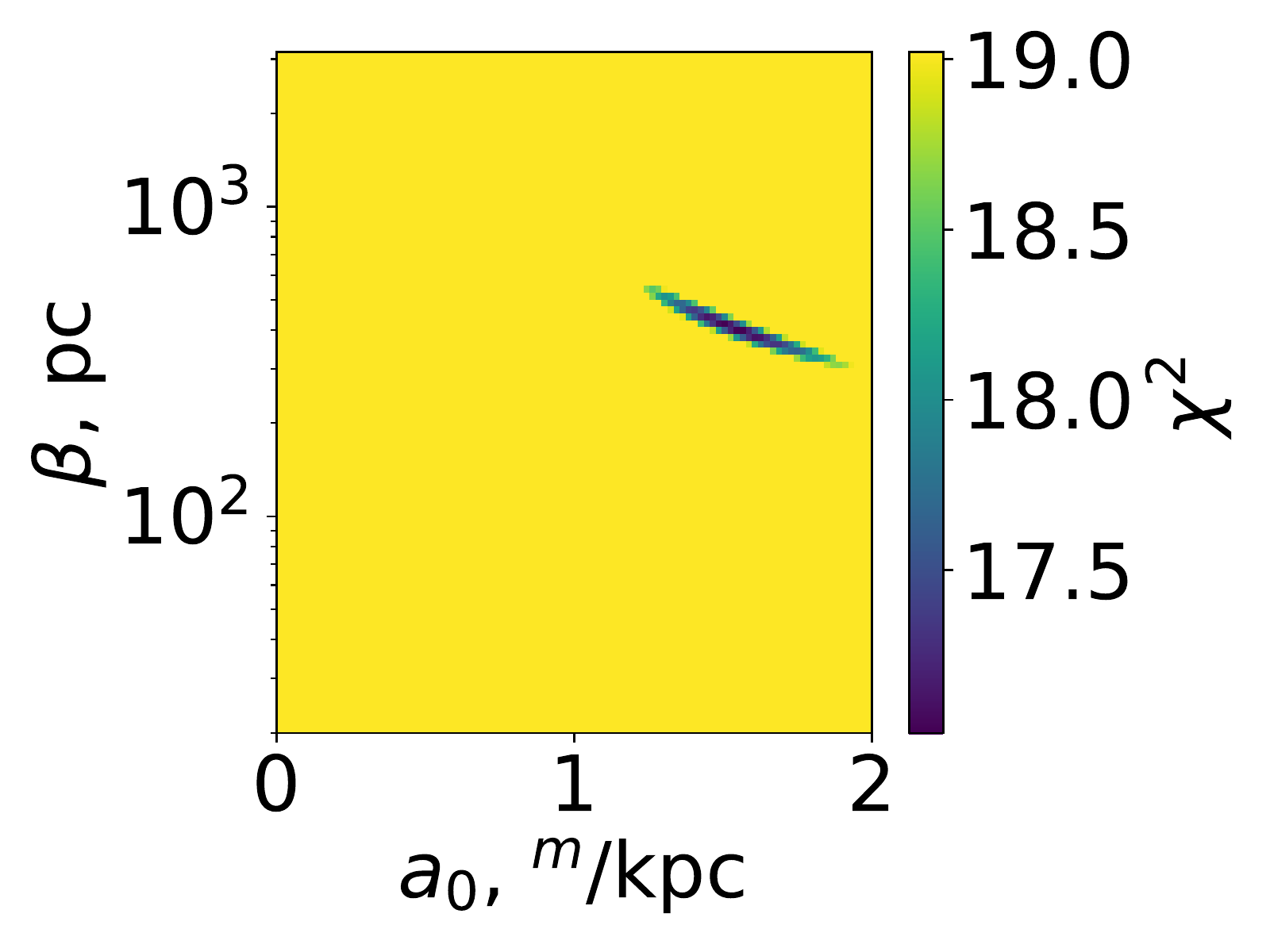}
 }
\caption{$\chi^2$ scan for 149151, 151503, 152806.}
\end{figure}

\begin{figure}[h]
\resizebox{1.0\columnwidth}{!}{%
  \includegraphics{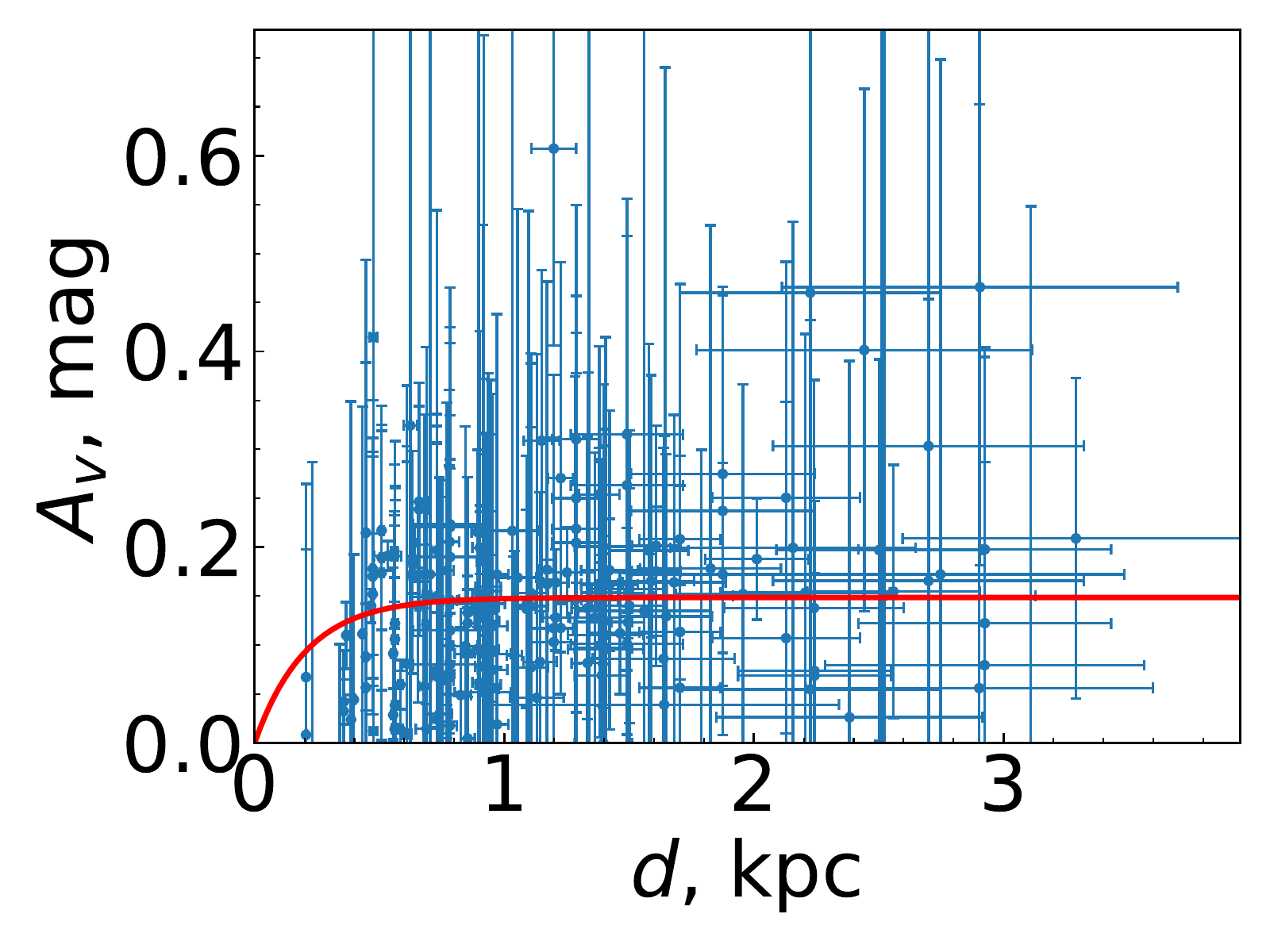}
  \includegraphics{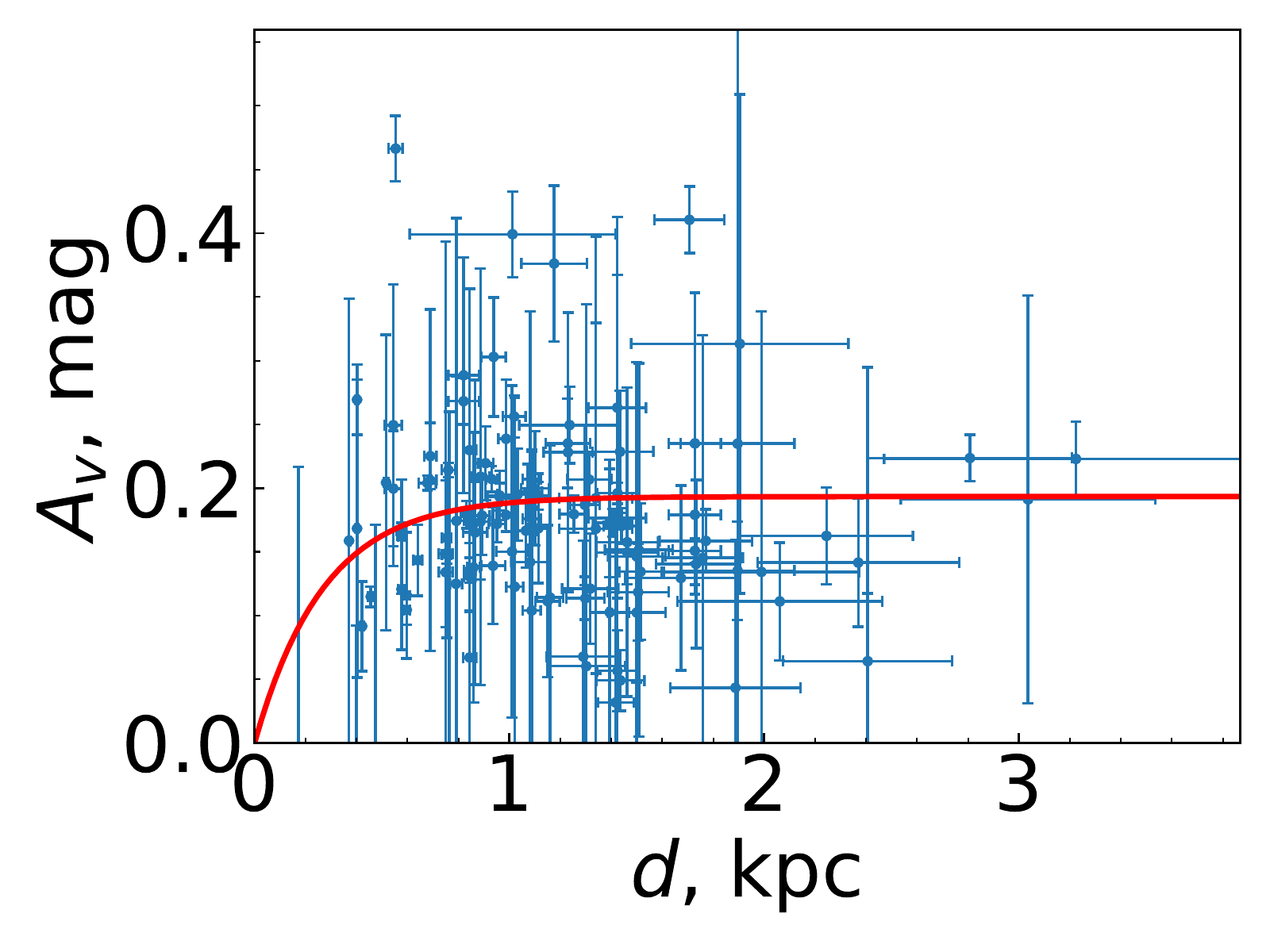}
  \includegraphics{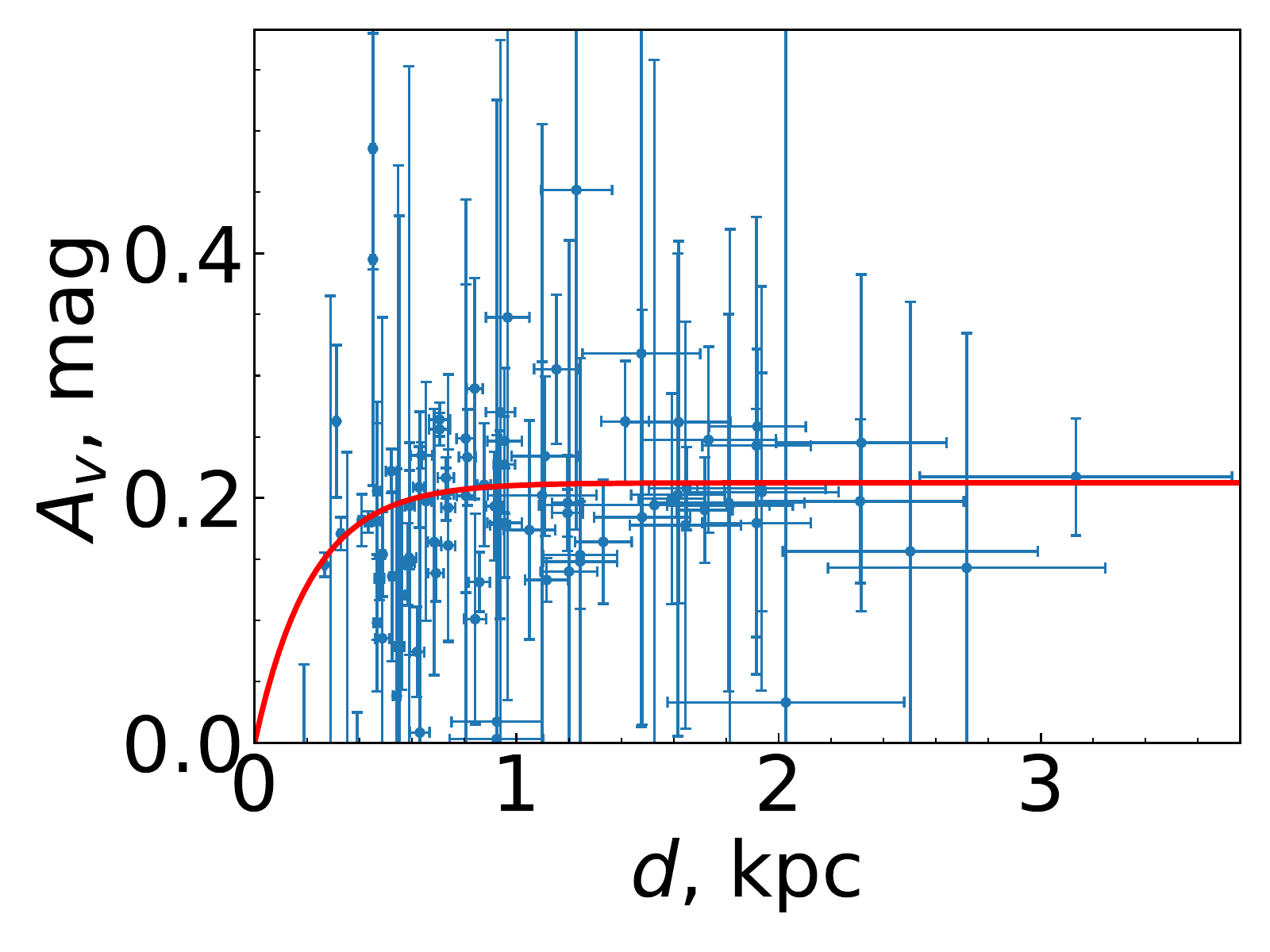}
 }
\caption{Best-fit for 160342, 161428, 162401.}
\end{figure}

\begin{figure}[h]
\resizebox{1.0\columnwidth}{!}{%
  \includegraphics{160342_chi2_map.pdf}
  \includegraphics{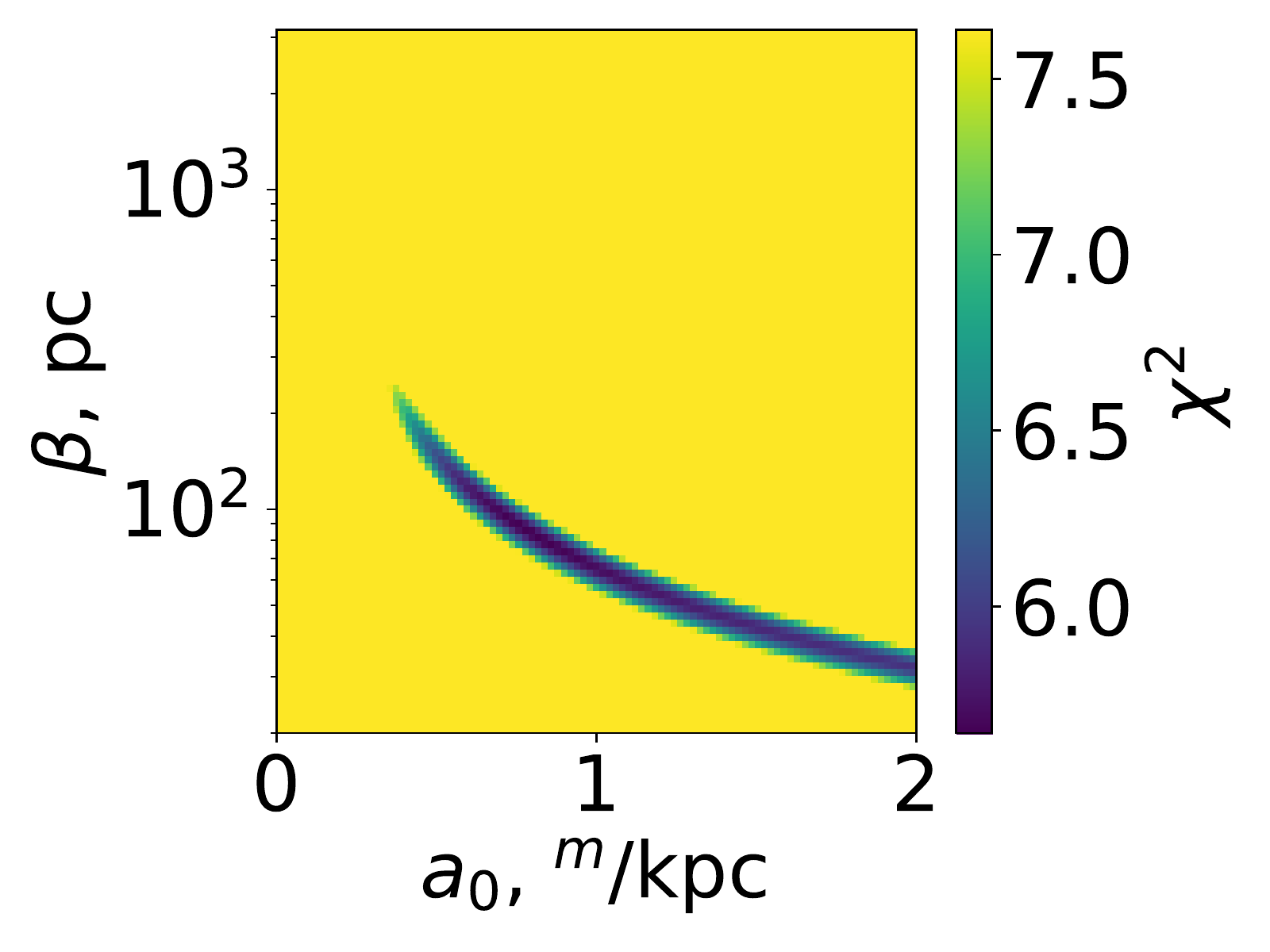}
  \includegraphics{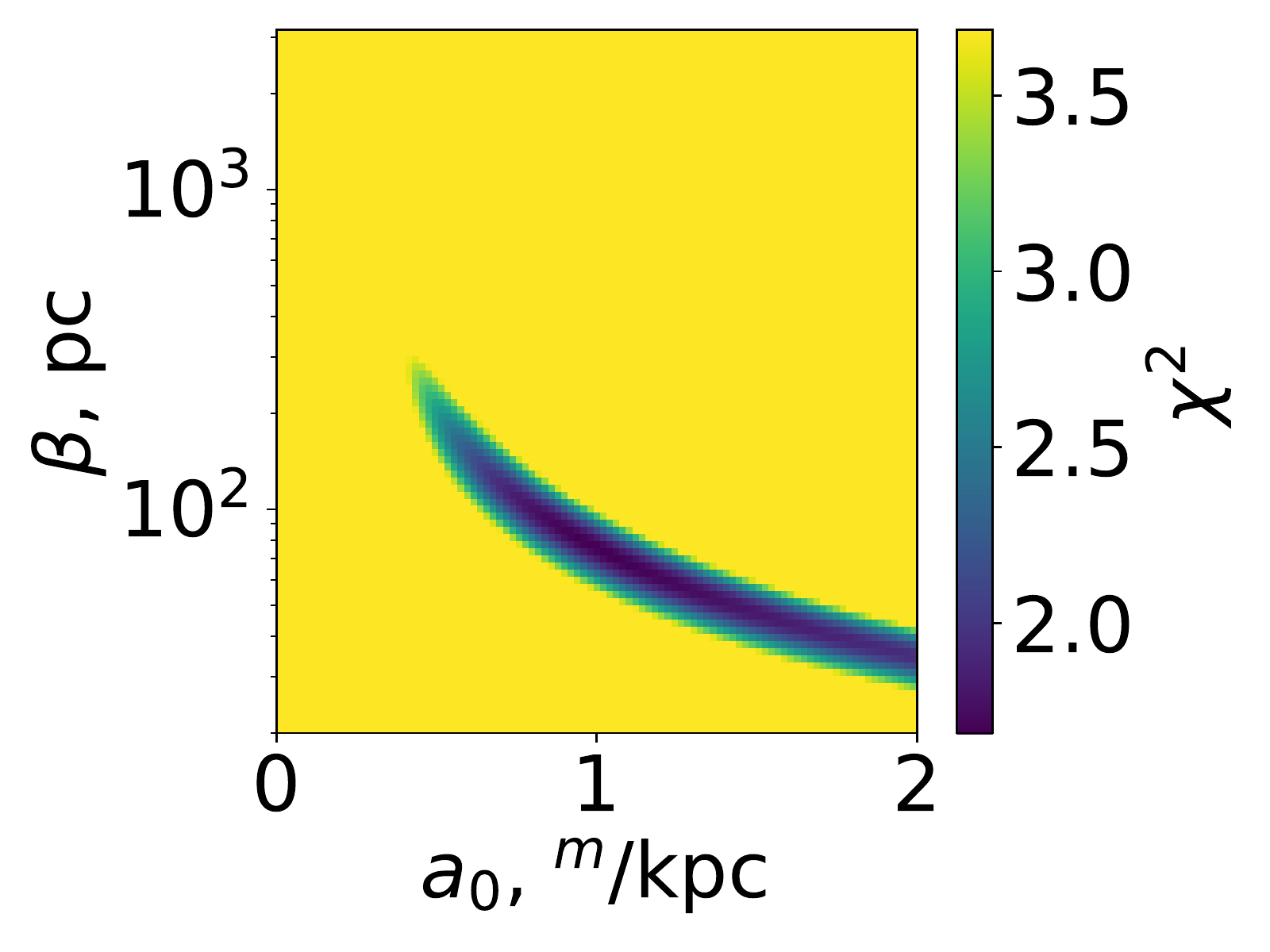}
 }
\caption{$\chi^2$ scan for 160342, 161428, 162401.}
\end{figure}

\begin{figure}[h]
\resizebox{0.66\columnwidth}{!}{%
  \includegraphics{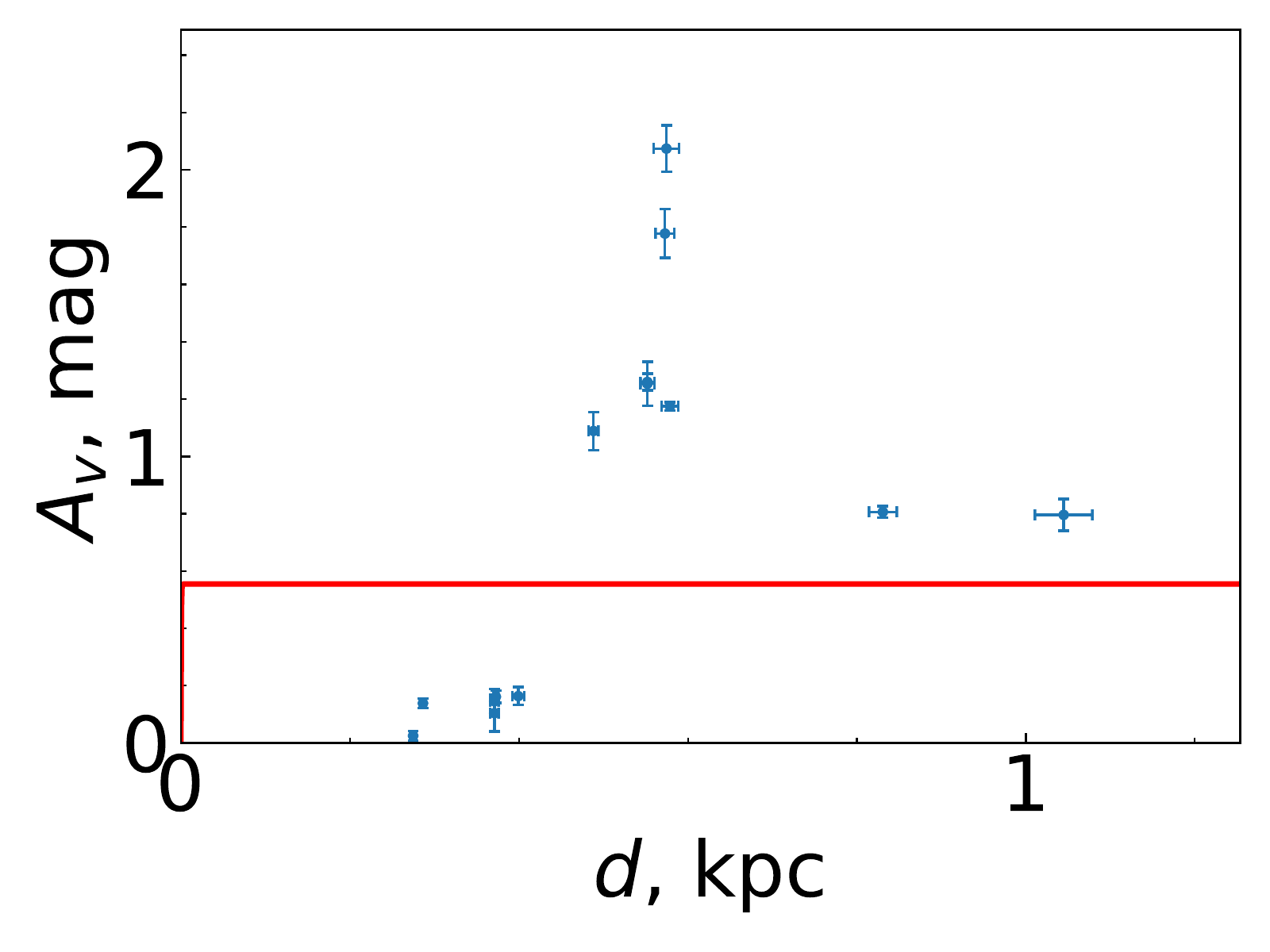}
  \includegraphics{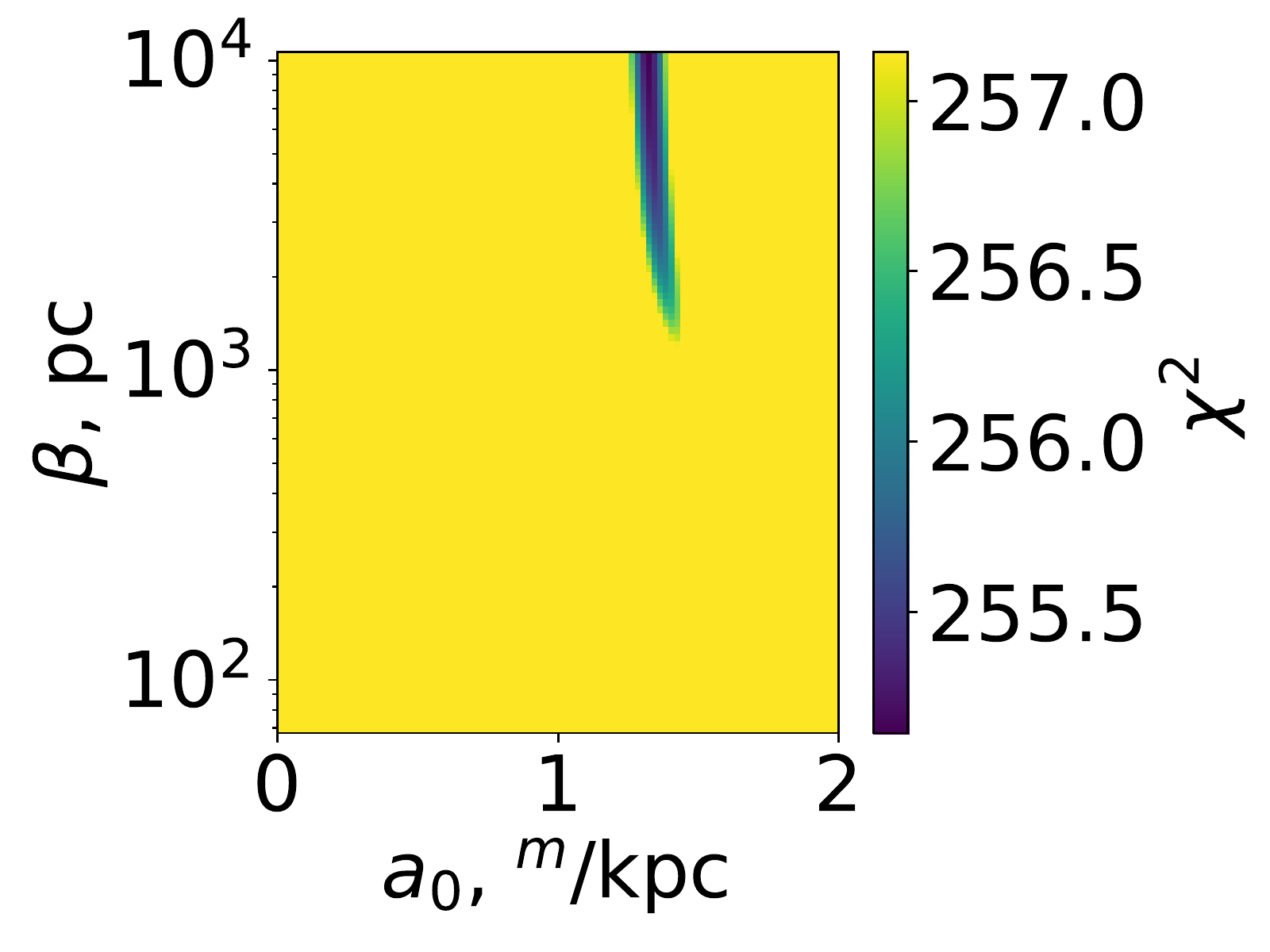}
 }
\caption{Best-fit and $\chi^2$ scan for 178840.}
\end{figure}


\end{appendix}

\end{document}